\documentclass[12pt]{article} 
\usepackage{amsfonts,amssymb,latexsym,amsmath, amsxtra, eufrak, hyperref} 
\usepackage{epsf,epsfig} 
 
\makeatletter 
\renewcommand\section{\@startsection {section}{1}{\z@}% 
                                   {-3.5ex \@plus -1ex \@minus -.2ex}%nn 
                                   {2.3ex \@plus.2ex}% 
                                   {\normalfont\large\bfseries}} 
\renewcommand\subsection{\@startsection{subsection}{2}{\z@}% 
                                     {-3.25ex\@plus -1ex \@minus -.2ex}% 
                                     {1.5ex \@plus .2ex}% 
                                     {\normalfont\bfseries}} 
 
\makeatother 
 
\let\non\nonumber 
 
\newcommand{\sectiono}[1]{\section{#1}\setcounter{equation}{0}} 
 
\newcommand{\bea}{\begin{eqnarray}} 
\newcommand{\eea}{\end{eqnarray}} 
\newcommand{\be}{\begin{equation}} 
\newcommand{\ee}{\end{equation}}

\newcommand{\sgn}{\mathrm{sgn}}

\newcommand{\im}{\mathrm{Im}}

\let\non\nonumber

\newcommand{\bF}{\mathbb{F}} 
\newcommand{\bP}{\mathbb{P}} 
\newcommand{\bZ}{\mathbb{Z}}  
 
\newcommand{\half}{\textstyle{\frac{1}{2}}} 
\newcommand{\ba}{{\bf a}} 
\newcommand{\bb}{{\bf b}} 
\newcommand{\bc}{{\bf c}} 
\newcommand{\bd}{{\boldsymbol d}} 
 
\newcommand{\bg}{{\boldsymbol g}} 
\newcommand{\bfk}{\mathbf{k}}   
\newcommand{\bmu}{{\boldsymbol \mu}}

\newcommand{\CL}{{\cal L}} 
\newcommand{\CM}{{\cal M}} 
\newcommand{\CN}{{\cal N}} 
\newcommand{\CO}{{\cal O}}

\begin{document} 
\begin{titlepage} 
 
\begin{center}

%\hfill 
 
\vskip 2 cm  
{\Large \bf Quantum geometry of elliptic Calabi-Yau manifolds}

\vskip 
1.25 cm {Albrecht Klemm$^1$, Jan Manschot$^{1,2}$ and Thomas Wotschke$^1$}\\ 
{\vskip 0.5cm $^1$Bethe Center for Theoretical Physics, Physikalisches Institut,\\ Universit\"at Bonn, 53115 Bonn, Germany \\}
{\vskip .5cm $^2$Max Planck Institute for Mathematics, 53111 Bonn, Germany}

\end{center} 
 
\vskip 2 cm 
 
\begin{abstract} 
\baselineskip=18pt 
 
We study the quantum geometry of the class of Calabi-Yau 
threefolds, which are elliptic fibrations over a 
two-dimensional toric base. A holomorphic anomaly equation 
for the topological string free energy is proposed, which is iterative 
in the genus expansion as well as in the curve classes in the base. $T$-duality on the fibre implies that the topological string free energy also captures 
the BPS-invariants of $D4$-branes wrapping the elliptic fibre and a class in the 
base. We verify this proposal by explicit computation of the BPS 
invariants of $3$ $D4$-branes on the rational elliptic surface.     

\end{abstract}

\end{titlepage} 
 
\pagestyle{plain} 
\baselineskip=19pt

%%%%%%%%%%%%%%%%%%%%%%%%%%%%%%%%%%%%%%%%%%%%%%%%%%%%%%%%%%%%%%%%%%%%%%%%%%%%%% 
\tableofcontents 
\newpage 
\section{Introduction}

\setcounter{equation}{0} 
 
Topological string theory on local Calabi-Yau manifolds has  
been a remarkable success story. It counts the open and closed  
instantons corrections to topological numbers, which can be seen  
as an extension from classical geometry to quantum geometry.  
By now we can solve it in very different ways, namely by localisation,  
by direct integration of the holomorphic anomaly equations, by the topological  
vertex \cite{Aganagic:2003db} or by the matrix model techniques in the
remodeled B-model \cite{Bouchard:2007ys}.  
The system gives deep insights in the interplay between large  
N gauge theory/string theory duality, mirror duality, the theory  
of modular forms and knot theory and is by geometric 
engineering~\cite{Katz:1996fh} intimately  
related to the construction of effective $N=2$ and $N=1$ rigid  
supersymmetric theories in four dimension. 
 
On global Calabi-Yau manifolds, which give rise to $N=2$ and $N=1$  
effective supergravity theories in four dimensions, the situation is  
less understood. Direct integration extends the theory of  
modular objects to the Calabi-Yau spaces and establishes  
that closed topological string amplitudes can be written as  
as polynomial in modular objects, but the boundary  
conditions for the integration are differently  
than in the local case not completely known. As an example,  
on the quintic surface the closed topological string  
can be solved up to genus 51~\cite{Huang:2006hq}.  
 
In~\cite{Chiang:1999tz} mirror symmetry was made local  
in the decompactification limit of Calabi-Yau threefolds.  
Here we want to do the opposite and study how the quantum  
geometry extends from the local to the global case, when a  
class of local Calabi-Yau geometries is canonically  
compactified by an elliptic Calabi-Yau fibration  with  
projection $\pi:M\rightarrow B$. This easy class of  
local to global pairs, will be described   
to a large extend by complete intersections in explicit  
toric realizations.  As we review in section \ref{classgeom},  
if the elliptic fibration has only $I_1$ fibres the classical cohomology of $M$ is completely  
determined by the classical intersection of the  
base $B$ and the number of sections, which  
depends on the Mordell Weyl group of the elliptic  
family.   
 
The decisive question to which extend this holds for the quantum  
geometry is addressed in section \ref{quantgeom} using mirror  
symmetry. The instanton numbers are counted by (quasi)-modular 
forms of congruence subgroups of ${\rm SL}(2,\mathbb{Z})$ capturing 
curves with a fixed degree in the base for all degrees 
in the fibre. The weights of the forms depend on the genus 
and the base class. This structure has been discovered  
for elliptically fibred surfaces in~\cite{Klemm:1996hh}  
and for elliptically fibred threefolds in~\cite{Klemm:2004km}.  
We establish here a holomorphic anomaly equation (\ref{eq:holannew}) 
based on the non-holomorphic modular completion of the quasimodular 
forms which is iterative in the genus, as in \cite{Bershadsky:1993cx}, 
and also in the base classes generalizing~\cite{Minahan:1998vr, Hosono:1999qc}.

Our construction can be viewed also as a step to  a better 
understanding of periods and instanton corrections in F-theory 
compactifications and a preliminary study using the data 
of~\cite{Mayr:1996sh}\cite{Klemm:1996ts}\cite{Klemm:2007in} 
reveals that the structure at the relevant generera 
$g=0,1$ extends.

A holomorphic anomaly equation is also known to appear for generating
functions of BPS invariants of higher dimensional $D$-branes, in 
particular $D4$-branes on a surface \cite{Vafa:1994tf,
Manschot:2009ia, Alim:2010cf, Manschot:2011dj}. Interestingly, on
elliptic Calabi-Yau fibrations, double $T$-duality on the elliptic fibre (or Fourier-Mukai
transform) \cite{Minahan:1998vr, Andreas:2000sj,Andreas:2001ve, Bena:2006qm} transforms
 $D2$-branes wrapped on base classes into $D4$-branes which also wrap
 the elliptic fibre and vice versa. 
The $D4$-brane holomorphic anomaly is therefore related to
the one of Gromov-Witten theory for these geometries. Moreover, 
the mirror periods provide predictions for $D4$-brane BPS
invariants which correspond to those of (small) black holes in supergravity.

We discuss higher dimensional branes on
Calabi-Yau elliptic fibrations in sections \ref{sec:sugrapartition} and
\ref{sec:ellsurface}. We compare the predictions from the periods for
$D4$-brane BPS invariants with existing methods in the literature for
the computation of small charge BPS invariants
\cite{Gaiotto:2006wm,Gaiotto:2007cd,Denef:2007vg, Collinucci:2008ht, 
  deBoer:2008zn, Manschot:2010qz, Manschot:2011xc}. The predictions of the
periods are in many cases compatible with these methods. We leave a
more precise study of $D4$-brane BPS states on general elliptic fibrations to future work.
 
Section \ref{sec:ellsurface} specializes to the elliptic fibration
over the Hirzebruch surface $\mathbb{F}_1$. The periods of its mirror
geometry provide the BPS invariants of $D4$-branes on the rational elliptic
surface (also known as $\half K_3$) as proposed originally by
Minahan {\it et al.} \cite{Minahan:1998vr}. We revisit and extend the verification of this proposal
for $\leq 3$ $D4$-branes using algebraic-geometric techniques
\cite{Gottsche:1990, Yoshioka:1994, Yoshioka:1996, Gottsche:1996, Manschot:2010nc,
  Manschot:2011dj, Manschot:2011ym}.

\subsection*{Acknowledgements}
\vspace{-.4cm}
We would like to thank Babak Haghighat for useful discussions.
Also we would  like to thank Marco Rauch for collaboration
in an initial state of the project. Part of the work of J. M. 
was carried out as a postdoc of the Institute de 
Physique Th\'eorique of the CEA Saclay. 
A. K. and T.W. are grateful to acknowledge support by the 
DFG to the project KL2271/1-1. T.W. is supported by the 
Deutsche Telekom Stiftung.  We would thank Murad Alim and 
Emanuel Scheidegger for informing us about their related 
work.

\section{Classical geometry of elliptic fibred Calabi-Yau spaces} 
\label{classgeom} 
 
In this section we study the classical geometry of elliptically  
fibered Calabi-Yau three manifolds $M$ with base $B$ and projection map  
$\pi:M\rightarrow B$. Such  elliptic fibrations might  
be described locally by a Weierstrass form  
\begin{equation}  
 y^2= 4 x^3 - x w^4 g_2({\underline u}) - g_3({\underline u}) w^6,  
\label{weierstrass} 
\end{equation}where ${\underline u}$ are coordinates on the base $B$.  
A global description can be defined by an embedding  
as a hypersurface or complete intersection in an ambient  
space $W$. Explicitly we consider cases, which allow a  
representation as a hypersurface or complete intersection  
in a toric ambient space. We restrict our attention to the case   
where the fiber degenerations are only of Kodaira type $I_1$,  
which means that the discriminant  $\Delta=g_2^3-27 g_3^2$ of (\ref{weierstrass})  
has only simple zeros on $B$, which are not simultaneously zeros 
of $g_2$ and $g_3$. Of course this is  
not enough to address immediately phenomenological interesting  
models in F-theory. However we note that these examples have a  
particular large number of complex moduli. Adjusting the latter  
and blowing up the singularities, not necessarily torically,   
is a more local operation, i.e. at least of co-dimemsion one in the  
base, which can be addressed in a second step.

\subsection{The classical geometrical data of elliptic fibrations}    
Denote by $\mathbb{P}^2(w_1,\ldots,w_r)$ a (weighted)  
projective bundle $W$ over the base $B$. We consider  
four choices of weights $(w_1,\ldots,w_r)= 
\{(1,2,3),(1,1,2),(1,1,1),(1,1,1,1)\}$ leading to  
three hypersurfaces and one complete intersection.  
In the case of rational elliptic surfaces these fiberes lead   
to $E_8, E_7,E_6$, and $D_5$ del Pezzo surfaces, named so as  the cohomology  
lattice of the surface has the intersection  form of  
the corresponding Cartan-matrix. We keep the names  
for the fibration types.     
 
Let us discuss the first case. This leads canonically to  
an embedding with a single section, however most of the discussion  
below applies to the other cases with minor  
modifications. Denote by $\alpha={\cal O}(1)$ the line bundle on  
$W$ induced by the hyperplane class of the projective  
fibre and $K=-c_1$ the canonical bundle of the base.  
 
The coordinates $w,x,y$ are sections of  
${\cal O}(1)$, ${\cal O}(1)^{2}\otimes  K^{-2}$ and  
${\cal O}(1)^{3} \otimes K^{-3}$ while $g_2$ and $g_3$  
are section of $K^{-4}$ and $K^{-6}$ respectively   
so that (\ref{weierstrass}) is a section of  
${\cal O}(1)^{6} \otimes K^{-6}$. The corresponding  
divisors $w=0,x=0,y=0$ have no intersection, i.e.  
$\alpha( \alpha + c_1)( \alpha + c_1)=0$ in the  
cohomology ring of $W$ and  
\begin{equation} 
\alpha( \alpha +  c_1)=0 
\label{relationdivs} 
\end{equation} 
in the cohomology ring of $M$. Let us assume that the  
discriminant $\Delta$ vanishes for generic complex moduli  
only to first order in the coordinates of $B$ at locii, 
which are not simultaneously zeros of $g_2$ and $g_3$.  
In this case its class  must satisfy  
\begin{equation}  
[\Delta]=c_1(B)=-K\   
\end{equation} 
to obey the Calabi-Yau condition and the fiber  
over the vanishing locus of the discriminant is  
of Kodaira type $I_1$. For this generic fibration, 
the properties of $M$ depend only on the properties  
of $B$.

For example using the adjunction formula  
and the relation (\ref{relationdivs}) to reduce to linear  
terms in $\alpha$ allows to write the total Chern class as\footnote{In the $D5$ complete intersection case $d_1=d_2=2$.  
One has to add a factor $(1+ \alpha+ c_1)$ in the numerator and a factor $(1+2 \alpha+ 2 c_1)$ in the denominator.} 
\begin{equation}  
{\cal C} =\left(1+\sum_{i=1}^{n-1}c_i\right) \frac{(1+\alpha)(1+w_2 \alpha+ w_2 c_1) (1+w_3 \alpha+w_3 c_1)}{1 + d \alpha + d c_1}\ .      
\label{chern} 
\end{equation} 
The Chern forms ${\cal C}_k$ of $M$ are the coefficients  
in the formal expansion of (\ref{chern}) of the degree $k$ in terms of  
$a$ and the monomials of the Chern forms  $c_i$ of base $B$.  
 The formulas (\ref{relationdivs}) and  (\ref{chern}) apply for all projectivisations.  
 
\begin{center} 
\begin{table}          
$$\footnotesize 
  \begin{array}{|c|ccc|} 
  \hline 
 \text{Fibre}& {\cal C}_2  & {\cal C}_3 & {\cal C}_4 \\[1mm] 
  \hline  
 E8  &  12 \alpha c_1+(11 c_1^2+c_2)& -60 \alpha c_1^2-(60 c_1^3+c_2c_1-c_3)&  12 \alpha c_1(30 c_1^2+c_2)\\[1mm] 
 E7  &  6 \alpha c_1+(5 c_1^2+c_2)& -18 \alpha c_1^2-(18 c_1^3+c_2 c_1-c_3)&   6 \alpha c_1(12 c_1^2+c_2)\\[1mm] 
 E6  &  4 \alpha c_1+(3 c_1^2+c_2)&-8 \alpha c_1^2-(8 c_1^3+c_2 c_1-c_3)&4 \alpha c_1 (6 c_1^2+c_2)\\[1mm] 
 D5  &  3 \alpha c_1+(2 c_1^2+c_2)&-4 \alpha c_1^2-(4 c_1^3+c_2 c_1-c_3)&3 \alpha c_1 (3 c_1^2+c_2)\\[1mm] 
 \hline 
  \end{array} 
 $$ 
  \caption{\label{tab:elliptic} Chern classes ${\cal C}_i$ of regular  
  elliptic Calabi-Yau manifolds. Integrating $\alpha$  over the fibre  
  yields a factor $a=\frac{\prod_i {d_i}}{\prod_i{w_i}}$, i.e. the number of sections $1,2,3,4$ for the three  
  fibrations in turn.}                
\end{table} 
\end{center} 
 
For $n=2$ one gets from table \ref{tab:elliptic} by integrating over the fibre  
in all cases $\chi(M)=12 \int_B c_1$ and $\mathbb{P}^1$ is the only admissible base.  
Similar for $n=3$  one gets for the different projectivisations  $\chi(M)=  
-60 \int_B c_1^2$, $\chi(M)= -36 \int_B c_1^2$, $\chi(M)= -24 \int_B c_1^2$  
and $\chi(M)= -16 \int_B c_1^2$.

The following discussion extends to all dimensions. For the sake  
of brevity we specialize to Calabi-Yau threefolds.  
Let $K_i$ denote the K\"ahler cone and ${\cal C}_i$  
the classes of the curves in the cone dual  
of the two dimensional base. Let $K_i K_j=c_{ij}$   
be the intersection form on the base. 
We expand the canonical class of the base as  
\begin{equation}  
\label{def:aa} 
K=-c_1= -\sum_i a^i K_i= -\sum a_i C^i\ , 
\end{equation} 
with $a_i$ and $a^i$ in $\mathbb{Z}$.   
We denote by ${\cal K}_a$ the divisors of the total  
space of the elliptic fibration  and distinguish  
between ${\cal K}_e$ the divisor dual to the elliptic  
fibre curve and  ${\cal K}_i$, $i=1,\ldots,b$, which  
are $\pi^*(C^i)$   
\begin{equation}  
\begin{array}{rl} 
\label{eq:intersection1} 
{\cal K}_e^3&=a \int_B c_1^2,\\   
{\cal K}_e^2 {\cal K}_i&=a  a_i, \\ 
{\cal K}_e {\cal K}_i {\cal K}_j&= a c_{ij}\  .
\end{array}  
\end{equation} 
Here $a$ denotes the number of sections, see Tab.1.
The intersection with the second Chern class of the total  
space can be calculated using table 1 as  
\begin{equation} 
\label{eq:2ndchernint} 
\begin{array}{rl} 
\displaystyle{\int_M c_2 J_e}&=\left\{\begin{array}{rl}   
                           \int_B (11 c_1^2+ c_2) & \qquad E_8,\\ [1 mm] 
                         2 \int_B (5 c_1^2+ c_2) & \qquad E_7,\\[1 mm] 
                         3 \int_B (3 c_1^2+ c_2) & \qquad E_6,\\[1 mm] 
                         4 \int_B (2 c_1^2+ c_2) & \qquad D_5,\\ [1 mm] 
                 \end{array}\right. \\  
                 &\\ 
\displaystyle{\int_M c_2 J_i}&= 12 a_i.  
\end{array} 
\end{equation}  
Here we denoted by $J_i$ the basis of harmonic $(1,1)$ forms  
dual to  the ${\cal K}_i$. 
 
Let us note two properties about the intersection numbers. They can  
be proved using the properties of almost Fano bases $B$ and  
$(\ref{eq:intersection1})$. For the first define the matrix 
\begin{equation} 
C_e=\left( 
\begin{array}{cc}  
\int_B c_1^2& a_1,\ldots,a_b\\ 
a_1& \\ 
\vdots&   c_{ij} \\  
 a_b&  
\end{array}\right)\ , 
\end{equation} 
then 
\begin{equation} 
\label{noninvertible} 
{\rm det}(C_e)=0\ . 
\end{equation} 
A further property concerns a decoupling limit between base  
and fibre in the K\"ahler moduli space.  Generally we   
can make a linear change in the basis of Mori vectors, which  
results in corresponding linear change in dual spaces of the K\"ahler  
moduli and the divisors 
\begin{equation} 
\label{noninvertible} 
\tilde l_i= m_{ij} l_j,\qquad \tilde t_i=m^T_{ij} t_j\ . 
\end{equation} 
To realize a decoupling between the base and the fibre we want to  
find a not necessarily integer basis change, which  
eliminates the couplings ${\tilde {\cal K}}_e^2 \tilde {\cal K}_i$  
and leaves the  couplings $\tilde {\cal K}_e \tilde {\cal K}_i \tilde  
{\cal K}_j$ invariant. It follows from  
(\ref{def:aa}, \ref{eq:intersection1}) and the obvious transformation 
of the triple  intersections  that there is a unique solution 
\begin{equation}  
\label{eq:coordredef}
m=\left( 
\begin{array}{cccc}  
1& \frac{a^1}{2}& \ldots &\frac{a^b}{2}\\ 
0& 1&0\ldots &0 \\ 
\vdots&& \vdots &\\  
0& 0& \ldots 0 &1  
\end{array}\right)\ , 
\end{equation} 
such that  
\begin{equation}  
\begin{array}{rl} 
\label{eq:intersection1} 
\tilde{\cal K}_e^3&=a (\int_B c_1^2-\frac{3}{2} a_i a^i+ \frac{3}{4} c_{ij}a^i a^j) \\   
\tilde{\cal K}_e^2 \tilde{\cal K}_i&=0 \\ 
\tilde{\cal K}_e \tilde{\cal K}_i \tilde{\cal K}_j&= a c_{ij}\ . 
\end{array}  
\end{equation}   

As we have seen the classical topological data of the total space  
of the elliptic fibration follows from simple properties of the  
fibre and the topology of the base. We want to extend these result 
in the next section to the quantum cohomology of the elliptic fibration.  
We focus on the Calabi-Yau threefold case, where the instanton  
contributions to the quantum cohomology is richest. To actually 
calculate quantum cohomology we need an explicit realisation 
of a class of examples, which we discuss in the next subsection.

\subsection{Realizations in toric ambient spaces}  
 
To have such a concrete algebraic realization  we use  
hypersurfaces or complete intersections in toric ambient  
spaces. 
 
Possible toric bases $B$ leading to the above described   
elliptic fibrations with only $I_1$ singularities of the  
Calabi-Yau $d$-fold are defined by reflexive polyhedra  
$\Delta_B$ in $d-1$ dimensions~\cite{BatyrevI}, as was observed  
in~\cite{Klemm:1996ts}. For the threefold case one has the   
following possibilities of 2-dimensional polyhedra. 
\begin{figure}[htb] 
\begin{center} 
\includegraphics[width=.8\textwidth]{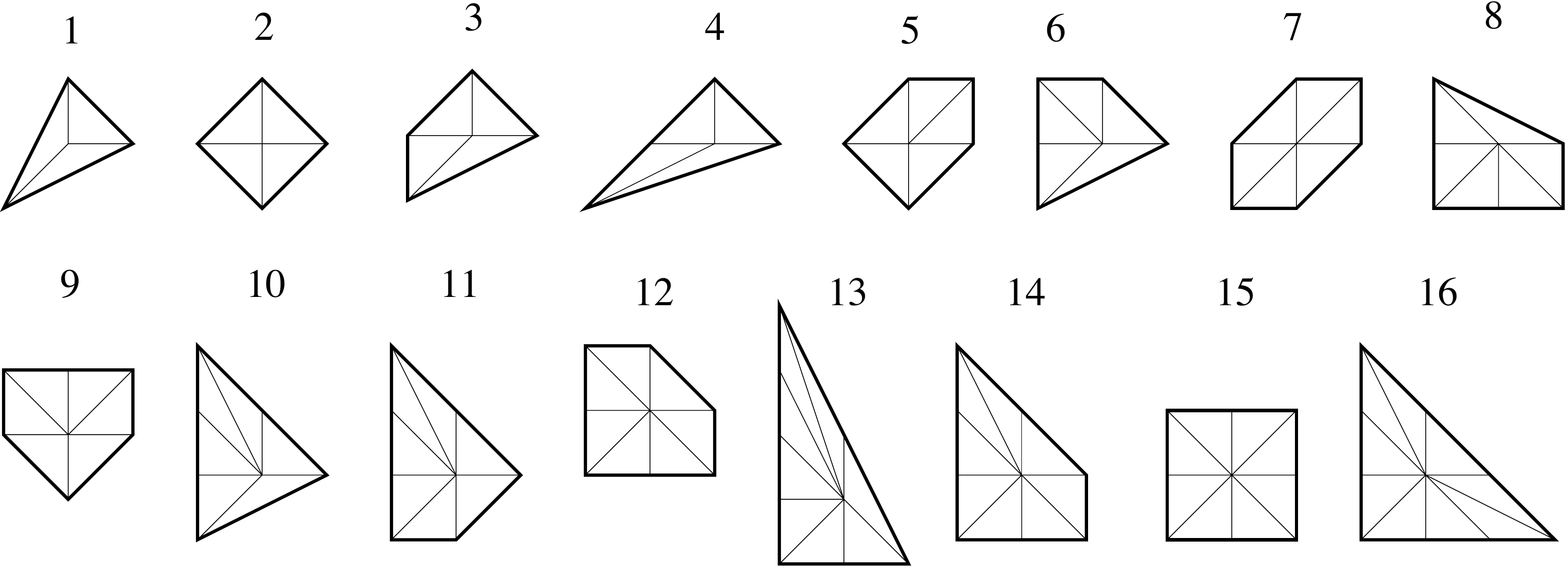} 
\begin{quote} 
\caption{These are the 16 reflexive polyhedra $\Delta_B$ in two dimensions, which  
build $11$ dual pairs $(\Delta_B,\Delta^*_B)$. Polyhedron $k$ is dual to  
polyhedron $17-k$ for $k=1,\ldots,5$. The polyhedra $6,\ldots,11$  
are selfdual. \vspace{-1.2cm}} \label{poly} 
\end{quote} 
\end{center} 
\end{figure}

The toric ambient spaces, which allow for smooth  
Calabi-Yau hypersurfaces as section of the canonical  
bundle, can be described by pairs of reflexive  
polyhedra $(\Delta,\Delta^*)$. Together with a  
complete star triangulation of $\Delta$, they define a  
complex family of Calabi-Yau threefolds. The mirror  
family is given by exchanging the role of $\Delta$  
and $\Delta^*$. A complete triangulation divides  
$\Delta$ in simplices of volume  
$1$. In a star triangulation all simplices contain
the unique inner of the reflexive polyhedron. 
Let us give first two examples for toric smooth ambient  
spaces in which the canonical hypersurface leads  
to the $E_8$ elliptic fibration over $\mathbb{P}^2$   
and over the Hirzebruch surface  $\mathbb{F}_1$.  
The polyhedron for the $E_8$ elliptic fibration  
over $\mathbb{P}^2$ with $\chi=-540$ is given by the  
following data  
\begin{equation} 
\label{datap2} 
 \footnotesize 
 \begin{array}{ccrrrrr|rrl} 
    \multicolumn{7}{c}{\nu_i }    &l^{(e)}& l^{(1)}&\\ 
    D_0    &&     1&     0&   0&   0&   0&     -6&  0& \\ 
    D_1    &&     1&     1&   0&   -2&   -3&      0&  1& \\ 
    D_2    &&     1&     0&   1&   -2&   -3&      0&  1& \\ 
    D_3    &&     1&    -1&  -1&   -2&   -3&      0&  1& \\ 
    D_z    &&     1&     0&   0&   -2&   -3&      1& -3& \\ 
    D_x    &&     1&     0&   0&   1&   0&     2&  0& \\ 
    D_y    &&     1&     0&   0&   0&   1&     3&  0 & 
  \end{array}.    
\end{equation} 
Here we give the relevant  points $\nu_i$ of the four  
dimensional convex reflexive polyhedron $\Delta$ embedded  
into a hyperplane in a five dimensional space and the  
linear relations $l^{(i)}$. This model has an unique star  
triangulation, see (\ref{canonicaltriangulationstar}),   
with the intersection ring  
\begin{equation}   
{\cal R}=9 J_e^3 + 3 J_e^2 J_1 + J_e J_1^2 \ . 
\end{equation}  
as follows from (\ref{eq:intersection1}) with $a=1$  
The evaluation of $c_2$ on the basis of the K\"ahler cone 
is follows from (\ref{eq:2ndchernint}) as   
$\int_M c_2 J_e=102$ and $\int_M c_2 J_1=36$.

The  polyhedron for the $E_8$ elliptic fibration  
over $\mathbb{F}_1$ with $\chi=-480$ reads   
\begin{equation} 
 \label{dataf1} 
 \footnotesize 
\begin{array}{ccrrrrr|rrrl|rrrl} 
    \multicolumn{7}{c}{\nu_i }    &l^{(e)}& l^{(1)}& l^{(2)} &&l^{(e)}+l^{(2)}&l^{(1)}+l^{(2)}&-l^{(2)}&\\ 
    D_0    &&     1&     0&   0&   0&   0&       -6&  0& 0& &    -6&  0& 0& \\ 
    D_1    &&     1&     1&   0&   -2&   -3&      0&  0& 1& &     1&  1& -1&\\ 
    D_2    &&     1&     0&   1&   -2&   -3&      0&  1& 0& &     0&  1& 0& \\ 
    D_3    &&     1&    -1&  -1&   -2&   -3&      0&  0& 1& &     1&  1& -1&\\ 
    D_4    &&     1&     0&  -1&   -2&   -3&      0&  1&-1& &    -1&  0& 1& \\ 
    D_z    &&     1&     0&   0&   -2&   -3&      1& -2&-1& &     0& -3& 1& \\ 
    D_x    &&     1&     0&   0&   1&   0&        2&  0& 0& &     2&  0& 0& \\ 
    D_y    &&     1&     0&   0&   0&   1&        3&  0& 0& &     3&  0& 0&  
  \end{array} .   
\end{equation} 
This example shows that there are two Calabi-Yau phases  
possible over $\mathbb{F}_1$, which are related by flopping a  
$\mathbb{P}^1$ represented by $l^{(2)}$. This transforms the  
half K3 to a del Pezzo eight surface, which can be shrunken to  
a point. In the first phase, the triangulation is described by  
(\ref{canonicaltriangulationstar})  the intersection ring and $\int_M c_2 J_i$  
follows by  (\ref{eq:intersection1}, \ref{eq:2ndchernint}) as 
\begin{equation}   
\label{eq:QF1} 
{\cal R}= 8 J_e^3 + 3 J_e^2 J_1 + J_e J_1^2 + 2 J_e^2 J_2 + J_1 J_2 J_3\ . 
\end{equation}  
and $\int_M c_2 J_e=92$, $\int_M c_2 J_1=36$ and $\int_M c_2 J_3=24$. 
For the second phase we flop the $\mathbb{P}^1$ that corresponds to the  
Mori cone element  $l^{(2)}$. Generally if we flop the curve ${\cal C}$ this  
changes the triple intersection of the divisors  
${\cal K}_i {\cal K}_j {\cal K}_k$~\cite{Witten:1996qb} by  
\begin{equation}  
 \Delta_{ijk}=-({\cal C}\cdot {\cal K}_i)({\cal C}\cdot {\cal K}_j)({\cal C}\cdot {\cal K}_k)\ . 
\end{equation} 
Now the intersection of the curves ${\cal C}_i$ which correspond  
to the mori cone vector  $l^{(i)}$ with the toric divisors  
$D_k$ is given by $({\cal C}_i\cdot D_k)=l^{(i)}_k$. On the other  
hand the ${\cal K}_k$  are combinations of  $D_k$ restricted  
to the hypersurface so that  $({\cal K}^k \cdot {\cal C}_i)=\delta_i^k$. 
 
In addition one has to change the basis in order to maintain  
positive intersection numbers\footnote{This is one criterion that holds  
in a simplicial K\"ahler cone. The full specification is that  
$\int_{\cal C} J>0$, $\int_{\cal D} J\wedge J>0$ and  $\int_M J\wedge J\wedge J>0$ 
for $J$ in the K\"ahler cone and ${\cal C}$, ${\cal D}$ curves and divisors.  
E.g. if the latter is simplicial and generated by $J_i$ then  
$J=\sum d_i J_i$  with $d_i>0$.}  $\tilde l^{(e)}=l^{(e)}+l^{(2)}$, $\tilde l^{(1)}=l^{(1)}+l^{(2)}$ and $\tilde l^{(2)}=-l^{(2)}$. For the $(1,1)$  
forms  $J_i$, which transform dual to the curves, we get then the intersection  
ring in the new basis of the K\"ahler cone 
\begin{equation}   
{\cal R}=  8 \tilde J_e^3 + 3 \tilde J_e^2 \tilde J_1 +  \tilde J_e \tilde J_1^2  
+ 9 \tilde J_e^2  \tilde J_2 + 3 \tilde J_e \tilde J_1 \tilde J_2 + \tilde J_1^2 \tilde J_2 + 
 9 \tilde J_e  \tilde J_2^2 + 3 \tilde J_1 \tilde J_2^2 +  
 9 \tilde J_2^3\ . 
\end{equation}  
The intersections with $c_2$ are not affected by the flop, only the basis  
change  has to be taken into account. In the second phase the triangulation  
of the base is given in the the middle of figure 2 and the triangulation  
of $\Delta$ is specified by  (\ref{canonicaltriangulationgen}). In this 
phase a $E_8$ del Pezzo surface can be shrunken to get to the elliptic  
fibration over $\mathbb{P}^2$. This identifies  the classes of the latter example as   
$J_e=\tilde J_2$, $J_1=\tilde J_1$, while the divisor dual  
to $\tilde J_e^3$ is shrunken.  
 
\begin{figure}[htb] 
\begin{center} 
\includegraphics[width=.4\textwidth]{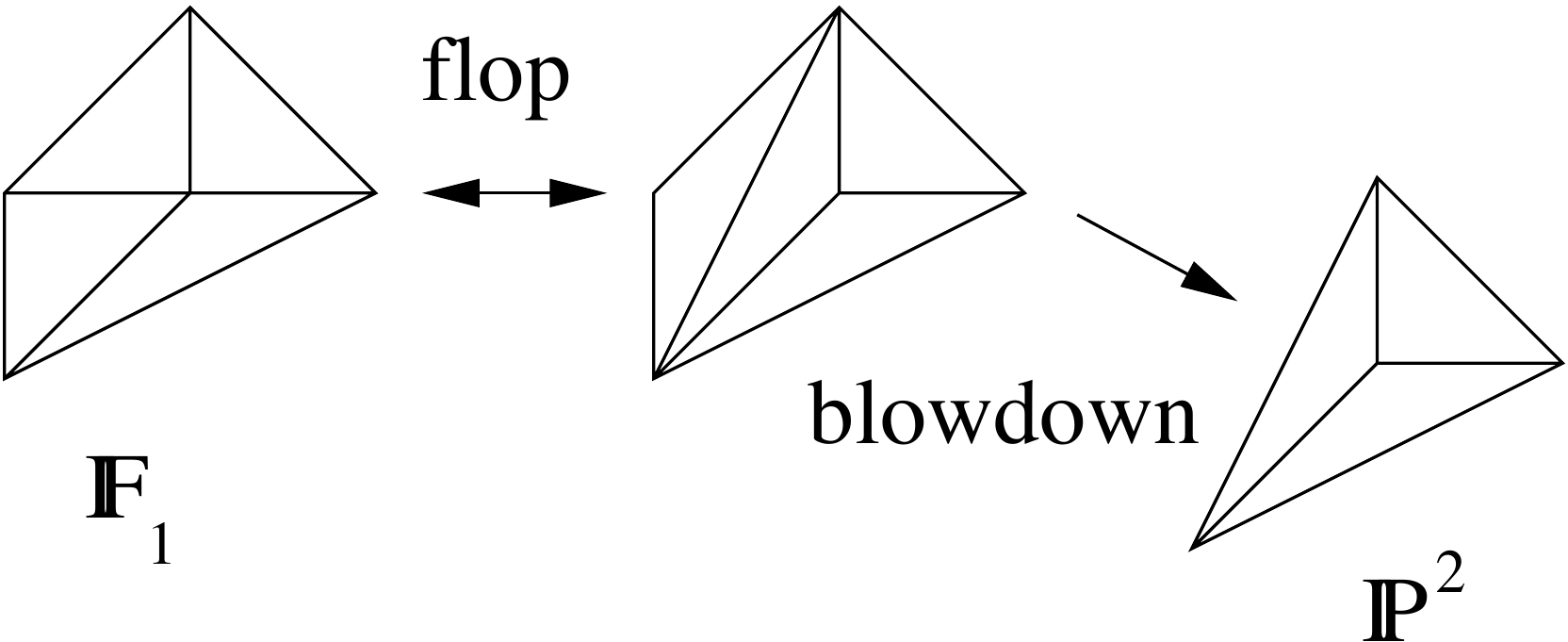} 
\begin{quote} 
\caption{The base triangulation for the flop in second example and the blowdown  
of an $E_8$ del Pezzo surface \vspace{-1.2cm}} \label{flopandblowdown} 
\end{quote} 
\end{center} 
\end{figure}

With $\Delta_B$ the toric polyhedron for the base and specifying  
by  
$$\{(e_1,e_2)\}=\{(-2,-3),(-1,-2),(-1,-1)\}$$   
toric data for the  $E_8,E_7,E_6$ fibre respectively it  
is easy to see that all toric hypersurface with the  
required fibration have the general form of the polyhedron $\Delta$.    
\begin{equation} 
\label{datageneral} 
  \footnotesize 
  \begin{array}{ccrrrrr|rrrrcl} 
    \multicolumn{7}{c}{ \nu_i }    &l^{(e)}&l^{(1)}&\ldots&l^{(b)}  &&\\ 
    D_0    &&     1&     0&   0&   0&0 & \sum_i e_i-1& 0&\ldots &0&\\ 
    D_1    &&     1&      &   &   e_1& e_2  &  0 &    *&\ldots&*&&\\ 
    \vdots    &&     1& \multicolumn{2}{c}{\Delta_B}&\vdots &\vdots& \vdots&*&\ldots&* &&\\ 
    D_r    &&     1&      &    &   e_1& e_2& 0&   *& \ldots&   *&\\ 
    D_z    &&     1&     0&   0&   e_1& e_2 &   1&  -\sum*& \ldots&  -\sum *\\  
    D_x    &&     1&     0&   0&   1&   0&      -e_1&    0&\ldots &0&&\\ 
    D_y    &&     1&     0&   0&   0&   1&      -e_2&    0&\ldots &0&&\\ 
     
  \end{array} 
\end{equation} 
We note that the fibre elliptic curve is realized  
in a two dimensional toric variety, which can be defined  
also by a reflexive 2 dimensional polyhedron $\Delta_F$.  
It  is embedded into $\Delta$ so that the inner of  
$\Delta_F$ is also the origin of $\Delta$. Its corners  
are  
$$\{(0,0,e_1,e_2),(0,0,1,0),(0,0,0,1)\}\ . $$  
The $E_6,E_7$ and $E_8$  fibre types correspond to  
the polyhedra in figure 1  with  numbers $1,4$ and $10$.  
To check the latter equivalence requires an change of  
coordinates in $SL(2,\mathbb{Z})$. The dual reflexive  
polyhedron $\Delta^*$ contains $\Delta_F^*$ embedding  
likewise in the coordinate plane spanned the 3rd  
and 4th axis.    
 
A triangulation of $\Delta_B$ as in figure 1 or 2 lifts in an  
universal way to a star triangulation of $\Delta$ as follows. 
To set the conventions denote by $(\nu^B_i,e_1,e_2)$ the points  
of the embedded base polyhedron $\Delta_B$ and label them as the  
points of $\Delta_B$ starting with the positive x-axis, which points  
to the right in the figures, and label points of $\Delta_B$ counter  
clockwise from $1,\ldots,r$. The inner point in  $\Delta_B$, $(0,0,e_1,e_2)$  
is labelled $z$. The two remaining points of $\Delta$; $(0,0,1,0)$ and  $(0,0,0,1)$  
are labelled by $x$ and $y$.  
 
Denote the k-th d-dimensional simplex in $\Delta_B$ by the labels of  
its vertices, i.e.  
$${\rm sim}^{(d)}_k:=(\lambda^k_1 \ldots, \lambda^k_{d+1})\ $$ 
and in particular denote the outer edges of $\Delta_B$ by 
$$\{{\rm ed}_k|k=1,\ldots,r\}:=\{(1,2),\ldots,(r,1)\}\ . $$ 
Any triangulation  of $\Delta_B$ is lifted to a star triangulation  
of $\Delta$, which is spanned by the simplices containing beside  
the inner point $(0,0,0,0)$ of $\Delta$ the points with the labels  
\begin{equation}  
\label{canonicaltriangulationgen} 
{\rm Tr}_\Delta= 
\{({\rm sim}^{(2)}_k,x),({\rm sim}^{(2)}_k,y)|k=1,\ldots, p\}\cup  
\{({\rm ed}_k,x,y)|k=1,\ldots r\}\ .  
\end{equation} 
 
In particular for star triangulations of $\Delta_B$ one has   
\begin{equation}  
\label{canonicaltriangulationstar} 
{\rm Tr}_\Delta=\{({\rm ed}_k,z,x),({\rm ed}_k,z,y), 
({\rm ed}_k,x,y)|k=1,\ldots r\} 
\end{equation} 
and  generators of the Mori cone for the elliptic phase contain   
the Mori cone generators $l^{(1)},\ldots,l^{(b)}$, which  
correspond to a star triangulation of the base  
polyhedron, which is the one in figure 1.   
We list here the mori cones  first seven case 
\begin{equation*} 
 \label{dataf1} 
 \footnotesize 
\begin{array}{c|r|rr|rr|rr|rrr|rrr|rrrrrr|} 
   \Delta_B& \multicolumn{1}{c}{1(1)} & \multicolumn{2}{c}{2(2)} & \multicolumn{2}{c}{3(2)} & \multicolumn{2}{c}{4(3)}  & \multicolumn{3}{c}{5(3)} & \multicolumn{3}{c}{6(3)} & \multicolumn{6}{c}{7(4)} \\ 
   \nu^B_i & l^{(1)}& l^{(1)}&l^{(2)}& l^{(1)}& l^{(2)}&  l^{(1)}&l^{(2)}&   l^{(1)}&l^{(2)}&l^{(3)}&   l^{(1)}&l^{(2)}&l^{(3)}&  l^{(1)}&l^{(2)}&l^{(3)}&l^{(4)}&l^{(5)}&l^{(6)}  \\ 
    z    &-3&  -2&-2   &-2&-1   &0&-2    &-1&-1&-1   &-1&-1&0   &-1&-1&-1&-1&-1&-1  \\    
    1    & 1&   1&0    &1&0     &0&1     &-1&1&0     &1&0&0     &-1& 1& 0&0 &0&1  \\        
    2    & 1&   0&1    &0&1     &1&0     &1&-1&1     &-1&1&0    & 1&-1&1 &0 &0&0 \\        
    3    & 1&   1&0    &1&-1    &-2&1    &0&1&-1     &1&-1&1    & 0& 1&-1&1 &0&0 \\        
    4    &  &   0&1    &0&1     &1&0     &0&0&1      &0& 1&-2   & 0& 0&1 &-1&1&0  \\        
    5    &  &   & &    & &      &        &1&0&0      &0& 0&1    & 0& 0&0 &1 &-1&1\\     
    6    &  &   & &    & &      &  &     & &         & & &      & 1& 0&0 &0 &1&-1\\     
         &  &   & &    & &      &  &     & &         & & &      &&&&&&\\     
    ex   & -&   &-&    &1&      & -&     & & 4&      & & 3      &&&&&17&\\   
 
 \end{array}    
\end{equation*} 
The remaining 9 cases are given  in the appendix~\ref{app:data}.  
We indicate in the brackets behind the model the number of  
K\"ahler moduli. If the latter is  smaller then the number of  
Mori generators the Mori and the dual K\"ahler cone are non  
simplicial.  This is the case for the models 7,9 and for 11-16. 
In  the last column we list  the number of extra triangulations. 
The corresponding phases  involve non-star triangulations of  
$\Delta$ and can be reached by flops. By the rules discussed  
above we can find the intersection ring and the mori cone 
in phases related by flop. We understand also the blowing  
down of one model. Non reflexivity posses a slight technical  
difficulty in providing the data for the calculation of  
the instantons. The fastest way to get the data for  
all cases is to provide for the models 15 and 16  
a simplicial  K\"ahler cone and reach all other cases\footnote{Except for 13 which  
is available  on request.} by flop and blowdowns. We will do this in the  
appendix~\ref{app:data}.

\section{Quantum geometry of elliptic fibrations}    
\label{quantgeom} \setcounter{equation}{0} 
 
From the data provided in the last section, namely the Mori cone  
and the intersection numbers, follow differential equations as  
well as particular solutions, which allow to calculate the instanton  
numbers as established mathematically for genus zero by Givental, Lian and Yau. 
These can be calculated very efficiently using the program  
described in~\cite{Hosono:1994ax}. In the cases  
at hand one can evaluate the genus one data using the genus zero results,  
the holomorphic anomaly equation for the Ray Singer Torsion, boundary  
conditions provided by the evaluation of $\int_M J_i c_2$ and the behaviour of the  
discriminant at the conifold to evaluate the elliptic instantons.  
 
The  higher genus curves are less systematically studied on compact  
3-folds. However if the total space of the elliptic fibration over  
a base class is a contractable rational surface, one can shrink the  
latter and obtain a local model on which the modular structure  
of higher genus amplitudes have been intensively studied. The  
explicit data suggest that that this structure is maintained   
for all classes in the base.  
 
We summarize in the next subsection the strategy to obtain the  
instanton data and based on the results we propose a general form of the  
instantons corrected amplitude in terms of modular forms coming   
from the elliptic geometry of the fibre and a simple and   
general holomorphic anomaly formula, which govern the all genus  
instanton corrected amplitudes  for the above discussed class  
of models. 
 
In the following subsection we use the B-model to   
prove some aspects of the proposed statements. This can  
establish the A-model results for genus 0 and 1, since  
mirror symmetry is proven and the B-model techniques apply. 
Higher genus B-model calculations have been first 
extended to compact multi-moduli Calabi-Yau manifolds 
in \cite{Haghighat:2009nr}.
 
\subsection{Quantum cohomology, modularity and the anomaly equations}    
 
The basic object, the instanton corrected triple  
intersections $C_{abc}(q^\beta)$ are due to special geometry  
all derivable from the holomorphic prepotential, which  
reads at the point of maximal unipotent 
monodromy~\cite{Candelas:1990rm}\cite{Hosono:1994ax}  
\begin{equation}  
F^{(0)}=(X^0)^2\left[ -{\kappa_{abc} t^a t^b t^c\over 3!} +A_{ab}  
t^a t^b +c_a t^a + \chi {\zeta(3)\over 2 (2 \pi i)^3} + \sum_{\beta\in H_2(M,\mathbb{Z})} n^\beta_{(0)} {\rm Li}_3 (q^\beta)\right] 
\label{eq:prepot} 
\end{equation} 
where $q_\beta=\exp(2 \pi i \sum_{a=1}^{h_2} \beta_a t^a)$,  
$c_a={1\over 24}\int_M c_2 \omega_a$ and  $\chi$ is the Euler number of $M$. 
By $\omega_a$, $a=1,\ldots, h_2(M)$, we denote harmonic $(1,1)$, which form a  
basis of the K\"ahler cone and the complexified K\"ahler parameter  
$t^a=\int_{{\beta_a}} (i \omega + b)$, where  $C_\beta$ is a curve  
class in the Mori cone dual to the K\"ahler cone and $b$ is the  
Neveu-Schwarz $(1,1)$-form $b-$field.  The real coefficients $A_{ab}$  
are not completely fixed. They are unphysical in the sense that  
$K(t,\bar t)$ and $C_{abc}(q)$ do not depend on them. The upper  
index ${(0)}$ on the $F^{(0)}$ indicates the genus of the instanton  
contributions. The triple couplings receive only contributions of  
genus $0$. The classical topological data provide us at the point of  
maximal unipotent monodromy with the $B$-model period integrals  
$\Pi=(F_I,X^I)=\left(\int_{B^I} \Omega,\int_{A_I} \Omega\right)^T$ over  
an integral symplectic basis of 3-cycles $(A_I, B^I)$, $I=0, \dots, h_{21}(W)$.    
This is achieved by matching the $b_3(W)$ solutions to the Picard-Fuchs  
equation with various powers of $\log(z_a)\sim t^a$, with the  
expected form of the $A$-model period vector     
\begin{equation} 
\Pi=X^0 \left(\begin{array}{c} 
2 {\cal F}^{(0)}- t^a \partial_{t^a}{\cal F}^{(0)}\\ 
\partial_{t^a} {\cal F}^{(0)}\\ 
1\\  
t^a\\ \end{array}\right)= 
X^0\left(\begin{array}{c} 
{\kappa_{abc} t^a t^b t^c\over 3!} + c_a t^a - i \chi{\zeta(3)\over (2 \pi)^3} + 2 f(q)-t^a \partial_{t^a} f(q)\\ 
-{\kappa_{abc}  t^b t^c\over 2}+ A_{ab}  t^b+ c_a+\partial_{t^a} f(q)\\ 
1\\  
t^a\\  
\end{array}\right)\ , 
\label{eq:periodsatinfinty} 
\end{equation} 
where the lower case indices run from $a=1,\ldots,h_{21}(W)=h_{11}(M)$.

One can define a generating function for the free energy in terms of a genus expansion in the coupling $g_s$ 
\begin{equation}  
F(g_s,q)=\sum_{g=0}^\infty g_s^{2g-2} F^{(g)}({\underline {q}}), 
\label{eq:allgenusgenartingfunction} 
\end{equation} 
where the upper index $F^{(g)}({\underline {q}})$ indicates as before  
the genus.   
 
According to the split of the cohomology $H_2(M,\mathbb{Z})$ into  
the base and the fibre cohomology, we define $q_B^\beta=\prod_{k=1}^{b_2(B)}  
\exp(2 \pi i \int_{\beta} i\omega+b)$, where now by a slight abuse of notation  
$\beta\in H_2(B,\mathbb{Z})$ and $q= \exp(2 \pi i \int_{f} i\omega+b)$, where  
$f$ is the curve representing the fibre.   Now we define  
the following objects  
\begin{equation} 
\label{eq:defFgn} 
F^{(g)}_{\beta}(q)={\rm Coeff}(F^{(g)}({\underline {q}}), q_B^\beta)\ . 
\end{equation}      
We have the following universal sectors 
\begin{eqnarray} 
F^{(0)}_{0}(q)&=& 
\left(\int_B c_1^2\right) \frac{t^3}{3!}+  
\chi {\zeta(3)\over 2 (2 \pi i)^3}-\chi\sum_{n=1}^\infty {\rm Li}_3(q^n),\\ 
F^{(1)}_{0}(q)&=&\left(\frac{\int_B c_2}{24}\right){\rm Li}_{1}(q),\quad  
F^{(g>1)}_{0}(q)=(-1)^g \frac{\chi}{2}\frac{|B_{2g}B_{2g-2}|}{2g(2g-2)(2g-2)!} . 
\end{eqnarray} 
We note that it follows from the expression for $ F^{(0)}_{0}(q)$  
that  
\begin{equation}  
C_{\tau\tau\tau}= \int_B c_1^2 + \frac{\chi}{2} \zeta(-3)- \frac{\chi}{2} \zeta(-3) E_4(q).    
\end{equation}  
The $F^{(g)}_\beta(q)$ have distinguished modular properties, which 
we describe now. We note that the general form $F^{(g)}_\beta(q)$  is as follows  
\begin{equation}  
\label{eq:Fg} 
F^{(g)}_\beta=\left(\frac{q^{\frac{1}{24}}}{\eta}\right) 
^{12 \sum_i a_i \beta^i} P_{2g +6 \sum_i a_i \beta^i-2} (E_2,E_4,E_6) 
\end{equation} 
with $P_{2g +6 \sum_i a_i \beta^i-2} (E_2,E_4,E_6)$ a (quasi)-modular 
form of weight $2g+6 \sum_i a_i \beta^i-2$ \cite{Kaneko:1995}.  
 
For the sectors $\beta>0$, which describe non-trivial dependence  
on the K\"ahler class of the base, we have the following recursion  
condition   
\begin{equation} \label{eq:holannew} 
\frac{\partial F^{(g)}_{\beta}(q)}{\partial E_2} =  
\frac{1}{24} \sum_{h=0}^g \sum_{\beta' + \beta''=\beta}  \left(\beta'\cdot\beta''\right) F^{(h)}_{\beta'} \, F^{(g-h)}_{\beta''}+ \frac{1}{24} \beta\cdot(\beta-K_B) \,  F^{(g-1)}_{\beta}\ .  
\end{equation}  
For the other types of elliptic fibrations $E_7$, $E_6$, \& $D_5$, the right-hand side is
divided by $a=2,3$ \& 4 respectively. Eq. (\ref{eq:holannew})
generalizes a similar equation due to \cite{Hosono:1999qc}, to
arbitrary classes in the base and types of fibres.  In 
particular if one restricts on elliptic fibrations over 
the blow up of $\mathbb{P}^2$  to a Hirzebruch surface 
$B=\mathbb{F}_1$ to the rational fibre class in the base  
(\ref{eq:holannew}) becomes the equation of~\cite{Hosono:1999qc}
counting curves of higher genus on the $E_8$, $E_7$, $E_6$, 
\& $D_5$ del Pezzo surfaces. The form (\ref{eq:Fg}) and 
its  relation to~\cite{Hosono:1999qc} has been observed 
in~\cite{Klemm:2004km} for the Hirzebruchsurface $\mathbb{F}_0$  
as base. A derivation of the equation (\ref{eq:holannew}) 
is given in section~\ref{sec:derivation}.

\subsection{The B-model approach to elliptically fibred Calabi-Yau spaces}\label{sec:bmodel} 
 
In this section we assume some familarity with the formalism 
developped in~\cite{Hosono:1993qy}\cite{Hosono:1994ax} and 
concentrate on features relevant and common to the 
B-model geometry of elliptic fibrations and how they 
emerge from the topological data of the $A$-model 
discussed in section~\ref{classgeom}. 

The vectors $l^{(i)}$ are the generators of the mori cone, i.e.  
the cone dual to the K\"ahler cone. As such they reflect  
classical properties of the K\"ahler moduli space and  
the classical intersection numbers, like the Euler number  
and the evaluation of $\int_M c_2 \omega_a$ on the basis of K\"ahler  
forms on the elliptic fibration.  
 
On the other hand the differential operators 
\begin{equation} 
\label{scalingeq}  
\left(\prod_{l_i^{(r)}>0} \partial^{l^{(r)}_i}_{a_i}- 
\prod_{l_i^{(r)}<0} \partial^{-l^{(r)}_i}_{a_i} \right)\tilde \Pi=0, 
\end{equation}  
annihilate the periods $\tilde \Pi=\frac{1}{a_0}\Pi$ of the  
mirror $W$. Here the $a_i$ are the coefficients of the monomials  
in the equation defining $W$. They are related to the natural  
large complex structure variables of $W$  by  
\begin{equation} 
 z_r=(-1)^{l^{r}_0} \prod_{i} a_i^{l^{r}_i}\ . 
\end{equation}

Note that $\Pi$ is well defined on $W$, while  
$\tilde \Pi$ is not an invariant definition of periods  
on $W$. However by commuting out $a^{-1}_0$ one can rewrite the equations  
(\ref{scalingeq}) so that they annihilate $\Pi$. Further 
they can be expressed in the independent complex  
variables $z_r$ using the gauge condition $\theta_{a_i}= 
\sum_r l^{k}_i \theta_{z_r}$, where $\theta_x= x \frac{d}{d x}$  
denotes the log derivative.  The equations (\ref{scalingeq})  
reflect symmetries of the holomorphic $(3,0)$ form and   
every positive $l$ in the Mori cone (\ref{scalingeq})  
leads a differential operator annihilating  $\Pi$. 
The operators  obtained in this way are contained  
in the left differential ideal annihilating $\Pi$, but  
they do not generate this ideal. There is however  
a factorisation procedure, basically factoring  
polynomials $P(\theta)$ to the left, that leads  
in our examples to a finite set of generators  
which determines linear combinations of  
periods as their solutions. It is referred to  
as a complete set of Picard-Fuchs operators. 
In this way properties of the instanton corrected  
moduli space of $M$, often called the quantum K\"ahler  
moduli space are intimately related to the $l^{(r)}$  
and below we will relate some of it properties  
to the topology of $M$.      
 
In particular the  mori generator $l^{(e)}$  
determines to a large extend the geometry  
of the elliptic fibre modulus. As one sees from  
(\ref{datageneral}) the mixing between the base  
and the fibre is encoded in the $z$ row of $l^{(i)}$,  
$i=1,\ldots,h_{11}(B)$ and  $l^{(e)}$ in  
(\ref{datageneral}). Let us call this the  
$z$-component of $l^{(i)}$ and the corresponding  
variable $a_z$. 
 
Following the procedure described above one obtains  
after factorizing from $l^{(e)}$ a second order  
generator Picard Fuchs operator. For the fibrations  
types introduced before it is given by 
\begin{equation} 
\label{fibrediffop}  
{\cal L}^{k}_e=\theta_e(\theta_e-\sum_{i} a^i \theta_i) - {\cal D}^{K}   
\end{equation} 
where $k=E8,E7,E6,D5$ refers to the fibration type and  
${\cal D}^{K}$ contains the dependence on the type  
\begin{equation}  
\begin{array}{rlrl} 
{\cal D}^{E8}& = 12 (6 \theta_e-1)(6 \theta_e-5)z_e, &  {\cal D}^{E7}&= 4 (4 \theta_e-1)(4 \theta_e-3)z_e,\\ 
{\cal D}^{E6}& = 3  (3 \theta_e-1)(3 \theta_e-2)z_e, &   {\cal D}^{D5}&= 4 (2 \theta_e-1)^2z_e\ .  
\end{array} 
\end{equation}  
Formally setting $\theta_i=0$ corresponds to the large base limit.  
Then the equation (\ref{fibrediffop}) becomes the Picard-Fuchs operator,  
which annihilates the periods over the standard holomorphic differential  
on the corresponding family of elliptic curves. 
 
In limit of large fibre one gets as local model  
the total space of the canonical line bundle  
${\cal O}(K_B)\rightarrow B$ over the  
Fano base $B$. Local mirror symmetry associates to  
such  noncompact Calabi-Yau manifolds a genus one     
curve with a meromorphic $1$-form $\lambda$ that  
is the limit of the holomorphic $(3,0)$-form.  
The local Picard-Fuchs system ${\cal L}^B_i$ annihilating  
the periods $\Pi_{loc}$  of $\lambda$ can be obtained  
as a limit of the compact Picard-Fuchs system for  
$l^{(i)}$, $i=1,\ldots,h_{11}(B)$ by formally setting  
$\theta_e=0$. It follows directly from  
(\ref{scalingeq}), since the Mori generators of the base  
have vanishing first entry and commuting out $a_0^{-1}$  
becomes trivial. Differently then for the elliptic curve  
of the fibre these Picard-Fuchs operators do not annihilate 
the periods over holomorphic differential one form of  
the elliptic curve, which are  $\frac{1}{a_z}\Pi_{loc}$.   
Given the local Picard-Fuchs system the dependence on  
$\theta_e$ can be restored by replacing $\theta_{a_z}$ by   
$\theta_e-\sum_{i} a_i \theta_i$ instead of  
$-\sum_{i} a_i \theta_i$. Since $l^{(i)}$ is negative 
$\theta_e$ appears in ${\cal L}_b^i$ only multiplied by  
at least one explicit $z_i^b$ factor. 
 
There are important conclusions that follow already  
from the  general form of the Picard-Fuchs system. To see  
them it is convenient to rescale $x_e=c_k z_e$, where  
$c_{E8}=432$, $c_{E7}=64$, $c_{E6}=27$, $c_{D5}=16$.  
It is often useful to also rescale the $z_i$ and call  
them $x_i$.
  
The effect of this is that the symbols of the  
Picard-Fuchs system become the same for all  
fiber types. From this we can conclude that  
for all fibre types the Yukawa-couplings and the  
discriminants are identical in the rescaled variables.  
 
The second conclusion is that the Picard-Fuchs equation of  
the compact Calabi-Yau is invariant under the $\mathbb{Z}_2$  
variable transformation 
\begin{equation}  
\label{changeofvar} 
x_e\rightarrow (1-x_e),\qquad x_i \rightarrow  \left(-\frac{x_e}{1-x_e}\right)^{a^i} x_i \ .  
\end{equation} 
This means that there is always a $\mathbb{Z}_2$ involution  
acting on the moduli space parametrized by $(x_e,x_i)$,  
which must be divided out to obtain the truly  
independent values of the parameters.  
 
Another consequence of this statement is 
that the discriminants $\Delta_i(x_j)$ of the base Picard-Fuchs system determine the  
discriminant locus of the global system apart from $\Delta(x_e)$ components.  
The former contains always a conifold component $\Delta_c(x_j)$ and only that one,  
if there are no points on the edges of the 2d polyhedron. Points on the edges  
correspond to $SU(2)$ or $SU(3)$ gauge symmetry enhancement  
discriminants which contain only $x_i$ variables dual  
to K\"ahler classes, whose $a^i=0$. They are therefore  
invariant under (\ref{changeofvar}). Moreover the lowest  
order term in the conifold discriminant is a constant and  highest terms  
are weighted monomials of degree $\chi(B)$ with weights for the $x_i$    
$a^i$ or $1$ if $a^i=0$. It follows by (\ref{changeofvar}) that the  
transformed conifold discriminant $\Delta'_c(x_j)\sim (1-x_e)^{\chi(B)}+O(x_i)$. 
   
\subsubsection{Examples: elliptic fibrations over $\mathbb{P}^2$ and $\mathbb{F}_1$} 

Let us demonstrate the above general statements with a 
couple of examples. We discuss the $E8$ elliptic fibration with base $\mathbb{P}^2$ and with base $\mathbb{F}_1$. \\ For the first example the Mori vectors are given as
\begin{equation}
\begin{split}
l^{(e)} &=(-6, 3, 2, 1, 0, 0, 0),\\
l^{(2)} &=(0, 0, 0, -3, 1, 1, 1).
\end{split}
\end{equation}
Form this we can derive the following set of Picard Fuchs equations, where we denote $\theta_i = z_i \partial_{z_i}$.
\begin{equation}
\begin{split}
\CL_1 &= \theta_e (\theta_e-3 \theta_2) - 12 z_e (6 \theta_e-5) ( 6\theta_e-5),\\
\CL_2 &= \theta_2^3 - z_2 (\theta_e-3\theta_2)(\theta_e-3\theta_2-1)(\theta_e-3\theta_2-2).
\end{split}
\end{equation}
The Yukawa couplings for this example read as follows, where we use $z_1=\frac{x_1}{432}$, $z_2 = \frac{x_2}{27}$ 
and the discriminants $\Delta_1=1-3 x_1+3 x_1^2-x_1^3-x_1^3 x_2$ and $\Delta_2 = 1+ x_2$\\
\begin{equation}
\begin{split}
C_{eee} &=\frac{9}{x_1^3 \Delta_1}, \\
C_{ee2}&=-\frac{3 (-1+x_1)}{x_1^2 x_2 \Delta_1},\\
C_{e22}&=\frac{(-1+x_1)^2}{x_1 x_2^2 \left(\Delta_1\right)},\\
C_{222}&=\frac{1-3 x_1+3 x_1^2}{3 x_2^2 \Delta_1 \Delta_2}.
\end{split}
\end{equation}\\
The second example over $\mathbb{F}_1$ has the following three generators of the Mori cone
\begin{equation}\label{eq:Morivectors}
\begin{split}
l^{(e)}&=(-6\ |\ 3,\phantom{-}2,\phantom{-}1,\phantom{-}0,\phantom{-}0,\phantom{-}0,\phantom{-}0),\\
l^{(2)}&=(\phantom{-}0\ |\ 0,\phantom{-}0,-1,-1,\phantom{-}0,\phantom{-}1,\phantom{-}1),\\
l^{(3)}&=(\phantom{-}0\ |\ 0,\phantom{-}0,-2,\phantom{-}1,\phantom{-}1,\phantom{-}0,\phantom{-}0),
\end{split}
\end{equation}
and gives rise to the following Picard-Fuchs equations 
\begin{equation}\label{eq:FullPF}
\begin{split}
{\cal L}_1&=\theta_1(\theta_1-2\theta_3-\theta_2)-12z_1(6\theta_1+5)(6\theta_1+1),\\
{\cal L}_2&=\theta_2^2-z_2(\theta_2-\theta_3)(2\theta_3+\theta_2-\theta_1),\\
{\cal L}_3&=\theta_3(\theta_3-\theta_2)-z_3(2\theta_3+\theta_2-\theta_1)(2\theta_3+\theta_2-\theta_1+1).
\end{split}
\end{equation}
This example contains the rational elliptic surface, which we discuss in detail 
in section \ref{sec:ellsurface}. Furthermore we focus on this example to give 
a proof of the holomorphic anomaly at genus zero by using mirror symmetry in section~\ref{sec:ellf1}.

\subsection{Modular subgroup of monodromy group} 
\label{sec:monodromy} 
 
\begin{figure}[h!] 
\begin{center} 
\includegraphics[width=10cm]{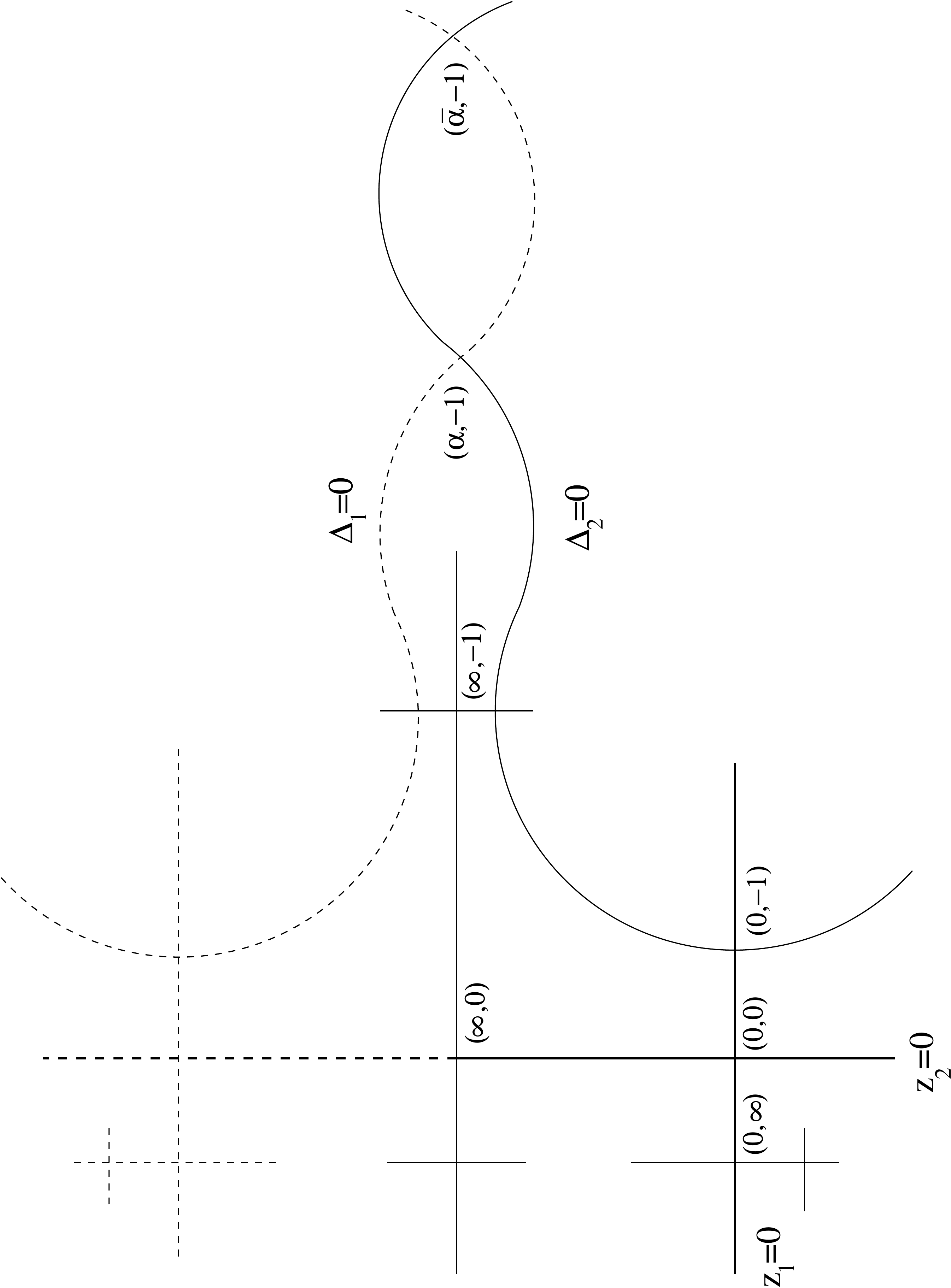} 
\begin{quote} 
\caption{The moduli space for the elliptic fibration Calabi-Yau space  over $\mathbb{P}^2$.  
\vspace{-1.2cm}} \label{moduli11169} 
\end{quote} 
\end{center} 
\end{figure}

The deeper origin of the appearance of modular forms is the 
monodromy group of the Calabi-Yau. Ref. \cite{Candelas:1994hw} 
explains that in the large volume limit of $X_{18}(11169)$,
the monodromy group reduces to an  $SL_2(\mathbb{Z})$ monodromy
group. This section recalls the  
appearance of this modular group and how it generalizes to other 
elliptic fibrations. The moduli space of $X_{18}(11169)$ with the degeneration loci is 
portrayed in Fig. \ref{moduli11169}. 
 
We continue by recalling the monodromy for the model in 
\cite{Candelas:1994hw} adapted to our discussion.  
The fundamental solution is given by: 
\begin{eqnarray} 
\label{eq:periodX18} 
w_0(x,y)&=&\sum_{m,n=0}^\infty 
\frac{(18n+6m)!}{(9n+3m)!\,(6n+2m)!\,(n!)^3\,m!}\,x^{3n+m} \,y^m\\ 
&=&\sum_{k=0}^\infty\frac{(6k)!}{k!\,(2k)!\,(3k)!}\,x^k\,U_k(y).\non 
\end{eqnarray} 
with 
\begin{eqnarray} 
U_\nu(y)&=&y^\nu\sum_{n=0}^\infty \frac{\nu!}{(n!)^3\,\Gamma(\nu-3n+1)}\,y^{-3n}\\ 
&=&y^\nu\sum_{n=0}^\infty 
\frac{\Gamma(3n-\nu)}{\Gamma(-\nu)\,(n!)^3}\, y^{-3n}, \non 
\end{eqnarray} 
which is a finite polynomial for positive integers $\nu$, since 
$\Gamma(\nu-3n+1)=\infty$ for sufficiently large $n$.  The translation to the parameters in 
  \cite{Candelas:1994hw} is $(x,y)=(\,(18\psi)^{-6},-3\phi)$.  The natural 
coordinates obtained from toric methods are $z_1=xy$ and 
$z_2=y^{-3}$. Note that the second line (\ref{eq:periodX18}) makes 
manifest the presence of the elliptic curve in the geometry. 
For this regime of the parameters one can easily find logarithmic 
solutions by taking derivates to $k$ and $n$ \cite{Hosono:1993qy}: 
\begin{eqnarray} 
\label{eq:CYperiods} 
2\pi i\, w^{(1)}_{e}(x,y)&=&\log(xy)\,w_0+\dots .\\ 
2\pi i\, w^{(1)}_{1}(x,y)&=&-3\log(y)\,w_0+\dots, \non 
\end{eqnarray} 
The periods are defined by $\tau=w^{(1)}_e/w_0$ and 
$t_1=w^{(1)}_1/w_0$ and $q=e^{2\pi i \tau}$, $q_1=e^{2\pi it_1}$. 
 
The two monodromies which generate the modular group are: 
\begin{eqnarray} 
&M_0: & (x,y)\to (e^{2\pi i}x,y),\,\, x\,\,\mathrm{small},\, 
y\,\,\mathrm{large}, \non \\ 
&M_\infty: & (x,y)\to (e^{2\pi i}x,y),\,\, x\,\, \mathrm{large},\, y\,\, \mathrm{large}.\non 
\end{eqnarray} 
The monodromy around $x=0$ follows directly from (\ref{eq:CYperiods}), it acts 
as: 
\be 
\label{eq:M0} 
{\bf M}_0=\left(\begin{array}{cc} 1 & 1\\0 & 1 \end{array}\right) 
\ee 
on $(w^{(1)}_e,w_0)^\mathrm{T}$. To determine the action on the 
periods of $M_\infty$, we need to analytically continue $w_0$ and 
$w^{(1)}_1$ to large $x$. To this end, we write $w_0$ as a Barnes integral: 
\be 
\label{eq:barnes} 
w_0(x,y)=\frac{1}{2\pi i}\int_C ds\, 
\frac{\Gamma(-s)\,\Gamma(6s+1)}{\Gamma(2s+1)\, \Gamma(3s+1)}\, e^{\pi i s} \,x^s\,U_s(y) 
\ee 
where $C$ is the vertical line from $-i\infty-\varepsilon$ to 
$i\infty-\varepsilon$. For small $|x|$ the contour can be deformed to 
the right giving back the expression in (\ref{eq:periodX18}). For large $|x|$ 
one instead obtains the expansion:   
\be 
w_0(x,y)=\frac{1}{6\pi^2}\sum_{r=1,5} \sin(\pi r/3)\sum_{k=0}^\infty\, a_r(k)\,(-x)^{-k-\frac{r}{6}}\,U_{-k-r/6}(y), 
\ee 
with $$a_r(k)=(-1)^k\frac{\Gamma(k+r/6) \Gamma(2k+r/3) 
  \Gamma(3k+r/2)}{\Gamma(6k+r)}.$$ The logarithmic solution 
$w^{(1)}_e$ is given similarly by: 
\begin{eqnarray} 
\label{eq:barnes2} 
w^{(1)}_e(x,y)&=&\frac{1}{2\pi i}\int_C ds\,\frac{\Gamma(-s)^2\,\Gamma(6s+1)\,\Gamma(s+1)}{ \Gamma(2s+1)\, \Gamma(3s+1)} 
\, e^{2\pi i s} \,x^s\,U_s(y),  \\ 
&=&\frac{1}{6\pi^2i}\sum_{r=1,5} e^{-\pi i r/6}\cos(\pi r/6) \sum_{k=0}^\infty a_r(k)\,(-x)^{-k-\frac{r}{6}}\,U_{-k-r/6}(y).\non 
\end{eqnarray} 
To determine the action of $M_\infty$, we define the basis $f_r(x,y)=\sum_{n=0}^\infty 
a_r(k)\, (-x)^{-k-\frac{r}{6}}\,U_{-k-r/6}(y)$ for $r=1,5$, and the 
matrix ${\bf A}$ which relates to the bases $(w^{(1)}_e,w^{(1)}_1)^\mathrm{T}={\bf A}\,(f_1,f_5)^\mathrm{T}$. Clearly, 
$M_\infty$ acts diagonally on the $f_r$: ${\bf T}=\mathrm{diag}(\alpha^{-1},\alpha^{-5})$ with $\alpha=e^{2\pi 
  i/6}$, which gives for ${\bf M}_\infty$ 
\be 
\label{eq:Minfty} 
{\bf M}_\infty={\bf ATA}^{-1}=\left(\begin{array}{cc} 0 & -1\\ 1 & 1 \end{array}\right)\in SL_2(\mathbb{Z}). 
\ee  
This gives for the monodromy around the conifold locus: 
\be 
{\bf M}_1={\bf M}_0{\bf M}_\infty^{-1}=\left(\begin{array}{cc} 1 & 0\\1 & 1 \end{array}\right). 
\ee 
The generator $S= \left(\begin{array}{cc} 0 & 1\\-1 & 
    0 \end{array}\right)$ of $SL_2(\mathbb{Z})$ corresponds to ${\bf 
  M}_0{\bf M}^{-1}_\infty$. 
 
The large volume limit is such that $r=q^{3/2}q_1\to 0$. We see that $M_0$ and 
$M_\infty$ map small $r$ to small $r$. The monodromies act on $r$ by \cite{Candelas:1994hw}: 
\be 
M_0 r=-r, \qquad M_\infty r=r 
\ee 
Thus we have established an 
action of $SL_2(\mathbb{Z})$ on the boundary of the moduli space. 

The above analysis can be extended straightforwardly to the other types of
fibrations using the expansions (\ref{eq:periodexps}). The matrix ${\bf M}_0$ is for all fibre types the same. We find that ${\bf
  M}_\infty=\left(\begin{array}{cc} 1-a &   -1\\ a &
    1 \end{array}\right)\in \Gamma_0(a)$ for $a=2,3$ and $4$ corresponding
to the fibre types $E_7$, $E_6$ and $D_5$. Note that ${\bf M}_\infty$ 
has order 4 and 3 for $a=2$ and 3 respectively, while the order is infinite for $a=4$.
Generalization to other base surfaces $B$ is also straightforward. In
case of multiple 2-cycles in the base, it is natural to define parameters for
each base class: $r_i=q^{a_i/2}q_i$, $i=1,\dots,b_2(B)$. This is
precisely the change of parameters given by
(\ref{eq:coordredef}). These transform as:
\be 
M_0 r_i=(-)^{a_i}r_i, \qquad M_\infty r_i=r_i.
\ee 

\section{Derivation of the holomorphic anomaly equation} 
\label{sec:derivation}

 \setcounter{equation}{0} 
\subsection{The elliptic fibration over $\mathbb{F}_1$}  
\label{sec:ellf1}

In the following we try to derive the holomorphic anomaly equation at genus zero  
by adapting the proof which appeared in ref.~\cite{Hosono:1999qc} for a similar geometry. 
We start by studying the Picard-Fuchs operator associated to the elliptic fiber 
$X_6[1,2,3]$ only. Denoting by $\theta_e=x_e\partial_{x_e}$ the Picard-Fuchs operator can be written as 
\begin{equation} 
{\cal L}=\theta^2_e-12x(6\theta_e+5)(6\theta_e+1). 
\end{equation} 
One can immediately write down two solutions as power series expansions around $x_e=0$. They are given by 
\begin{equation} 
\label{eq:ellcurveperiods} 
\phi(x_e)=\sum_{n\,\geq\,0}a_n x_e^n,\quad \tilde{\phi}(x_e)=\log(x_e)\phi(x_e)+\sum_{n\,\geq\,0}b_n x_e^n, 
\end{equation} 
with 
\begin{equation} 
a_n=\frac{(6n)!}{(3n)!(2n)!n!},\quad  
b_n=a_n(6\psi(1+6n)-3\psi(1+3n)-2\psi(1+2n)-\psi(1+n)), 
\end{equation} 
where $\psi(z)$ denotes the digamma function. The mirror map is thus given by 
\begin{equation} 
2\pi i \tau =\frac{\tilde{\phi}(x_e)}{\phi(x_e)}. 
\end{equation} 
Using standard techniques from the Gauss-Schwarz theory for the Picard-Fuchs equation (cf.~\cite{Lian-Yau}) one observes 
\begin{equation} 
j(\tau)=\frac{1}{x_e(1-432x_e)}, 
\end{equation} 
which can be inverted to yield 
\begin{equation} 
x_e(\tau)=\frac{1}{864}(1-\sqrt{1-1728/j(\tau)})=q-312q^2+{\cal O}(q^3). 
\end{equation} 
Further, the polynomial solution $\phi(x_e)$ can be expressed in terms of modular forms as 
\begin{equation} 
\phi(x_e)={}_2 F_1(\frac{5}{6},\frac{1}{6},1;432x_e)=\sqrt[4]{E_4(\tau)}, 
\end{equation} 
from which one can conclude that 
\begin{equation}\label{eq:modrel} 
\begin{split} 
E_4(\tau)&=\phi^4(x_e),\\ 
E_6(\tau)&=\phi^6(x_e)(1-864x_e),\\ 
\Delta(\tau)&=\phi^{12}(x_e) x_e(1-432x_e),\\ 
\frac{1}{2\pi i}\frac{dx_e}{d\tau}&=\phi^2(x_e) x_e(1-432x_e). 
\end{split} 
\end{equation}

Let us now examine the periods of the mirror geometry $Y$ in the limit that the fiber $F$ of the Hirzebruch surface becomes small. Due to the special structure of the Picard-Fuchs system which is found in eq.~(\ref{eq:FullPF}) the first three period integrals in the notation of \cite{Hosono:1999qc} read 
\begin{equation} 
\begin{split} 
w_0(x_e,y,0)&=\phi(x_e),\\ 
w_1^{(1)}(x_e,y,0)&=\tilde{\phi}(x_e),\\ 
w_2^{(1)}(x_e,y,0)&=\log(y)\phi(x_e)+\xi(x_e)+\sum_{m\,\geq\,1}({\cal L}_m \phi(x_e))y^m, 
\end{split} 
\end{equation} 
with 
\begin{equation} 
\xi(x_e)=\sum_{n\,\geq\,0}a_n(\psi(1+n)-\psi(1))x_e^n, 
\end{equation} 
and 
\begin{equation} 
{\cal L}_m=\frac{(-)^{m}}{m(m!)}\prod_{k=1}^m(\theta_{x_e}-k+1). 
\end{equation} 
This can be obtained by applying the Frobenius method to derive the period integrals, see e.g.~\cite{Hosono:1994ax}. The mirror map reads 
\begin{equation}\label{eq:mirrormap} 
2\pi i t_i = \frac{w^{(1)}_i(x_e,y,0)}{w_0(x_e,y,0)},\quad i=1,2. 
\end{equation} 
Comparing this with our previous discussion about the Picard-Fuchs operator of the elliptic fiber we see that for $t_1=\tau$ there is nothing left to discuss. Hence, let's study the mirror map associated to $t_2=t$. We observe that by formally inverting, the inverse mirror map can be determined iteratively through the relation 
\begin{equation}\label{eq:mirrory} 
y(q,p)=p \zeta e^{-\sum_{m\geq1}c_m(x_e)y^m}, 
\end{equation} 
where $\zeta=e^{-\frac{\xi(x_e)}{\phi(x_e)}}$ and 
\begin{equation} 
c_m(x_e)=\frac{{\cal L}_m\phi(x_e)}{\phi(x_e)}. 
\end{equation} 
Using eq.~(\ref{eq:modrel}) $c_1(x_e)$ is given by 
\begin{equation}\label{eq:c1} 
\begin{split} 
c_1(x_e)&=-\frac{1}{12}(f_1-2)-\frac{f_1}{12}\frac{E_2(\tau)}{\phi^2(x_e)}\\ 
&=-\frac{1}{\phi^6}\frac{f_1}{12}(E_2E_4-E_6), 
\end{split} 
\end{equation} 
where we introduced $f_1=(1-432x_e)^{-1}$. In order to obtain the other $c_m(x_e)$ one uses 
\begin{equation}\label{eq:thetaxder} 
\begin{split} 
\theta_{x_e} f_1&=f_1(f_1-1),\\ 
\theta_{x_e}\left(\frac{E_2}{\phi^2}\right)&=-\frac{1}{\phi^8}\frac{f_1}{12}\left(E_2^2E_4-2E_2E_6+E_4^2\right),\\ 
\theta_{x_e}\left(\frac{E_6}{\phi^6}\right)&=-\frac{1}{\phi^{12}}\frac{f_1}{12}\left(6E_4^3-6E_6^2\right), 
\end{split} 
\end{equation} 
and finds the following kind of structure. One can show inductively that 
\begin{equation}\label{eq:cmstruct} 
c_m(x_e)=\frac{1}{\phi^{6m}}\left(\frac{f_1}{12}\right)^m Q_{6m}(E_2,E_4,E_6), 
\end{equation} 
where $Q_{6m}$ is a quasi-homogeneous polynomial of degree $6m$ and type $(2,4,6)$, i.e. 
$$Q_{6m}(\lambda^2 x_e,\lambda^4 y, \lambda^6 z)=\lambda^{6m}Q_{6m}(x_e,y,z).$$ 
Also by induction, it follows from (\ref{eq:c1}) and (\ref{eq:thetaxder}) 
that $Q_{6m}$ is linear in $E_2$. This allows to write a second structure which is analogous to the one appearing in ref.~\cite{Hosono:1999qc} and given by 
\begin{equation} 
c_m(x_e)=B_m \frac{E_2}{\phi^2}+D_m, 
\end{equation} 
where the coefficients $B_m$, $D_m$ obey the following recursion relation 
\begin{equation}\label{eq:recursion} 
\begin{split} 
B_{m+1}&=-\frac{m}{(m+1)^2}\left[(\theta_{x_e}-m)B_m+D_1B_m-B_1D_m\right],\\ 
D_{m+1}&=-\frac{m}{(m+1)^2}\left[(\theta_{x_e}-m)D_m-D_1D_m+B_1B_m\right], 
\end{split} 
\end{equation} 
with $B_1=-\frac{f_1}{12}$ and $D_1=-\frac{1}{12}(f_1-2)$. A formal solution to the recursion relation (\ref{eq:recursion}) can be given by 
\begin{equation} 
\begin{split} 
B_m&=-\frac{f_m}{12},\\ 
D_m&=\frac{1}{f_1}\left[\frac{(m+1)^2}{m}f_{m+1}+(\theta_{x_e}-m-\frac{1}{12}(f_1-2))f_m\right], 
\end{split} 
\end{equation} 
where we define $f_m$ to be 
\begin{equation} 
f_m(x_e)=\tilde{\phi}(x_e){\cal L}_m\phi(x_e)-\phi(x_e){\cal L}_m\tilde{\phi}(x_e). 
\end{equation}  
Due to the relations (\ref{eq:thetaxder}) we conclude, that the $f_m$ as well as $B_m$ and $D_m$ are polynomials in $f_1$. Since $f_1$ is a rational function of $x_e$, it transforms well under modular transformations. Therefore modular invariance is broken only by the $E_2$ term in $c_m$. We express this via the partial derivative of $c_m$ 
\begin{equation}\label{eq:dercm} 
\frac{\partial c_m(x_e)}{\partial E_2}=-\frac{1}{12}\frac{f_m(x_e)}{\phi^2(x_e)}. 
\end{equation} 
 
In order to prove the holomorphic anomaly equation (\ref{eq:holannew})
one first shows using the general results about the period integrals
in \cite{Hosono:1994ax} that the instanton part of the prepotential
can be expressed by the functions $f_m (x_e)$. A tedious calculation
reveals  
\begin{equation}\label{eq:derF0} 
\frac{1}{2\pi i}\frac{\partial}{\partial t}F^{(0)}(\tau,t)=\sum_{m\,\geq\,1}\frac{f_m (x_e)}{\phi^2(x_e)}y^m. 
\end{equation} 
Using the inverse function theorem and eqs.~(\ref{eq:dercm}), (\ref{eq:mirrory}) yields 
\begin{equation} 
\frac{\partial y}{\partial E_2}=\frac{1}{12}\left(\frac{1}{2\pi i}\frac{\partial y}{\partial t}\right)\left(\frac{1}{2\pi i}\frac{\partial F^{(0)}}{\partial t}\right). 
\end{equation} 
Now, we have 
\begin{equation} 
\frac{\partial}{\partial E_2}\left(\frac{1}{2\pi i}\frac{\partial 
    F^{(0)}}{\partial t}\right)=\frac{1}{12}\left(\frac{\partial^2 
    F^{(0)}}{\partial (2\pi it)^2}\right)\left(\frac{1}{2\pi 
    i}\frac{\partial F^{(0)}}{\partial t}\right), 
\end{equation} 
which implies that up to a constant term in $p$ one arrives at 
\begin{equation} 
\frac{\partial F^{(0)}}{\partial E_2}=\frac{1}{24}\left(\frac{1}{2\pi 
    i}\frac{\partial F^{(0)}}{\partial t}\right)^2. 
\end{equation} 
By definition of $F^{(0)}_n$, Eq.~(\ref{eq:defFgn}), we have $\frac{1}{2\pi 
  i}\frac{\partial}{\partial t} F^{(0)}(\tau,t)=\sum_{m\,\geq\,1}m\, F^{(0)}_m \,p^m$ and hence obtain by resummation 
\begin{equation}\label{eq:DZnDE2} 
\frac{\partial F^{(0)}_n}{\partial E_2}=\frac{1}{24}\sum_{s=1}^{n-1}s(n-s) F^{(0)}_s  F^{(0)}_{n-s}. 
\end{equation} 
This almost completes the derivation of (\ref{eq:holannew}). We still need to determine the explicit form of $F^{(0)}_n$. To achieve this we proceed inductively. Using (\ref{eq:modrel}), (\ref{eq:derF0}) and (\ref{eq:mirrory}) one obtains 
\begin{equation} 
F^{(0)}_1=\frac{\zeta f_1}{\phi^2}=q^{\frac{1}{2}}\frac{E_4}{\eta^{12}}. 
\end{equation} 
Employing the structure (\ref{eq:cmstruct}) one can evaluate (\ref{eq:derF0}) and calculate that 
\begin{equation} 
\begin{split} 
F^{(0)}_n&=\frac{\zeta^n f_1^n}{\phi^{6n}}P_{6n-2}(E_2,E_4,E_6),\\ 
&=\left(\frac{\zeta f_1}{\phi^2}\right)^n \frac{1}{\phi^{4n}}P_{6n-2}(E_2,E_4,E_6),\\ 
&=\frac{q^{\frac{n}{2}}}{\eta^{12n}}P_{6n-2}(E_2,E_4,E_6), 
\end{split} 
\end{equation} 
where $P_{6n-2}$ is of weight $6n-2$ and is decomposed out of (parts 
of) $Q_m$'s. This establishes a derivation of the holomorphic anomaly 
equation (\ref{eq:holannew}) at genus zero for the elliptic fibration 
over Hirzebruch surface $\mathbb{F}_1$ with large fibre class. We collect some results for the other fibre types in appendix \ref{app:fibref1}.

\subsection{Derivation from BCOV}  
The last section provided a derivation of the anomaly
equation (\ref{eq:holannew}) for genus 0 from the mirror
geometry. More fundamental is a derivation purely within the
context of moduli spaces of maps from Riemann surfaces to a Calabi-Yau
manifold. This is the approach taken by BCOV~\cite{Bershadsky:1993cx}
to derive holomorphic anomaly equations for genus $g$ $n$-point correlation function with
$2g-2+n>0$. The correlation functions are given by covariant
derivatives to the free energies $F^{(g)}$: 
$C^{(g)}_{i_1i_2\dots i_n}=D_{i_1}\dots D_{i_n} F^g$, with $D_i$
covariant derivatives of for sections of the bunde
$\mathcal{L}^{2-2g}\otimes \mathrm{Sym}^n\,T$, with $T$ the tangent
bundle of the coupling constant moduli space, and $\mathcal{L}$ a line
bundle over this space whose Chern class correponds to $G_{i\bar j}$.

The holomorphic anomaly equation reads for the $n$-point functions 
\begin{equation}\label{eq:BCOVholan} 
\begin{split} 
\bar \partial_i C^{(g)}_{i_1 \dots i_n} &= \frac{1}{2} \bar C_{\bar i \bar j \bar k} e^{2K} G^{j \bar j} G^{k \bar k} C^{(g-1)}_{j k i_1 \dots i_n} + \\ 
&+ \frac{1}{2} \bar C_{\bar i \bar j \bar k} e^{2K} G^{j \bar j} G^{k \bar k} \sum_{r=0}^g \sum_{s=0}^n \frac{1}{s! (n-s)!} \sum_{\sigma \in S_n} F^{(r)}_{j i_{\sigma(1)} \dots i_{\sigma(s)}} C^{(g-r)}_{k i_{\sigma(s+1)} \dots \sigma(n)}\\ 
& - (2g-2+n-1) \sum_{s=1}^n G_{i \bar i_s} C^{(g)}_{i_1 \dots i_{s-1} i_{s+1} \dots i_n}. 
\end{split} 
\end{equation} 
This equation can be summarized in terms of the generating function:
\be
F(\lambda,x^i;t^i)=\sum_{g=0}^\infty\sum_{n=0}^\infty
\lambda^{2g-2}\frac{1}{n!} C^{(g)}_{i_1\dots i_n} x^{i_1}\dots x^{i_n}
+(\frac{\chi}{24}-1)\log \lambda.
\ee
Contrary to~\cite{Bershadsky:1993cx}, we take the terms with
$2g-2+n\leq 0$ as given by $D_1\dots D_nF^{(g)}$ instead of
setting them to 0. Eq. (\ref{eq:BCOVholan}) implies that $F$ satisfies
\begin{eqnarray}
\label{eq:mastereq}
\bar \partial_i \exp(F)=\left[\frac{\lambda^2}{2} \bar F_{\bar i
    \bar j \bar k} e^{2K} G^{j \bar j} G^{k \bar
    k}\frac{\partial^2}{\partial x^j\partial x^k}-G_{\bar i
    j}x^j\left(\lambda\frac{\partial}{\partial \lambda}+x^k\frac{\partial}{\partial x^k} \right) \right]\exp(F).
\end{eqnarray}
To relate (\ref{eq:mastereq}) to the holomorphic anomaly 
Eq. (\ref{eq:holannew}) for this geometry, we split the $t^i$ into a
fibre parameter $\tau$ and base parameters $t^i$. Then we
write $F(\lambda,x;\tau,t)$ as a Fourier expansion instead of a Taylor
expansion in $x^i$:
\be
F(\lambda,x;\tau,t)=\sum_{g=0}^\infty \lambda^{2g-2}
F^{(g)}_\beta(\tau)\,f^{(g)}_\beta(x^i,t^i)\,e^{2\pi i \beta x}
p^\beta + (\frac{\chi}{24}-1)\log \lambda,
\ee
with $p^\beta=e^{2\pi i t\beta}$, and $f^{(g)}_\beta(x^i,t^i)$ are functions such that
$D_iF\vert_{x=0}=\partial_{x_i}F\vert_{x=0}$ and $f^{(g)}_\beta(0,t^i)=1$. 
In the large volume
limit, the covariant derivatives $D_i$ become flat
derivatives $\frac{\partial}{\partial t^i}$ and thus $f^{g}_\beta(x^i,t^i)\to 1$. Therefore, to deduce (\ref{eq:holannew})
from (\ref{eq:mastereq}) we can set $x^i=0$ and replace the
$\frac{\partial}{\partial x^i}$ by $\frac{\partial}{\partial t^i}$.

Eq. (\ref{eq:holannew}) follows now by considering
$\frac{1}{2\pi i}\partial_{\bar\tau}\exp(F)$ on the right hand side of
(\ref{eq:mastereq}). As discussed earlier, all $\bar \tau$ dependence arises from
completing the weight 2 Eisenstein series: $\widehat
E_2(\tau)=E_2(\tau)-\frac{3}{\pi \tau_2}$, which gives:
\be
\label{eq:diffE2}
\frac{\partial}{\partial E_2}=\frac{4\pi^2 
  \tau_2^2}{3}\frac{\partial}{2\pi i\partial \bar \tau}.
\ee

We first discuss how  the right-hand side of (\ref{eq:holannew}) can
be derived from Eq. (\ref{eq:mastereq}) for the geometry $X_{18}(11169)$.
We use the
basis (\ref{eq:intersection1}), and choose as parameters the ``base''
parameter $t=b+iJ$ (which is related to $r$ of
Subsec. \ref{sec:monodromy} by $r=e^{2\pi it}$) and the fibre parameter
$\tau=\tau_1+i\tau_2$. We are interested in the large volume limit $\tau\to i\infty$, $t\to i\infty$
in such a way that $J\gg \tau_2$. In this limit, the
K\"ahler potential is well approximated by the polynomial form:
\be
K\approx -\log(\textstyle{\frac{4}{3}}\,\mathcal{\tilde K}_{ijk} J^{i}J^{j}J^{k})= -\log(\frac{4}{3}(\alpha \tau_2^3+3\tau_2 J^2))
\ee
with $\alpha=\tilde{\cal K}_e^3$ (\ref{eq:intersection1}). This gives
for the metric:
\be
\left( \begin{array}{cc} G_{\tau\bar \tau} &  G_{t \bar
      \tau}\\G_{\tau\bar t} &  G_{t \bar t}\end{array}\right)\approx \left( \begin{array}{cc}\frac{1}{4\tau_2^2} & \frac{\alpha\tau_2}{3J^3}\\ \frac{\alpha\tau_2}{3J^3} & \frac{1}{2J^2} \end{array}\right),\non
\ee
which gives for the matrix $e^KG^{i\bar j}$:
\be
e^KG^{-1}\approx\left( \begin{array}{cc} \frac{1}{J^2} &
    -\frac{2\alpha\tau_2^2}{3J^3}\\ -\frac{\alpha\tau_2^2}{3J^3} &
    \frac{1}{2\tau_2} \end{array}\right)
\ee
Thus in the limit $J\to \infty$, one finds that only $e^KG^{t\bar
  t}\approx\frac{1}{2\tau_2}$ does not vanish. Therefore, $\bar C_{\bar \tau
    \bar j \bar k} e^{2K} G^{j \bar j} G^{k \bar k}\frac{\partial^2}{\partial x^j\partial x^k}\approx \frac{1}{4\pi^2}
  \frac{1}{4\tau_2^2}\frac{\partial^2}{\partial x^t \partial x^t}$.\footnote{The
    factor $\frac{1}{4\pi^2}$ appears due to a factor $-2\pi
    i$ between the moduli in
    \cite{Bershadsky:1993cx} and ours.}
Using (\ref{eq:diffE2}), this shows that (\ref{eq:mastereq}) reduces to:
\be
\label{eq:deranom}
\frac{\partial}{\partial E_2}
\exp(F)=\frac{\lambda^2}{24} \left(p\frac{\partial}{\partial p} \right)^2\exp(F).
\ee
Expansion of both sides in $p$ and taking the $p^n$ coefficient gives
a holomorphic anomaly equation as (\ref{eq:holannew}) for $g=0$. It
also gives the correct (\ref{eq:holannew}) for $g>0$ except for the
appearance of $K_B$. We belief that a more thorough analysis of the
covariant derivatives will explain this term.  Assuming the form 
$f^{(g)}_\beta(x,t)\to 1+x^2\,\beta\cdot K_B+\dots$ would give 
the shift in (\ref{eq:holannew}).

The derivation is very similar for the other types of fibres discussed
in Section \ref{classgeom}. The right hand side of
Eq. (\ref{eq:deranom}) is simply divided by $a$, in agreement with
\cite{Hosono:1999qc}.

\section{T-duality on the fibre}
\label{sec:sugrapartition}\setcounter{equation}{0} 

One can perform two T-dualities around the circles of the elliptic
fibre. Due to the freedom in choosing the circles, this leads to an 
$SL_2(\mathbb{Z})$ (or a congruence subgroup) group of dualities
mapping  IIA branes to IIA branes. This duality group is equal to
the modular subgroup of the monodromy group which leave invariant the $F_g$'s
 discussed in Sec. \ref{sec:monodromy}.

Let $D2_{f/\beta}$ be a $D2$-brane wrapped either on the elliptic fibre $f$ or on
a class $\beta$ in the base. Moreover, we denote by $D4_{f}$ a D4-brane wrapped around the
base and $D4_\beta$ is a D4-brane wrapped around the cycle $\beta$ in
the base and the fibre $f$. The double T-duality on both circles of the elliptic fibre transforms pairs of D-brane charges heuristically in the following way:
\begin{eqnarray}
&&\left( \begin{array}{l} D6 \\ D4_f\end{array} \right)=\gamma
\left(\begin{array}{l} \tilde{D6} \\ \tilde{D4}_f\end{array}
\right),\non \\
&& \left( \begin{array}{l} D4_\beta \\ D2_{\beta}\end{array} \right)=\gamma \left(\begin{array}{l} \tilde{D4}_\beta \\ \tilde{D2}_{\beta}\end{array} \right),\\
&& \left( \begin{array}{l} D2_{f} \\ D0\end{array} \right)=\gamma \left(\begin{array}{l} \tilde{D2}_{f} \\ \tilde{D0}\end{array} \right),\non
\end{eqnarray}
with $\gamma$ in $SL_2(\mathbb{Z})$ or a congruence subgroup.
See for more a more formal treatment of T-duality on Calabi-Yau's
\cite{Andreas:2000sj, Andreas:2001ve}. 
T-duality is not valid for every choice of the K\"ahler parameter. One
way to see this is that the BPS invariants of $D2$ branes do not depend on the
choice of the K\"ahler moduli but those of $D4$ and $D6$ branes
do through wall-crossing. The choice where the two are related by T-duality is sufficiently
close to the class of the elliptic fibre, this is called a suitable
polarization in the literature \cite{Huybrechts:1996}. Sufficiently close means that no wall
is crossed between the fibre class and the suitable polarization.

The equality of invariants of $D0$ branes and $D2$ branes
wrapping the fibre can be easily verified. The BPS invariant of D0
branes is known to be equal to the Euler number \cite{Kontsevich:2008}:
\be
\label{eq:D0index}
\Omega(\,(0,0,0,n),X)=-\chi(X).
\ee
One can verify in for example \cite{Hosono:1993qy} that these equal the BPS
invariants of D2 branes wrapping the $E_8$ elliptic fibre of $X$. If the modular
group is a congruence subgroup of level $n$ then only the BPS index
corresponding to $0\mod n$ $D2$ branes wrapping the fibre equals (\ref{eq:D0index}).   

Our interest is in the $D4$-branes which can be obtained from $D2_{\gamma}$ with
$\gamma=\beta+nf$ by T-duality. These $D4$-branes wrap classes
in the bases times the fibre, and have $D0$ brane charge $n$. 
$D4$-branes on Calabi-Yau manifolds correspond 
to black holes in 4-dimensional space-time and are well studied
\cite{Maldacena:1997de}, in particular M-theory relates the degrees of freedom of
$D4$-brane black holes to those of a $\CN=(4,0)$ CFT with left and
right central charges: 
\be
c_L= P^2+\half c_2\cdot P, \qquad c_R=P^3+c_2\cdot P,
\ee
with $P$ the 4-cycle wrapped by the D4-brane, and $c_2$ the second
Chern class of the Calabi-Yau. Typically, the number of 2-cycles in
the D4-brane is larger than the number of 2-cycles in the Calabi-Yau.

In the following, we will use the notation of \cite{Manschot:2009ia}. The homology class
$P$ gives naturally rise to a quadratic form $D_{ab}=d_{abc}P^c$
which has signature $(1,b_2(X)-1)$. Let $\Lambda$ be the lattice
$\mathbb{Z}^{b_2}$ with quadratic form $D_{ab}$. The dual lattice with quadratic form $D_{ab}^{-1}$ is
denoted by $\Lambda^*$. The K\"ahler modulus $J$ gives the
projection of a vector $\bfk\in \Lambda$ to the positive definite
subspace of $\Lambda\otimes \mathbb{R}$:
\be
\bfk_+=\frac{\bfk\cdot J}{J^2}\, J,
\ee
with $J^2=d_{abc}P^aJ^bJ^c$. 
%The partition function of a conformal
%field theory has integer coefficients and transforms as a modular
%form. It is however not necessary that it can be written as a sum of
%holomorphic times anti-holomorphic terms. For example, sigma models
%whose target space is non-compact. 

The supergravity partition function of D4-branes takes generically the following
form \cite{deBoer:2006vg, Manschot:2009ia}:
\begin{eqnarray}
\label{eq:sugrapartition}
\mathcal{Z}_P(C,\tau;t)&=&\sum_{Q_0,Q} \bar \Omega(\Gamma;t)\,(-1)^{P\cdot Q}\\
&&\times e\left(-\bar \tau \hat Q_{\bar 0}+\tau (Q-B)^2_+/2 +\bar \tau
(Q-B)^2_-/2 + C\cdot (Q-B/2)\right).\non
\end{eqnarray}
with $t=B+iJ$ the complexified K\"ahler modulus and $\hat Q_{\bar 0}=-Q_0+\half
Q^2$.  $\mathcal{Z}_P(C,\tau;t)$ transforms as a modular form of weight
$(\frac{1}{2},-\frac{3}{2})$. The invariants $\bar \Omega(\Gamma;t)$
are rational invariants and related to the integer invariants
$\Omega(\Gamma;t)$ by the multi-cover formula: 
\be
\bar \Omega(\Gamma;t)=\sum_{m|\Gamma}\frac{\Omega(\Gamma/m;t)}{m^2}
\ee
Note that the multi cover contributions come here with a factor $m^{-2}$,
whereas in Gromov-Witten theory they are multiplied with $m^{-3}$.
The invariants $\Omega(\Gamma;t)$ are related to the Euler number of the appropriate moduli space $\CM_t(\Gamma)$
by:
\be
\Omega(\Gamma;t)=(-1)^{\dim_\mathbb{C}\CM_t(\Gamma)}\,\chi(\CM_t(\Gamma)).
\ee

If the $B$-field decouples from the stability condition, $\mathcal{Z}_P(C,\tau;t)$ allows a theta function
decomposition:
\begin{eqnarray}
\label{eq:thetadecomp}
\mathcal{Z}_P(C,\tau;t)&=&\sum_{\mu} \overline{h_{P,\mu}(\tau)}\,\Theta_{P,\mu}(\tau,C,B),
\end{eqnarray}
and $h_{P,\mu}(\tau)$ is a vector valued modular form of weight
$-1-b_2(X)/2$ given by:
\be
h_{P,\mu}(\tau)=\sum_{Q_0}\Omega_P(\hat Q_{\bar 0})\,q^{\hat Q_{\bar 0}}
\ee
with $\hat Q_{\bar 0}=-Q_0+\half \mu^2$.  This symmetry is also present in the MSW conformal
field theory which arises in the near horizon geometry of a single
center $D4$-brane black hole \cite{deBoer:2006vg}. We refer to
\cite{Manschot:2009ia} for a discussion of the relation between the
supergravity partition and the CFT partition function.
In terms of the central charges of $c_{L/R}$ of this conformal field theory, $h_{P,\mu}(\tau)$
typically takes the form:
\be
h_{P,\mu}(\tau)=\frac{f_{P,\mu}(\tau)}{\eta(\tau)^{c_R}}
\ee
with $f_{P,\mu}(\tau)$ a vector-valued modular form of weight
$-1-b_2(X)/2+c_R/2$. Precisely this structure is also found for the
genus $0$ amplitudes obtained from the mirror periods, see Eq. (\ref{eq:Fg}) combined with
Eq. (\ref{eq:2ndchernint}). The prediction from the mirror periods is
obtained from the large base limit, and corresponds to $h_{P,{\bf 0}}(\tau)$

The triple intersection $P^3$ vanishes for the D4-branes obtained by T-duality from
the periods. These are therefore not large black holes, but we nevertheless obtain
detailed knowledge about the spectrum of ``small'' black holes using mirror symmetry. 
For small $D0$ and $D4$-brane charge, the BPS invariants can be
computed either from the microscopic $D$-brane perspective or the supergravity context 
\cite{Gaiotto:2006wm,Gaiotto:2007cd,Denef:2007vg, Collinucci:2008ht,
  deBoer:2008zn, Manschot:2010qz, Manschot:2011xc}.  
For example from the microscopic point of view, the moduli space of a single $D4$-brane is given by projective space
$\mathbb{P}^n$. Using index theorems one can compute that  $n=\frac{1}{6}P^3+\frac{1}{12}c_2\cdot
P-1$ \cite{Maldacena:1997de}. Therefore, the first coefficient of
$h_{P,0}(\tau)$ is expected to be
\be 
\label{eq:D4index}
\Omega_P(-\textstyle{\frac{1}{24}}c_R)=\frac{1}{6}P^3+\frac{1}{12}c_2\cdot P.
\ee
The second coefficient corresponds to adding a unit of (anti)
$D0$-brane charge.  Now the linear system for the divisor of the
$D4$-brane is constrained to pass through the $D0$-brane. This gives with
Eq. (\ref{eq:D0index}) \cite{Gaiotto:2006wm}:
\be
\Omega_P(1-\textstyle{\frac{1}{24}}c_R )\cong\chi(X) (\frac{1}{12}c_2\cdot P-1).
\ee
Here we have written a ``$\cong$'' instead of ``$=$'' since if
$1-\textstyle{\frac{1}{24}}c_R \geq 0$ gravitational degrees of freedom might start
contributing which are less well understood. 

Continuing with two units of $\bar D0$ charge, one finds:
\be
\label{eq:D42D0index}
\Omega_P(2-\textstyle{\frac{1}{24}}c_R)\cong \half\chi(X)(\chi(X)+5) (\frac{1}{12}c_2\cdot P-2).
\ee
One can in principle continue along these lines, which becomes
increasingly elaborate since
\begin{itemize}
\item[-] effects of $D2$-branes become important,
\item[-] single center black holes contribute for $\hat Q_{\bar 0}>0$,
\item[-] the index might depend on the background moduli $t$.
\end{itemize}

We now briefly explain which bound states appear in the supergravity
picture for small $D0/4$-brane charge.   The first terms in the $q$-expansion cannot correspond to single center
black holes since $\hat Q_{\bar 0}<0$. The first terms correspond to bound states
of $D6$ and $\bar D6$-branes \cite{Denef:2007vg}. If $P$ is an
irreducible cycle (it cannot be written as $P=P_1+P_2$ with $P_1$ and $P_2$ effective classes) then the charges $\Gamma_1$ and $\Gamma_2$ of the constituents are
\be
\label{eq:rank1charges}
\Gamma_1=(1,P,\half P^2-\frac{c_2}{24},\frac{1}{6}P^3+\frac{c_2\cdot
  P}{24}), \qquad \Gamma_2=(-1,0,\frac{c_2}{24},0),
\ee
The index of a 2-center bound state is given by:
  $$\left<\Gamma_1,\Gamma_2\right>\,\Omega(\Gamma_1)\,\Omega(\Gamma_2),$$
with $\left<\Gamma_1,\Gamma_2\right>=-P_1^0Q_{0,2}+P_1\cdot
Q_2-P_2\cdot Q^1+P^0_2Q_{0,1}$ the symplectic inner product. Since the constituents are single D6-branes with a non-zero
  flux, their index is $\Omega(\Gamma_i;t)=1$. 
Therefore, $\Omega_P(-\textstyle{\frac{1}{24}}c_R)=\left<\Gamma_1,\Gamma_2\right>=\frac{1}{6}P^3+\frac{1}{12}c_2\cdot 
  P$, which reproduces Eq. (\ref{eq:D4index}).

 One can continue in a similar  fashion with adding other constituents to compute indices with
  higher charge. For example, BPS states with charge
  $\Gamma=(0,2P,0,\frac{1}{3}P^3+\frac{c_2\cdot P}{12})$ corresponds to
$\Gamma_1$ as in (\ref{eq:rank1charges}) and
\be
\Gamma_2=(-1,P,-\half P^2+\frac{c_2}{24},\frac{1}{6}P^3+\frac{c_2\cdot P}{24}),
\ee
One obtains then $\Omega_{2P}(-\textstyle{\frac{1}{24}}c_R)=\frac{8}{6}P^3+\frac{2}{12}c_2\cdot P$. Similarly, one could also
add $\bar D0$ charges, and find the right hand sides of
Eqs. (\ref{eq:D4index}) to (\ref{eq:D42D0index}) with $P$ replaced by $2P$.

\subsection*{Example: $X_{18}(9,6,1,1,1)$}
\vspace{-.3cm}
We now consider the periods for $X_{18}(9,6,1,1,1)$, i.e. a elliptic fibtration over
$\bP^2$ and compare with the above discussion. This Calabi-Yau has a 2-dimensional K\"ahler cone, and lends it self
well to studies of D4-branes. We consider D4-branes wrapping the
divisor whose Poincar\'e dual is the hyperplane class $H$ of the base
surface $\bP^2$. The number of wrappings is denoted by $r$.

The genus 0 Gromov-Witten invariants are well-studied \cite{Candelas:1994hw, Hosono:1993qy}. 
Adjusting for the different power in the multi-cover formula, one obtains the following predictions for $h_{nH}(\tau)$:
\begin{eqnarray}
h_H(\tau)&=&\frac{31E_4^4+113E_4E_6^2}{48\eta(\tau)^{36}}\non
\\
&=&q^{-3/2}(3-1080\,q+143770\,q^2+204071184\,q^3+\dots),\non\\
h_{2H}(\tau)&=&\frac{-196319E_4E_6^5-755906E_4^4E_6^3-208991E_4^7E_6}{221184\,\eta(\tau)^{72}}-\frac{1}{24} E_2h_H(\tau)^2+\frac{1}{8}h_H(2\tau)\non\\
&=&q^{-3}(-6+2700\,q-574560\,q^2+\cdots)+\frac{1}{4}h_H(2\tau),\non \\
h_{3H}(\tau)&=&q^{-9/2}(27-17280\,q+5051970\,q^2+\cdots) +\frac{1}{9}h_H(3\tau).\non 
\end{eqnarray}
We want to compare this to the expressions derived above from the
point of view of $D4$-branes. For $r=1$, we have 
\be
\Omega(\Gamma;J)=\frac{1}{12}c_2\cdot H=3,
\ee
in agreement with the first coeficient of $h_H(\tau)$.
The second term in the $q$-expansion corresponds to
\be
\Omega(1,\half H,-1)=\chi(X) (\frac{1}{12}c_2\cdot P-1)=1080,
\ee
which is also in agreement with the periods. For two $\bar D0$ branes
we find a small discrepancy, one finds:
\be
\half (\frac{1}{12}c_2\cdot P-2)\chi(X)\left(\chi(X)+5\right)=144450.
\ee
This is an access of $1080=-2\chi(X)$ states compared to the 3rd coefficient in $h_1(\tau)$. This
number is very suggestive of a bound state picture, possibly involving
$D2$ branes. Since $\hat Q_{\bar 0}>0$ one could argue that these
states are due to intrinsic gravitational degrees of freedom, but it
seems actually a rather generic feature if we consider other elliptic
fibrations (e.g. over $\mathbb{F}_1$).  

For $r=2$, also the first two coefficients of the spectrum match with
the $D4$-brane indices, and the 3rd differs by $-6\chi(X)$. 
Something non-trivial happens for $r=3$. We leave an interpretation of these
indices from multi-center solutions for a future publication, and continue with
the example of the local elliptic surface \cite{Minahan:1998vr}.

\section{BPS invariants of the rational elliptic surface}
\label{sec:ellsurface}\setcounter{equation}{0} 

This section continues with the comparison of the $D4$- and
$D2$-brane spectra for $E_8$ elliptic fibration over the Hirzebruch
surface $\bF_1$ which was first addressed by Refs. \cite{Minahan:1998vr, Yoshioka:1998ti}.  
Let $\sigma:\bF_1\to X$ be the embedding of $\bF_1$ into the Calabi-Yau
The surface $\bF_1$ is itself a fibration $\pi: \bF_1\to C\cong \bP^1$
with fibre $f\cong\bP^1$, with  intersections $C^2=-1$, $C\cdot f=1$
and $f^2=0$. The K\"ahler cone of $X$ is spanned by the elliptic fibre
class $J_1$, 
and the classes $J_2=\sigma_*(C+f)$ and $J_3=\sigma_*(f)$.  The Calabi-Yau
intersections and Chern classes are given by (\ref{eq:QF1}).  
  
A few predictions from the periods for the D4-brane partition functions
are:
\begin{eqnarray}
\label{eq:predictions}
h_{C}(\tau)&=&\frac{E_4(\tau)}{\eta(\tau)^{12}}=q^{-1/2}(1+252q+\dots),
\\ 
h_{f}(\tau)&=&\frac{2E_4(\tau)E_6(\tau)}{\eta(\tau)^{24}}\non\\
&=&-2q^{-1}+480+282888q+\cdots,\non\\ 
h_{2C}(\tau)&=&\frac{E_2(\tau)E_4(\tau)^2+2E_4(\tau)
  E_6(\tau)}{24\eta(\tau)^{24}}+\frac{1}{8}h_C(2\tau)\non \\ 
&=& -9252\, q-673760\,q^2+\dots+\frac{1}{4}h_C(2\tau), \non\\
h_{3C}(\tau)&=&\frac{54E_2^2E_4^3+216E_2E_4^2E_6+109E_4^4+197E_4E_6^2}{15552\eta^{36}}+\frac{2}{27}h_{C}(3\tau)\non\\
&=&848628\,q^{3/2}+115243155\,q^{5/2}+\dots+\frac{1}{9}h_{C}(3\tau).\non
%Z_4&=q^{2}\frac{24E_2^3E_4^3+144E_2^2E_4^3E_6+109E_2E_4^5+269E_2E_4^2E_6^2+272E_4^4E_6+154E_4E_6^3}{62208\eta^{48}}. 
\end{eqnarray}

Since $c_2(X)\cdot f=24$, explicit expressions in terms of modular
forms for the divisors $h_{C+nf}(\tau)$ become rather
lengthy. Interestingly, one finds that for this class the first
coefficients (checked up to $n=12$), are given by $1+2n$ in agreement
with Eq. (\ref{eq:D4index}). Moreover, the second and third
coefficients are respectively given by
$\chi(X)(\frac{1}{12}c_2\cdot P-1)$ and $\half
\chi\,(\chi+9)\,(\frac{1}{12}c_2\cdot P-2)$ as long as the
corresponding $\hat Q_{\bar 0}<0$. 

Another interesting class are $r$ $D4$ branes wrapped on the divisor $C$,
which is however not an ample
divisor since $C=J_2-J_3$.  The Euler number of this divisor is
$c_2\cdot C=12$, it is in fact the rational elliptic surface $\bF_9$, which
is the 9-point blow-up of the projective plane
$\bP^2$, or equivalently, the 8-point blow-up of $\bF_1$.

For $r$ $D4$ branes we have $P=rC$. Eq. (\ref{eq:QF1}) shows
that the quadratic form $D_{abc}P^c$ restricted to $J_1$ and $J_3$ is:  
\be
r\left( \begin{array}{cc} 1 & 1 \\ 1 & 0 \end{array} \right)
\ee
The other 8 independent classes of $H^2(P,\mathbb{Z})$ are not
``visible'' to the computation based on periods, since these 2-cycles of
$P$ do not pull back to 2-cycles of $X$. 
We continue by confirming the expressions found from the periods with
a computation of the Euler numbers of the moduli spaces of semi-stable
sheaves as in Refs. \cite{Yoshioka:1998ti, Minahan:1998vr}. The algebraic computations are more naturally
performed in terms of Poincar\'e polynomials, and thus give more
refined information about the moduli space
\cite{Manschot:2011ym}. Moreover, the 8 independent classes which are
not visible from the Calabi-Yau point of view, can be distinguished
from this perspective.

One might wonder whether the extra parameter appearing with
the Poincar\'e polynomial is related to the higher genus
expansion of topological strings. However, the refined information of
the genus expansion captures is different. Roughly speaking, the D2-brane moduli space is a
torus fibration over a base manifold \cite{Gopakumar:1998jq}. The genus expansion
captures the cohomology of the torus, whereas the D4-brane moduli
space gives naturally the cohomology of the total moduli space. For
$r=1$, Ref. \cite{Hosono:1999qc} argues that the torus fibration is also present
for moduli spaces of rank 1 sheaves on $\bF_9$, but it is non-trivial
to continue this to higher rank. Another approach to verify the Fourier-Mukai transform at a refined
level is consider the refined topological string partition function with parameters $\epsilon_1$ and $\epsilon_2$, and then take the Nekrasov-Shatashvili limit
$\epsilon_1=0$, $\epsilon_2\ll 1$ instead of the topological string
limit $\epsilon_1=-\epsilon_2=g_s$. 

The structure described in Sec. \ref{sec:sugrapartition} for
$D4$-brane partition functions simplifies
when one specializes to a (local) surface. The charge vector
$\Gamma$ becomes $(r,\mathrm{ch}_1,\mathrm{ch}_2)$  with $r$ the ranks
and $\mathrm{ch}_i$ the Chern characters of the sheaf. Other frequently occuring quantities are the determinant
$\Delta=\frac{1}{r}(c_2-\frac{r-1}{2r}c_1^2)$, and
$\mu=c_1/r\in H^2(S,\mathbb{Q})$. 
In terms of the Poincar\'e polynomial
$p(\CM,w)=\sum_{i=0}^{2\dim_\mathbb{C}(\CM)}b_i(\CM)\,w^i$ of the
moduli space $\CM$, the (refined) BPS invariant reads:
\be
\Omega(\Gamma,w;J):=\frac{w^{-\dim_\mathbb{C}\CM_J(\Gamma)}}{w-w^{-1}}\, p(\CM_J(\Gamma),w),\non
\ee
In the case of surfaces, a formula is available for the dimension of
the moduli space:
\be
\label{eq:dimM}
\dim_{\mathbb{C}} \CM_J(\Gamma)=2r^2\Delta-r^2\chi(\CO_S)+1.\non
\ee
One can verify that the Poincar\'e polynomials computed later in this section are in
agreement with this formula.

The rational invariant corresponding to $\Omega(\Gamma,w;J)$ is \cite{Manschot:2010nc}: 
\begin{eqnarray}
\label{eq:refw}
\bar \Omega(\Gamma,w;J)&=&\sum_{m|\Gamma}
\frac{\Omega(\Gamma/m,-(-w)^m;J)}{m} 
\end{eqnarray}
The numerical BPS invariant $\Omega(\Gamma;J)$ follows
from the $\Omega(\Gamma,w;J)$ by:
\be
\label{eq:owtoo}
\Omega(\Gamma;J)=\lim_{w\to -1}
(w-w^{-1})\,\Omega(\Gamma,w;J),
\ee
and similarly for the rational invariants $\bar \Omega(\Gamma;J)$. 

The generating function (\ref{eq:sugrapartition}) becomes for a complex surface $S$:
\begin{eqnarray}
\label{eq:genfunction}
\mathcal{Z}_r(\rho,z,\tau;S,J)&=&\sum_{c_1, c_2}\,\bar \Omega(\Gamma,w;J)\,(-1)^{rc_1\cdot K_S}\,\non\\
&&\times \bar q^{r \Delta(\Gamma) -\frac{r\chi(S)}{24}-\frac{1}{2r}(c_1+rK_S/2)^2_-} q^{
  \frac{1}{2r}(c_1+rK_S/2)^2_+} e^{2\pi i \rho \cdot (c_1+rK_S/2)},
\end{eqnarray}
with $\rho \in H^2(S,\mathbb{C})$, $w=e^{2\pi i z}$ and $q=e^{2\pi i
  \tau}$. 
Twisting by a line bundle leads to an isomorphism of moduli
spaces. It is therefore sufficient to determine $\Omega(\Gamma,w;J)$
only for $c_1 \mod r$, and it moreover implies that $\mathcal{Z}_r(\rho,z,\tau;S,J)$ allows a
theta function decomposition as in (\ref{eq:thetadecomp}): 
\be
\label{eq:thetadecomp2}
\mathcal{Z}_r(\rho,z,\tau;S,J)=\sum_{\mu\in \Lambda^*/\Lambda} \overline{h_{r,\mu}(z,\tau;S,J)}\,\Theta_{r,\mu}(\rho,\tau;S),
\ee
with
\be
h_{r,\mu}(z,\tau;S,J)=\sum_{c_2} \bar \Omega(\Gamma,w;J)\,q^{r\Delta(\Gamma)-\frac{r\chi(S)}{24}},
\ee
and
\be
\Theta_{r,\mu}(\rho,\tau;S)=\sum_{\bfk \in H^2(S,r\mathbb{Z}) +rK_S/2+\mu} (-1)^{r\bfk
  \cdot K_S} q^{\bfk^2_+/2r}\bar q^{-\bfk^2_-/2r} e^{2\pi i\rho\cdot
  \bfk }.\non
\ee
Note that $\Theta_{r,\mu}(\rho,\tau;S)$ depends on $J$ through
$\bfk_\pm$ and does not depend on $z$.  

The generating function of the numerical invariants $\Omega(\Gamma;J)$ follows simply from Eq. (\ref{eq:owtoo}):
\be
\mathcal{Z}_{r}(\rho,\tau;S,J)=\lim_{z\to \frac{1}{2}} \quad
(w-w^{-1})\, \mathcal{Z}_{r}(z,\rho,\tau;S,J). 
\ee
Physical arguments imply that this function transforms as a
multivariable Jacobi form of weight $(\frac{1}{2},-\frac{3}{2})$
\cite{Vafa:1994tf, Manschot:2008zb} with a non-trivial multiplier system. For
rank $>1$ this is only correct after the addition of a suitable non-holomorphic
term \cite{Vafa:1994tf, Minahan:1998vr}.

%\setcounter{equation}{0}
%\label{sec:explicit}
This section verifies the agreement of the BPS invariants obtained
from the periods and vector bundles for
$h_{r,c_1}(z,\tau;\mathbb{F}_9,J_{m,n})$ for $r\leq 3$. The results
for $r\leq 2$ are due to G\"ottsche \cite{Gottsche:1990} and Yoshioka \cite{Yoshioka:1998ti}
The computations apply notions and techniques from algebraic
geometry as Gieseker stability, Harder-Narasimhan filtrations and the
blow-up formula. We refer to \cite{Manschot:2010nc, Manschot:2011ym}
for further references and details. The most crucial difference between the computations for
$\mathbb{F}_9$ and those for Hirzebruch surfaces in \cite{Manschot:2010nc, Manschot:2011ym}
is that the lattice arising from $H^2(\bF_9,\bZ)$ is now 10
dimensional. We continue therefore with giving a detailed description of different
bases of $H^2(\bF_9,\bZ)$, gluing vectors and theta functions.

\subsection{The lattice $H^2(\bF_9,\bZ)$} 
\label{sec:lattice}
%\vspace{-.4cm}
The second cohomology $H^2(\bF_9,\mathbb{Z})$  gives
naturally rise to a unimodular basis, it is in fact the unique
unimodular lattice with signature $(1,9)$, which we denote by
$\Lambda_{1,9}$. For this paper 3 different bases ($\bf C$, $\bf D$ and $\bf E$) of $\Lambda_{1,9}$ are useful. The
first basis is the geometric basis $\bf C$, which keeps manifest that 
$\mathbb{F}_9$ is the 9-point blow-up of the projective plane $\bP^2$.  
The basis vectors of $\bf C$ are $H$ (the hyperplane class of $\bP^2$) and
$\bc_i$ (the exceptional divisors of the blow-up). \footnote{We
  will use in general boldface to parametrize vectors.} The quadratic form is diag$(1,-1,\dots,-1)$.
The canonical class $K_9$ of $\mathbb{F}_9$ is given in terms of this
basis by:
\be
K_9=-3H+\sum_{i=1}^9 {\bf c}_i.
\ee
One can easily verify that $K^2_9=0$. Note that $-K_9$ is numerically effective but not ample.

The second basis $\bf D$ parametrizes $\Lambda_{1,9}$ as a gluing of the two non-unimodular lattices
$A$ and $D$. The basis $\bf D$ is given in terms of $\bf C$ by:
\begin{eqnarray}
&&{\bf a}_1=-K_9, \qquad {\bf a}_2=H-{\bf c}_9, \non \\
&&{\bf d}_i={\bf c}_{i}-{\bf c}_{i+1}, \qquad 1\leq i \leq 7, \label{eq:E8basis}\\
&&{\bf d}_8=-H+{\bf c}_7+{\bf c}_8+{\bf c}_9 . \non
\end{eqnarray}
The $\bf a_i$ are basis elements of $A$ and $\bf d_i$ of $D$. Since $A$ and $D$ are not unimodular, integral lattice elements of
$\bf C$ do not correspond to integral elements of $D$. For example, $\bc_9$ is given by
\be
{\bf c}_9=\frac{1}{2}\left({\bf a}_1+{\bf a}_2+\sum_{i=1}^6 i{\bf d}_i+3{\bf d}_7+4{\bf d}_8\right).
\ee
The other ${\bf c}_i$ are easily determined using ${\bf c}_9$. The quadratic form $\mathcal{Q}_A$ of the lattice $A$ is:
\be
\label{eq:QA}
\mathcal{Q}_A=\left( \begin{array}{cc} 0 & 2 \\ 2 & 0 \end{array} \right),
\ee
and $\mathcal{Q}_D$ of the lattice $D$ is minus the $D_8$ Cartan matrix:
\be
\label{eq:QD}
\mathcal{Q}_D=-\mathcal{Q}_{D_8}=-\left(\begin{array}{cccccccc} 2 & -1 & 0 & 0 & 0 & 0 & 0 & 0 \\ -1 & 2 & -1
     & 0 & 0 & 0 & 0 & 0 \\ 0& -1 & 2 & -1 & 0 & 0 & 0 & 0  \\ 0 & 0 &
     -1 & 2 & -1 & 0 & 0 & 0 \\ 0&  0 & 0 &
     -1 & 2 & -1 & 0 & 0 \\ 0 & 0 & 0 & 0 &
     -1 & 2 & -1 & -1 \\  0 & 0 & 0 & 0 & 0 &
     -1 & 2 & 0 \\  0 & 0 & 0 & 0 & 0 & 
     -1 & 0 & 2 \end{array}\right).
\ee

Gluing of $A$ and $D$ to obtain $\Lambda_{1,9}$ corresponds to an isomorphism
between $A^*/A$ and $D^*/D$. This isomorphism is given by 4 gluing
vectors $\boldsymbol g_i$, since the discriminants of $A$ and $D$ are
equal to 4. We choose them to be:
\begin{eqnarray}
{\boldsymbol g}_0&=&{\bold 0}, \non \\
{\boldsymbol g}_1&=&\half (1,0,1,0,1,0,1,0,0,1), \non \\
{\boldsymbol g}_2&=&\half (0,1,0,0,0,0,0,0,1,1), \non \\
{\boldsymbol g}_3&=&\half (1,1,1,0,1,0,1,0,1,0). \non 
\end{eqnarray}

Theta functions which sum over $D$ will play an essential role later
in this section. The theta functions $\Theta_{rD_8,\bmu}(\tau)$ are
defined by:
\be
\label{eq:thetaD8}
\Theta_{rD_8,\mu}(\tau)=\sum_{\bfk = {\bmu} \mod r\mathbb{Z}} q^{\frac{\bfk^2}{2r}}.
\ee
Such sums converge rather slowly. Therefore, we also give
their expression in terms of unary theta functions
$\theta_i(\tau)=\theta_i(0,\tau)$ (defined in Appendix
\ref{app:modfunctions}). For $r=1$ and the glue vectors $\bg_i$ one has:
\begin{eqnarray}
\Theta_{D_8,\bg_0}(\tau)&=&\half\left(\,
  \theta_3(\tau)^8+\theta_4(\tau)^8\,\right),\non \\
\Theta_{D_8,\bg_1}(\tau)&=&\half\theta_2(\tau)^8 , \non \\
\Theta_{D_8,\bg_2}(\tau)&=& \half\left(\,\theta_3(\tau)^8-\theta_4(\tau)^8\,\right), \non \\
\Theta_{D_8,\bg_3}(\tau)&=&\half \theta_2(\tau)^8 . \non 
\end{eqnarray}

For $r=2$, the $\mu$ in the $\Theta_{2D_8,\mu}(\tau)$ take values in
$D/2D$. The $2^8$ elements are naturally grouped in 6 classes with
multiplicities 1, 56, 140, 1, 56 and 2 depending on the corresponding
theta function $\Theta_{2D_8,\mu}(\tau)$. We choose as representative for each class:
\begin{eqnarray}
{\boldsymbol d}_0&=&\bold 0,\non \\
\bd_1&=&(1,0,0,0,0,0,0,0), \non \\
\bd_2&=&(1,0,1,0,0,0,0,0), \non \\
\bd_3&=&(0,0,0,0,0,0,1,1),\non \\
\bd_4&=&(1,0,1,0,1,0,0,0),\non \\
\bd_5&=&(1,0,1,0,1,0,1,0).\non 
\end{eqnarray}
% 
%Is there any theory which gives these classes and multiplicities
%directly? Maybe along the lines of Vakil (9910077).
%
Elements $\bmu \in \bg_i +D/2D$ fall similarly in conjugacy classes
corresponding to their theta functions. We let $m_{i,j}$ denote the
number of elements in the class represented by $\bg_i+\bd_j$. The non-vanishing
$m_{i,j}$ are given in Table \ref{tab:multiplic}.
\begin{table}[h!]
\begin{center}
\begin{tabular}{l|rrrrrr}
$m_{i,j}$ & 0 & 1 & 2 & 3 & 4 & 5 \\
\hline
0 & 1 & 56 & 140 & 1 & 56 & 2  \\
1 & 128 &  &  & 128 &  &    \\
2 & 16 & 112 & 112 &  & 16 &  \\
3 & 128 &  &  & 128 &  &    
\end{tabular}
\caption{The number of elements $m_{i,j}$ in $\bg_i+D/2D$ with equal theta
  functions $\Theta_{2D_8,\bg_i+\bd_j}(\tau)$.}  
\label{tab:multiplic}
\end{center}
\end{table}

The corresponding theta functions are given by:
\begin{eqnarray}
\Theta_{2D_8,\bd_0}(\tau)&=&\half\left(\,
  \theta_3(2\tau)^8+\theta_4(2\tau)^8\,\right),\non \\
\Theta_{2D_8,\bd_1}(\tau)&=&\textstyle{\frac{1}{16}}\left(\,
  \theta_3(\tau)^8-\theta_4(\tau)^8\,\right)-\half \theta_2(2\tau)^6\theta_3(2\tau)^2, \\
\Theta_{2D_8,\bd_2}(\tau)&=&\textstyle{\frac{1}{32}} \theta_2(\tau)^8,\non \\
\Theta_{2D_8,\bd_3}(\tau)&=&\half\left(\,
  \theta_3(2\tau)^8-\theta_4(2\tau)^8\,\right),\non \\
\Theta_{2D_8,\bd_4}(\tau)&=&\half \theta_2(2\tau)^6\theta_3(2\tau)^2,\non \\
\Theta_{2D_8,\bd_5}(\tau)&=&\half \theta_2(2\tau)^8,\non 
\end{eqnarray}
For $\bg_1$: 
\begin{eqnarray}
\Theta_{2D_8,\bg_1}(\tau)\quad\,\, &=& \textstyle{\frac{1}{8}}
\theta_2(\tau)^4\,\left( \theta_3(2\tau)^4-\frac{1}{2}\theta_4(2\tau)^4 \right),\non \\
\Theta_{2D_8,\bg_1+\bd_3}(\tau)&=&\Theta_{2D_8,\bd_2}(\tau), \non
\end{eqnarray}
For $\bg_2$: 
\begin{eqnarray}
\Theta_{2D_8,\bg_2}(\tau) \quad\,\, &=& \textstyle{\frac{1}{4}} \theta_2(\tau)^2\,\theta_3(2\tau)^6,\non \\
\Theta_{2D_8,\bg_2+\bd_1}(\tau)&=&\textstyle{\frac{1}{16}}\theta_2(\tau)^6\,\theta_3(2\tau)^2, \\
\Theta_{2D_8,\bg_2+\bd_2}(\tau)&=&\textstyle{\frac{1}{16}}\theta_2(\tau)^6\,\left(\,\theta_3(2\tau)^2-\theta_4(\tau)^2\, \right),\non \\
\Theta_{2D_8,\bg_2+\bd_4}(\tau)&=&\textstyle{\frac{1}{4}}\theta_2(2\tau)^6\theta_2(\tau)^2,\non 
\end{eqnarray}
For $\bg_3$: 
\begin{eqnarray}
\Theta_{2D_8,\bg_3}(\tau) \quad\,\, &=& \Theta_{2D_8,\bg_1}(\tau),\non \\
\Theta_{2D_8,\bg_3+\bd_3}(\tau)&=&\Theta_{2D_8,\bd_1+\bd_1}(\tau), \non
\end{eqnarray}

The third basis is basis $\bf E$ corresponding to the representation of
$\Lambda_{1,9}$ as the direct sum of the two lattices $B$ and $E$,
whose basis vectors ${\bf b}_i$ and ${\bf e}_i$ are:
\begin{eqnarray}
&&{\bf b}_1=-K_9, \qquad {\bf b}_2={\bf c}_9, \non \\
&&{\bf e}_i={\bf c}_{i}-{\bf c}_{i+1}, \qquad 1\leq i \leq 7, \label{eq:E8basis}\\
&&{\bf e}_8=-H+{\bf c}_6+{\bf c}_7+{\bf c}_8 . \non
\end{eqnarray}

The element $H$ of basis $\bf C$ is in terms of this basis: $H=(3,3,3,6,9,12,15,10,5,2)$.
The intersection numbers for $\bb_i$ are $\bb_1^2=0$, $\bb_2^2=-1$ and
$\bb_1\cdot \bb_2=1$. The quadratic form $\mathcal{Q}_E$ for $E$ is minus the $E_8$ Cartan matrix,
which is given by:
\be
\left(\begin{array}{cccccccc} 2 & -1 & 0 & 0 & 0 & 0 & 0 & 0 \\ -1 & 2 & -1
     & 0 & 0 & 0 & 0 & 0 \\ 0& -1 & 2 & -1 & 0 & 0 & 0 & 0  \\ 0 & 0 &
     -1 & 2 & -1 & 0 & 0 & 0 \\ 0&  0 & 0 &
     -1 & 2 & -1 & 0 & -1 \\ 0 & 0 & 0 & 0 &
     -1 & 2 & -1 & 0 \\  0 & 0 & 0 & 0 & 0 &
     -1 & 2 & 0 \\  0 & 0 & 0 & 0 & -1 & 
     0 & 0 & 2 \end{array}\right),
\ee

The 256 elements in $E/2 E$ fall in 3 inequivalent Weil
orbits, and orbits with vectors of length 0, 2 and 4
with multiplicities $m_0=1$, $m_1=120$ and $m_2=135$ respectively. We choose as
representatives:
\begin{eqnarray}
{\boldsymbol e}_0&=&\bf 0, \non \\
{\boldsymbol e}_1&=& (1,0,0,0,0,0,0,0), \non \\
{\boldsymbol e}_2&=& (1,0,1,0,0,0,0,0). \non 
\end{eqnarray} 
The corresponding theta functions $\Theta_{rE_8,{\boldsymbol e}_0}$
are for $r=1,2$:
\begin{eqnarray}
&&\Theta_{E_8,{\boldsymbol e}_0}(\tau)\,\, = E_4(\tau), \non \\
&&\Theta_{2E_8,{\boldsymbol e}_0}(\tau)= E_4(2\tau), \non \\
&&\Theta_{2E_8,{\boldsymbol e}_1}(\tau)=\frac{1}{240}\left(\,E_4(\tau/2) -E_4(\tau/2+1/2) \,\right), \non \\
&&\Theta_{2E_8,{\boldsymbol e}_2}(\tau)=\frac{1}{15}\left(\,E_4(\tau)-E_4(2\tau)\,\right). \non 
\end{eqnarray}

\subsection{BPS invariants for $r\leq 3$}
\subsubsection*{Rank 1}
\vspace{-.4cm}
The results from the periods for $h_C(\tau)$ is (\ref{eq:predictions}):
\be
\label{eq:ris1}
h_{C}(\tau)=\frac{E_4(\tau)}{\eta(\tau)^{12}}.
\ee
This can easily be verified with the results for sheaves on surfaces. 
The result for $r=1$ and a surface $S$ is \cite{Gottsche:1990}: 
\be
h_{1,c_1}(z,\tau;S)=\frac{i}{\theta_1(2z,\tau)\,\eta(\tau)^{b_2(S)-1}}
\ee
The dependence on $J$ can be omitted for $r=1$ since all rank 1
sheaves are stable. If we specialize to $S=\bF_9$, take the limit
$w\to -1$, and sum over all $c_1\in E$ one obtains Eq. (\ref{eq:ris1}).

\subsubsection*{Rank 2} 
\vspace{-.4cm}
The prediction by the periods for $r=2$ is given by $h_{2C}(\tau)$ in (\ref{eq:predictions}).
This is a sum over all BPS invariants for $c_1\cdot \ba_i=0$,
$i=1,2$. In order to verify this result, it is useful to decompose
$h_{2C}(\tau)$ according to the three conjugacy classes of $E/2E$:
$h_{2C}(\tau)=\sum_{i=0,1,2}m_i\,h_{2,{\boldsymbol
    e}_i}(\tau)\,\Theta_{2E_8,{\boldsymbol e}_i}(\tau)$. One obtains \cite{Minahan:1998vr}:
\begin{eqnarray} 
\label{eq:Z2e}
h_{2,{\boldsymbol e}_0} (\tau)&=& \frac{1}{24\,\eta(\tau)^{24}}\left[ E_2(\tau) \,\Theta_{2E_8,{\boldsymbol e}_0}(\tau)
  + \left(
    \theta_3(\tau)^4\theta_4(\tau)^4-\textstyle{\frac{1}{8}}
    \theta_2(\tau)^8\right)\left(\theta_{3}(\tau)^4+\theta_4(\tau)^4
  \right)\right], \non \\
&&+\frac{1}{8}h_{C}(2\tau),\non \\
h_{2,{\boldsymbol e}_1}(\tau)&=&
\frac{1}{24\,\eta(\tau)^{24}}\,\left[  E_2(\tau)  \,\Theta_{2E_8,{\boldsymbol e}_1}(\tau)
    -\textstyle{\frac{1}{8}}E_4(\tau)\,\theta_2(\tau)^4 \right], \\
h_{2,{\boldsymbol e}_2}(\tau)&=&
\frac{1}{24\,\eta(\tau)^{24}}\,\left[  E_2(\tau)  \,\Theta_{2E_8,{\boldsymbol e}_2}(\tau)
   -\textstyle{\frac{1}{8}}\theta_2(\tau)^8\left(\theta_3(\tau)^4+\theta_4(\tau)^4
  \right) \right]. \non 
\end{eqnarray}

Verification of these expressions is much more elaborate then for $r=1$. We will use the
approach of \cite{Yoshioka:1994, Yoshioka:1995, Yoshioka:1998ti}. 
The main issues are:
\begin{itemize}
\item[-] determination of the BPS
invariants for a polarization close to the class ${\bf a}_2$ (a suitable
polarization) 
\item[-] wall-crossing from the suitable polarization to $J=-K_9={\bf a}_1$. 
\end{itemize}
These issues are dealt with for the Hirzebruch surfaces \cite{Yoshioka:1994, Yoshioka:1995}, and for
$\bF_9$  in \cite{Yoshioka:1998ti}.  The main difficulty for $\bF_9$
compared to the Hirzebruch surfaces is that that the class $f$ and
$K_9$ span the lattice $A$, which is related to $\Lambda_{1,9}$ by a
non-trivial gluing with the lattice $D$.  

Before turning to the explicit expressions, we briefly outline the
computation; we refer for more details about the used techniques to
\cite{Manschot:2011ym}. The polarization $J$ is parametrized by
$J_{m,n}=m\,\ba_1+n\,\ba_2$. In order to determine the BPS invariants for the suitable polarization
$J_{\varepsilon,1}$, view $\mathbb{F}_{9}$ as the 8-point blow-up of
the Hirzebruch surface $\mathbb{F}_1$:  $\phi: \mathbb{F}_{9} \to\mathbb{F}_1$. 
We choose to perform this blow-up for the polarization $J_{\bF_1}=f$, with
$f$ the fibre class of the Hirzebruch surface.  The pull back of this
class to $\bF_9$ is $\phi^*f=J_{0,1}$. The generating function of the
BPS invariants for this choice takes a relatively simple form: it
either vanishes or equals a product of eta and theta functions 
\cite{Yoshioka:1995, Manschot:2011ym} depending on the Chern
classes. This function represents the sheaves whose restriction to the rational curve ${\bf a}_2$ is
semi-stable. The generating function $h_{r,c_1}(z,\tau;\mathbb{F}_9,J_{0,1})$ is 
therefore this product formula multiplied by the factors due to
blowing-up the 8 points. To obtain the BPS invariants from this function, one has to
change $J_{0,1}$ to $J_{\varepsilon,1}$ and subtract the contribution
due to sheaves which became (Gieseker) unstable due to this change \cite{Manschot:2011ym}.
Consequently, we can determine the BPS invariants for any other choice
of $J$ by the wall-crossing formula \cite{Yoshioka:1996, 
  Kontsevich:2008, Joyce:2008}. In particular, we determine the
invariants for $J_{1,0}=-K_9$ and change to the basis $\bf E$ in order to
compare with the expression from the periods.  

We continue with determining the BPS invariants for $J=J_{0,1}$. The
BPS invariants vanish for $c_1\cdot \ba_2=1\mod 2$: 
\be
\label{eq:h21}
h_{2,c_1}(z,\tau;\bF_9,J_{\varepsilon,1})=0, \qquad c_1\cdot \ba_2=1\mod 2. 
\ee
Since BPS invariants depend on $c_1 \mod 2\Lambda_{1,9}$, we
distinguish further $c_1\cdot \ba_2=0\mod 4$ and $c_1\cdot \ba_2=2\mod 4$. For
these cases, we continue as in \cite{Manschot:2011ym} using the (extended) Harder-Narasimhan
filtration. A sheaf $F$ which is unstable for $J_{\varepsilon,1}$ but
semi-stable for $J_{0,1}$, can be described as a HN-filtration of
length $2$ whose quotients we denote by $E_i$, $i=1,2$. If we parametrize the first Chern class of $E_2$ by
$\mathbf{k}=(\, {\bf k}_A, {\bf k}_D \,)$, then the discriminant $\Delta(F)$ is by:
\be
\label{eq:deter2}
2\Delta(F)=\Delta(E_1)+\Delta(E_2)-\frac{1}{4}(2 {\bf k}_A-c_1\vert_A)^2-\frac{1}{4}(2 {\bf k}_D-c_1\vert_D)^2.
\ee
The choice of ${\bf k}_D$ does not have any effect on the
stability of $F$ as long as $J$ is spanned by $J_{0,1}$ and
$_{1,0}$. Therefore (\ref{eq:deter2}) shows that the sum over
${\bf k}_D$ gives rise to the theta functions $\Theta_{2D_8,\mu}(\tau)$.
The condition for semi-stability for $J_{0,1}$ but unstable for
$J_{\varepsilon,1}$ implies $(c_1(E_1)-c_1(E_2))\cdot \ba_2=0$. This
combined with $c_1\cdot \ba_2=0\mod 4$ gives for $c_1(E_i)=0\mod 2$,
which shows that $c_1(E_i)=\boldsymbol g_j\mod 2
\Lambda_{1,9}$ only for $j=0,2$. One obtains after a detailed analysis
for $c_1\cdot \ba_2=0 \mod 4$: 
\begin{eqnarray}
\label{eq:h22a}
h_{2,c_1}(z,\tau;\bF_9,J_{\varepsilon,1})&=&\frac{-i\,\eta(\tau)}{\theta_1(2z,\tau)^2\,\theta_1(4z,\tau)}\prod_{i=1}^8B_{2,\ell_i}(z,\tau)
\\
&&  +\left( \frac{w^{4\{(\frac{1}{2}\bg_0-\frac{1}{4}c_1)\cdot \ba_1
      \}}}{1-w^4}-\frac{1}{2}\delta_{0,\{(\frac{1}{2}\bg_0-\frac{1}{4}c_1)\cdot
    \ba_1\} }\right)\,\Theta_{2D_8,c_1-2\bg_0}(\tau) \,h_{1,\bf 0}(z,\tau)^2 \non \\
&&+\left( \frac{w^{4\{(\frac{1}{2}\bg_2-\frac{1}{4}c_1)\cdot
      \ba_1
      \}}}{1-w^4}-\frac{1}{2}\delta_{0,\{(\frac{1}{2}\bg_2-\frac{1}{4}c_1)\cdot
    \ba_1\} }\right) \,\Theta_{2D_8,c_1-2\bg_2}(\tau) \,h_{1,\bf 0}(z,\tau)^2,  \non
\end{eqnarray}
where $\{\lambda \}=\lambda-\lfloor\lambda \rfloor$ and
$\ell_i=c_1\cdot \bc_i$. The right hand
side on the first line correspond to the sheaves whose restriction to
$\ba_2$ are semi-stable. The functions $B_{2,\ell}(z,\tau)=\sum_{n\in \mathbb{Z}+\ell/2} q^{n^2}w^n/\eta(\tau)^2$
are due to the blow-up formula \cite{Yoshioka:1996, Gottsche:1998,
  Li:1999, Manschot:2011ym}. The second and third line are the
subtractions due to sheaves which are unstable for
$J_{\varepsilon,1}$. 

Similarly one obtains for $c_1\cdot \ba_2=2 \mod 4$: 
\begin{eqnarray}
\label{eq:h22b}
h_{2,c_1}(z,\tau;J_{\varepsilon,1})&=&\frac{-i\,\eta(\tau)}{\theta_1(2z,\tau)^2\,\theta_1(4z,\tau)}\prod_{i=1}^8B_{2,\ell_i}(z,\tau)
\\
&&  +\left( \frac{w^{4\{(\frac{1}{2}\bg_1-\frac{1}{4}c_1)\cdot \ba_1
      \}}}{1-w^4}-\frac{1}{2}\delta_{0,\{(\frac{1}{2}\bg_1-\frac{1}{4}c_1)\cdot
    \ba_1\} }\right)\,\Theta_{2D_8,c_1-2\bg_1}(\tau) \,h_{1,\bf 0}(z,\tau)^2 \non \\
&&+\left( \frac{w^{4\{(\frac{1}{2}\bg_3-\frac{1}{4}c_1)\cdot
      \ba_1
      \}}}{1-w^4}-\frac{1}{2}\delta_{0,\{(\frac{1}{2}\bg_3-\frac{1}{4}c_1)\cdot
    \ba_1\} }\right) \,\Theta_{2D_8,c_1-2\bg_3}(\tau) \,h_{1,\bf 0}(z,\tau)^2.  \non
\end{eqnarray}

What remains is to change the polarization $J$ from
$J_{\varepsilon,1}$ to $J_{1,0}$ and determine the change of the
invariants using wall-crossing formulas. For
$J=(m,n,\bold 0)\in A\oplus D$, we obtain the following expression:
\begin{eqnarray}
\label{eq:h2mn}
h_{2,c_1}(z,\tau;
J_{m,n})&=&\frac{-i\,\eta(\tau)}{\theta_1(2z,\tau)^2\,\theta_1(4z,\tau)}\prod_{i=1}^8B_{2,\ell_i}(z,\tau)
\\
&&+\sum_{j=0,\dots, 3} h^A_{2,c_1-2\bg_j}(z,\tau; J_{m,n})\,\Theta_{2D,c_1-2\bg_j}(\tau).\non
\end{eqnarray}
with 
\begin{eqnarray} 
h^A_{2,c_1}(z,\tau; J_{m,n})&=&   h^A_{2,c_1}(z,\tau; J_{\varepsilon,1})+\half\sum_{(a_1,a_2) \in A+c_1} \half \left(
  \,\sgn(a_1n+a_2m) - \sgn(a_1+a_2\varepsilon)\,\right)  \non \\
&&  \times\, (w^{4a_2}-w^{-4a_2})\,q^{-4a_1a_2}
h_{1,\bf 0}(z,\tau)^2, \non 
\end{eqnarray}
The functions  $h^A_{2,c_1}(z,\tau; J_{\varepsilon,1})$ are
rational functions in $w$ multiplied by $h_{1,\bf 0}(z,\tau)^2$  which can easily be read off from Eq. (\ref{eq:h22a}). 
 For $J=J_{1,0}$ the functions can be expressed in terms of modular functions.

Table \ref{tab:h20w} presents the BPS invariants for $J=J_{1,0}$. 
As expected, the Euler numbers are indeed in agreement with the predictions (\ref{eq:Z2e}). One can also verify
that for increasing $c_2$, the Betti numbers asymptote to those of
$r=1$ or equivalently the Hilbert scheme of points of $\mathbb{F}_9$.

\begin{center}
\begin{table}[h!]
\begin{tabular}{llrrrrrrrrrrr}
$c_1$ & $c_2$ & $b_0$ & $b_2$ & $b_4$ & $b_6$ & $b_8$ & $b_{10}$ & $b_{12}$ & $b_{14}$
& $b_{16}$ &  $\chi$ \\
\hline
$\boldsymbol{e}_0$ & 2 & 1 & 10 & 55 & & & & & & &   132  \\
& 3 & 1 & 11 & 76 & 396 & 1356 &  &  &  &  &  3680 \\
& 4 & 1 & 11 & 78 & 428 & 1969 & 7449 & 20124 & &  &    60120 \\ 
& 5 & 1 & 11 & 78 & 430 & 2012 & 8316 & 30506 & 95498 & 221132 &     715968\\ 
$\boldsymbol{e}_1$ & 1 & 1 & 9  & & & & & & & &  20  \\
& 2 & 1 & 11 & 75 & 309 &  & &  &  &  &  792 \\
& 3 & 1 & 11 & 78 & 426 & 1843 & 5525 &  & &  &   15768 \\
& 4 & 1 & 11 & 78 & 430 & 2010 & 8150 & 27777 & 68967 & &  214848 \\ 
$\boldsymbol{e}_2$ & 1 & 1 &  &  & & & & & & &  2  \\
& 2 & 1 & 11 & 60 &  &  &  & & &  &    144 \\
& 3 & 1 & 11 & 78 & 404 & 1386 & & & & &     3760  \\
& 4 & 1 & 11 & 78 & 430 & 1981 & 7495 & 20244 & & &  60480 
\end{tabular}
\caption{The Betti numbers $b_n$ (with $n\leq
  \dim_\mathbb{C} \mathcal{M}$) and Euler numbers $\chi$ of the moduli spaces of semi-stable sheaves
  on $\bF_9$ with $r=2$, $c_1={\boldsymbol e}_i$, and $1\leq c_2\leq 4$ for $J=J_{1,0}$.}  
\label{tab:h20w}
\end{table}
\end{center}

We define the functions $h^A_{2,c_1}(z,\tau):=h^A_{2,c_1}(z,\tau;J_{1,0})$, which only 
depend on $c_1|_A=\alpha_1\,\ba_1+\alpha_2\,\ba_2$ with $\alpha_1,\alpha_2 \in
0,\half, 1, \frac{3}{2}$. One finds for $\alpha_2=0 \mod 4$:  
\be
\frac{h^A_{2,c_1}(z,\tau)}{h_{1,0}(z,\tau)^2}=-\frac{1}{8}\frac{1}{2\pi i }
\frac{\partial}{\partial z} \ln \left(\, \theta_1(4\tau,4z+2\alpha_1)\,\theta_1(4\tau,4z-2\alpha_1)\,\right),
\ee
and for $\alpha_2\neq 0\mod 4$ using (\ref{eq:F}):
\be
\frac{h^A_{2,c_1}(z,\tau)}{h_{1,0}(z,\tau)^2}=\frac{i}{2}\frac{q^{-\alpha_1\alpha_2}\,\eta(4\tau)^3}{\theta_1(4\tau,2\alpha_2\tau)}\left(\frac{w^{-2\alpha_2}\,\theta_1(4\tau,4z+2(\alpha_1-\alpha_2)\tau)}{\theta_1(4\tau,4z+2\alpha_1\tau)}-\frac{w^{2\alpha_2}\,\theta_1(4\tau,-4z+2(\alpha_1-\alpha_2)\tau)}{\theta_1(4\tau,4z+2\alpha_1\tau)}   
\right).
\ee

To prove the agreement of the Euler numbers with the periods, we specialize to $w=-1$.
Let $D_k=\frac{1}{2\pi i}\frac{\partial}{\partial
  \tau}-\frac{k}{12}E_2(\tau)$ be the differential operator which maps
weight $k$ modular forms to modular forms of weight $k+2$. Then one can
write $h_{2,c_1}(\tau;J_{1,0})$ as:
\begin{eqnarray}
\label{eq:h2num10}
h_{2,c_1}(\tau;J_{1,0})&=&\frac{1}{\eta(\tau)^{24}} \left(\, \frac{1}{2}
\delta_{c_1\cdot {\bf a}_2,0}
D_4(\theta_3(2\tau)^m\theta_2(2\tau)^{8-m}) \right. \\
&&\left. \qquad +\sum_{i=0,\dots,3}
  f^A_{c_1-2\bg_j}(\tau)\,\Theta_{2D,c_1-2\bg_j}(\tau) \right), \non
\end{eqnarray}
with
\be
\begin{array}{ll}
f^A_{0,0}(\tau)& = \textstyle{\frac{1}{8}}\theta_3(2\tau)^4+\textstyle{\frac{1}{24}}E_2(\tau), \non \\
f^A_{-\frac{1}{2},\frac{1}{2}}(\tau)&= \half \theta_2(2\tau)\, \theta_3(2\tau)^3,  \non \\
f^A_{\frac{1}{2},\frac{1}{2}} (\tau)&= \half \theta_2(2\tau)^3\,
\theta_3(2\tau), \non \\
f^A_{1,0}(\tau)&= \textstyle{\frac{1}{12}}\theta_2(2\tau)^4-\textstyle{\frac{1}{24}}\theta_3(2\tau)^4+\textstyle{\frac{1}{24}} E_2(\tau),\non \\
f^A_{0,1}(,\tau)&=\textstyle{\frac{1}{24}}\theta_2(2\tau)^4-\textstyle{\frac{1}{12}}\theta_3(2\tau)^4, \non \\
f^A_{1,1}(z,\tau)&=-\textstyle{\frac{1}{8}}\theta_2(2\tau)^4. \non
\end{array}
\ee
If $c_1|_B=0$, this reproduces the
functions in \cite{Minahan:1998vr, Yoshioka:1998ti} depending on whether the classes
in lattice $E$ are even or odd.

\subsubsection*{Modularity}
\vspace{-.3cm}
Electric-magnetic duality of $\CN=4$ $U(r)$ Yang-Mills theory implies
modular properties for its partition function
\cite{Vafa:1994tf}. Determination of the modular properties gives
therefore insight about the quantum realization of electric-magnetic duality.

The expression in Eq. (\ref{eq:h2mn}) does not transform as a
modular form for generic choices of $J$. However, using the
theory of indefinite theta functions \cite{Zwegers:2002}, the
functions can be completed to a function $\widehat h_{2,c_1}(z,\tau;
J)$ which does transform as a modular form \cite{Manschot:2011dj}.   
Interestingly, Eq. (\ref{eq:h2num10}) shows that $h_{2,c_1}(z,\tau;J)$ becomes a quasi-modular form for $\lim_{J\to J_{1,0}}\widehat
h_{2,c_1}(z,\tau;J)$, i.e. it can be expressed in terms of modular
forms and Eisenstein series of weight 2. In some 
cases it becomes even a true modular form. This is due to
the special form of $\mathcal{Q}_A$. 

The transition from mock modular to quasi-modular can be made precise.
Due to the gluing vectors, the function $f_{2,c_1}(z,\tau;J)=h_{2,c_1}(z,\tau;J)/h_{1,c_1}(z,\tau)^2$ takes the form:
\begin{eqnarray}
f_{2,c_1}(z,\tau;J_{m,n})&=&\sum_{\mu}
f^A_{2,(c_1-2\mu)_A}(z,\tau;J_{m,n})\,\Theta_{2D,(2\mu-c_1)_D}(\tau)
\non \\
&& + \,\delta_{c_1\cdot \ba_2,0}\,\frac{i\,\eta(\tau)^3}{\theta_1(\tau,4z)}\theta_{3}(2\tau,2z)^k
\theta_{2}(2\tau,2z)^{8-k},
\end{eqnarray}
where $k$ is the number of $c_1\cdot \bc_i=1\mod 2$  for $1\leq i \leq
8$. 

The completed generating function $\widehat f_{2,c_1}(z,\tau;J)$  is a slight generalization of
 Eq. (22) in Ref \cite{Manschot:2011dj}:\footnote{Here we
   have used the equation for $\beta_{\frac{3}{2}}(x)$ in terms of
   $\beta_{\frac{1}{2}}(x)$ and $e^{-\pi x}$.}
\begin{eqnarray}
&&\widehat f_{2,c_1}(\tau;J)=f_{2,c_1}(\tau;J)+ \\ 
&&\quad \sum_{\mathbf{c}\in -c_1\atop + H^2(\Sigma_9, 
  2\mathbb{Z})}\left(\frac{K_{9}\cdot J}{4\pi \sqrt{J^2\,y}}\,e^{-\pi
  y \mathbf{c}_+^2}
-\frac{1}{4} K_{9}\cdot \mathbf{c} \,\sgn(\mathbf{c}\cdot
J)\,\beta_{\frac{1}{2}}(\mathbf{c}_+^2\,y)\right)\,(-1)^{K_{9}\cdot
\mathbf{c}} q^{-\mathbf{c}^2/4},  \non 
\end{eqnarray}

We parametrize $J$ by $a_0+t\,a_1$, and carefully
study the limit $t\to 0$ (this should correspond to $R\to \infty$ in
\cite{Minahan:1998vr}). In this limit, $J$ approaches
$-K_9$. Moreover, $J\cdot K_9=t$ and $J^2=t(2-t)$. If one parametrizes $\mathbf{c}$ by
$(n_0,n_1,\mathbf{c}_\perp)$, only terms with $n_1=0$ contribute to
the sum in the limit $t\to 0$. Therefore the term with
$\beta_{\frac{1}{2}}$ does not contribute to the anomaly. After a Poisson resummation
on $n_0$, one finds that the limit is finite and given by
\begin{eqnarray}
&&\widehat f_{2,c_1}(\tau;J_{1,0})=f_{2,c_1}(\tau;J_{1,0})+
\frac{\delta_{c_{1}\cdot \ba_1,0}}{4\pi\, \im \tau} \sum_{\mathbf{c}\in -c_{1,\perp}\atop + H^2_\perp(\Sigma_9,
  2\mathbb{Z})}q^{-\mathbf{c}_\perp^2/4}.
\end{eqnarray}
This reproduces the holomorphic anomaly equation discussed in Section
\ref{sec:quantumgeom} for the periods. Thus the non-holomorphic
dependence of $D4$-brane partition functions is consistent with the
one of topological strings when the systems can be related via $T$-duality.
 Note that for $c_1\cdot \ba_1=1 \mod 2$, the non-holomorphic dependence
of $f_{2,c_1}(\tau;J)$ vanishes in the limit $J\to J_{1,0}$, in
agreement with (\ref{eq:h2num10}).

%We define the following six functions:
%
%\begin{eqnarray}
%f^A_{0,0}(z,\tau)&=&-\frac{\theta_1'(4\tau,4z)}{4\,\theta_1(4\tau,4z)}
%\non \\
%f^A_{-\frac{1}{2},\frac{1}{2}}(z,\tau)&=&\frac{-iq^{-\frac{3}{4}}}{2}\frac{\eta(4\tau)^3}{\theta_1(4\tau,3\tau)}\left(\frac{w^3\,\theta_1(4\tau,4z+2\tau)}{\theta_1(4\tau,4z-\tau)}-\frac{w^{-3}\theta_1(4\tau,4z-2\tau)}{\theta_1(4\tau,4z+\tau)}
%\right)\non\\
%f^A_{\frac{1}{2},\frac{1}{2}} (z,\tau)&=&\frac{-iq^{-\frac{1}{4}}}{2}\frac{\eta(4\tau)^3\,\theta_1(4\tau,4z)}{\theta_1(4\tau,\tau)}\left(\frac{w}{\theta_1(4\tau,4z-\tau)}-\frac{w^{-1}}{\theta_1(4\tau,4z+\tau)}
%\right),\non \\
%f^A_{1,0}(z,\tau)&=&
% -\frac{\theta_1'(4\tau,4z+2\tau)}{4\,\theta_1(4\tau,4z+2\tau)}-\frac{1}{2}
%\non \\
%f^A_{0,1}(z,\tau)&=&\frac{-i}{2}\frac{\eta(4\tau)^3}{\theta_1(4\tau,2\tau)}\left(\frac{w^2\,\theta_1(4\tau,4z+2\tau)}{\theta_1(4\tau,4z)}-\frac{w^{-2}\theta_1(4\tau,4z-2\tau)}{\theta_1(4\tau,4z)}\right)\non\\  
%f^A_{1,1}(z,\tau)&=&
%\frac{-iq}{2}\frac{\eta(4\tau)^3}{\theta_1(4\tau,2\tau)}\left(\frac{w^2\,\theta_1(4\tau,4z+4\tau)}{\theta_1(4\tau,4z+2\tau)}-\frac{w^{-2}\theta_1(4\tau,4z-4\tau)}{\theta_1(4\tau,4z-2\tau)}\right)\non
%\end{eqnarray}
%
%where $'$ stands for $\frac{1}{2\pi i }\frac{\partial}{\partial z}$.

%

%Note that there is a natural candidate for the partition function of the conformal field theory for $r=2$:
%
%\be
%-2\frac{E_4E_6}{\eta(\tau)^{24}}=-2q^{-1}+480+282888q+\dots
%\ee
%
%Note that this gets the first and second coefficient right.

\subsubsection*{Rank 3}
\vspace{-.4cm}
Similarly as for $r=2$, Ref. \cite{Minahan:1998vr} also decomposes
$h_{3C}(\tau)$ into different Weyl orbits. We will restrict in the
following to the ${\boldsymbol e}_0=\bf 0$ orbit in $E/3E$ since the
expressions become rather lengthy. In order to present
$h_{3,{\boldsymbol e}_0}(\tau)$, define 
\be
b_{3,\ell}(\tau)=\sum_{m,n\in \mathbb{Z}+\ell/3} q^{m^2+n^2+mn}.
\ee
Then $h_{3,{\boldsymbol e}_0}(\tau)$ is given by \cite{Minahan:1998vr}:
\begin{eqnarray} 
h_{3,{\boldsymbol
    e}_0}(\tau)&=&\frac{1}{2592\,\eta^{36}}\left[\left(51\,b_{3,0}^{12}-184\,b_{3,0}^9 b_{3,1}^3+336\,b_{3,0}^6 b_{3,1}^6+288\, b_{3,0}^3
b_{3,1}^9+32\, b_{3,1}^{12}\right)\right.\non\\
&& + E_2b_{3,0}\left(\,36\,b_{3,0}^{9}-112\,b_{3,0}^6 b_{3,1}^3+32\,b_{3,0}^6 b_{3,1}^3-64\, b_{3,1}^9\right)\\
&&\left. + E_2^2 b_{3,0}^2\left(9\,b_{3,0}^{6}-16\,b_{3,0}^3 b_{3,1}^3+16\,b_{3,0}^6 \right)\right]\non
%&=&4068\,q^{3/2}+251235\,q^{5/2}+8037792\,q^{7/2}+\dots
\end{eqnarray}

In order to verify this expression, we extend the analysis for $r=2$
to $r=3$. For $c_1\cdot \ba_2=\pm 1\mod 3$ the BPS invariants vanish
for a suitable polarization:
\be
h_{3,c_1}(z,\tau;J_{\varepsilon,1})=0.
\ee

The HN-filtrations for the sheaves which are unstable for
$J_{\varepsilon,1}$ but semi-stable for $J_{0,1}$ have length 2 or 3. From those of length
2, one obtains rational functions in $w$ multiplied by
$h_{1,0}(z,\tau)\,h_{2,\bmu}(z,\tau)\,\Theta_{2D_8,\bmu}(\tau)$, with
$\bmu=\bf 0$, $\ba_2$, $\bd_i$ and $\bd_i+\ba_2$. The theta function
arising from the sum over the $D_8$ lattice is more involved for
filtrations of length 3. Instead of a direct sum, a ``twisted'' sum of 2 $D_8$-lattices
appears; we will denote this lattice by $D_8^\mathrm{t}$:
\begin{eqnarray}
\Theta_{2D_{8}^\mathrm{t};\bmu_1,\bmu_2}(\tau)&=&\sum_{\bfk_i\in
  D_8+\bmu_i,i=1,2}q^{\bfk_1^2+\bfk_1\cdot \bfk_2+\bfk_2^2} \\
&=&\sum_{i}m_i\,\Theta_{2D_8,\bmu_1+\bmu_2+\bd_i}(\tau)\,\Theta_{2D_8,\bmu_1-\bmu_2+\bd_i}(3\tau)
\end{eqnarray}
where  $m_i$ are the multiplicities of the theta characteristics
$\bmu_1+\bmu_2+\bd_i$, thus for $\bmu_1+\bmu_2\in D$, $i=1,\dots ,6$,
and for $\bmu_1+\bmu_2\in D/2$, $i=1,\dots ,4$. For numerical
computations the second line is considerably faster than the first
line.  We obtain after a careful analysis: 
\begin{eqnarray}
\label{eq:h30}
h_{3,0}(z,\tau; J_{\varepsilon,1})&=&\frac{i\eta(\tau)^{3}}{\theta_1(2z,\tau)^2\,\theta_1(4z,\tau)^2\,
  \theta_1(6z,\tau)}\,B_{3,0}(z,\tau)^8  \\
&&+2\left(\frac{1}{1-w^{12}}-\frac{1}{2}
\right)h_{1,0}(z,\tau)\sum_{i=0,3}h_{2,(0,0,\bd_i)}(z,\tau;J_{\varepsilon,1})\Theta_{2D_8,\bd_i}(3\tau)\non
\\
&&+2\left(\frac{w^6}{1-w^{12}}\right)h_{1,0}(z,\tau)\sum_{i=0,3} h_{2,(0,1,\bd_i)}(z,\tau;J_{\varepsilon,1})\Theta_{2D_8,\bd_i}(3\tau)\non
\\ 
&&+2\left(\frac{1}{1-w^{6}}-\frac{1}{2}\right)h_{1,0}(z,\tau)
\sum_{i=1,2,4,5} m_{0,i}\,h_{2,(0,0,\bd_i)}(z,\tau;J_{\varepsilon,1})\Theta_{2D_8,\bd_i}(3\tau)
\non\\
&&+2\left(\frac{w^3}{1-w^{6}}\right)h_{1,0}(z,\tau)
\sum_{i=0,1,2,4} m_{2,i}\,h_{2,\bg_2+\bd_i}(z,\tau;J_{\varepsilon,1})\Theta_{2D_8,\bg_2+\bd_i}(3\tau)
\non\\
&&-\left(\frac{1+w^{12}}{(1-w^8)(1-w^{12})}-\frac{1}{1-w^{12}}
  +\frac{1}{6}\right)\Theta_{2D_8^\mathrm{t};0,0}(\tau)\,
h_{1,0}(z,\tau)^3 \non \\
&&-2\left(\frac{w^{6}}{(1-w^4)(1-w^{12})}-\frac{w^6}{1-w^{12}}\right)\Theta_{2D_8^\mathrm{t};\bg_2,0}(\tau)\, 
h_{1,0}(z,\tau)^3\non \\
&&-\left(\frac{w^4+w^{16}}{(1-w^8)(1-w^{12})}\right)\Theta_{2D_8^\mathrm{t};0,0}(\tau)\, h_{1,0}(z,\tau)^3\non
\end{eqnarray}
The functions due to the blowing-up of 8 points are now given by \newline
$B_{3,k}(z,\tau)=\sum_{m,n \in \mathbb{Z}+k/3} q^{m^2+n^2+mn}w^{4m+2n}/\eta(\tau)^3.$
We have used in (\ref{eq:h30}) that $h_{2,c_1}(z,\tau;J_{m,n})$ only depends on the 
conjugacy class of $c_1$ in $D/2D$, and moreover that $h_{2,c_1}(z,\tau;J_{m,n})=h_{2,c_1'}(z,\tau;J_{m,n})$ if $c_1=(0,0,\bd_i)$ and
$c_1'=(0,1,\bd_i)$ for $i=1,2,4,5$ (but not for $i=0, 3$) and $c_1=(0,0,\bd_i)+\bg_2$
and $c_1'=(0,1,\bd_i)+\bg_2$. 

%
%For $c_1\cdot \ba_2=0\mod 6$:
%
%\begin{eqnarray}
%h_{3,c_1}(z,\tau;J_{\varepsilon,1})&=&\frac{i\eta(\tau)^{3}}{\theta_1(2z,\tau)^2\,\theta_1(4z,\tau)^2\,
%  \theta_1(6z,\tau)}\prod_{i=1}^8B_{3,k_i}(z,\tau)\\
%&& +
%\end{eqnarray}
%
%For $c_1\cdot \ba_2=3\mod 6$:
%
%\begin{eqnarray}
%h_{3,c_1}(z,\tau;J_{\varepsilon,1})&=&\frac{i\eta(\tau)^{3}}{\theta_1(2z,\tau)^2\,\theta_1(4z,\tau)^2\,
%  \theta_1(6z,\tau)}\prod_{i=1}^8B_{3,k_i}(z,\tau) \\
%&& +
%\end{eqnarray}
%
Having determined $h_{3,0}(z,\tau; J_{\varepsilon,1})$, what rests is
to perform the wall-crossing from $J_{\varepsilon,1}$ to $J_{1,0}$. To
this end we define:
\begin{eqnarray}
h^A_{3,c_1}(z,\tau;J)&=&\sum_{\ba = c_1|_A\mod 2A} \half \left( \sgn(a_1n+a_2 m) -\sgn(a_1+a_2\varepsilon )\right)\non\\
&&\left(w^{6a_2}-w^{-6a_2}\right)q^{-3a_1a_2}\,h_{2,(\ba,c_1|_D) }(z,\tau;J_{|a_1|,|a_2|})\,h_{1,0}(z,\tau),
\end{eqnarray}
with $\ba=(a_1,a_2)$. Then $h_{3,0}(z,\tau;J)$ is given by \cite{Manschot:2010xp, Manschot:2010nc}:
\be
h_{3,0}(z,\tau;J)=h_{3,0}(z,\tau;J_{\varepsilon,1})+\sum_{\ba\in
2A/A} m_{i,j}\, h^A_{3,\ba +\bg_i+\bd_j}(z,\tau;J)\,
\Theta_{2D_8,\bg_i+\bd_j}(3\tau). \non
\ee
The Betti numbers for $J=J_{1,0}$ and small $c_2$  are presented in Table
\ref{tab:h30w}, and indeed agree with the Euler numbers computed from the
periods.  
\begin{center}
\begin{table}[h!]
\footnotesize
\begin{tabular}{lrrrrrrrrrrrrrr}
$c_2$ & $b_0$ & $b_2$ & $b_4$ & $b_6$ & $b_8$ & $b_{10}$ & $b_{12}$ & $b_{14}$
& $b_{16}$ &  $b_{18}$ & $b_{20}$ & $b_{22}$ &   $\chi$ \\
\hline
3 & 1 & 10 & 65 & 320 & 1025 & 1226 & & & & &&&  4068  \\
4 & 1 & 11 & 77 & 417 & 1902 & 7372 & 23962 & 57452 & 68847 &&&& 251235
\\
5 & 1 & 11 & 78 & 429 & 2002 & 8260 & 30710 & 103867 & 316586 &
836221 & 1706023 & 2029416 & 8037792 \end{tabular}
\caption{The Betti numbers $b_n$ (with $n\leq
  \dim_\mathbb{C} \mathcal{M}$) and the Euler number $\chi$ of the moduli spaces of semi-stable sheaves
  on $\bF_9$ with $r=3$, $c_1=0$, and $3\leq c_2\leq 5$ for $J=J_{1,\varepsilon}$.}  
\label{tab:h30w}
\end{table}
\end{center}

One might wonder how to derive the modular properties
$h_{3,0}(z,\tau;J)$. The completion takes in general a very
complicated form due to the quadratic condition on the lattice
points \cite{Manschot:2010nc}. One can show however that for $J=J_{1,0}$ the
quadratic condition disappears from the generating function due to a
special symmetry of the lattice $A$, and therefore one again obtains
quasi-modular forms at this point.\footnote{We thank S. Zwegers for
  providing this argument.}

\appendix

\sectiono{Toric data for the elliptic hypersurfaces}\label{app:data} 
 
Here we collect the toric data necessary to treat all models discussed.  
We list the Mori cones in the star triangulation for the bases of  
model 8-15 of figure 1   
 
\begin{equation*} 
 \label{dataf1} 
 \footnotesize 
\begin{array}{c|rrrr|rrrrr|rrrr|rrrrrr| } 
   \Delta_B& \multicolumn{4}{c}{8(4)} & \multicolumn{5}{c}{9(4)} & \multicolumn{4}{c}{10(4)} & \multicolumn{6}{c}{11(5)}\\ 
   \nu^B_i &l^{(1)}&l^{(2)}&l^{(3)}&l^{(4)}&  l^{(1)}& l^{(1)}&l^{(2)}& l^{(1)}& l^{(2)}&  l^{(1)}&l^{(2)}&l^{(3)}&l^{(4)}&   l^{(1)}&l^{(2)}&l^{(3)}&l^{(4)}&l^{(5)}&l^{(6)}  \\ 
    z      &0&-1&0&-1    &-1&-1 &0 &-1&-1    &0 &-1&0&0    & 0&-1& 0& 0&-1&-1 \\    
    1      &0& 0&0&1     &-1& 1 &0 &0 & 0    &1 &0 &0 &0   & 1& 0& 0& 0&0 &1  \\        
    2      &1& 0&0&0     &1 &-1 &1 &0 & 0    &-2&1 &0 &0   &-2& 1& 0& 0&0 &0 \\        
    3      &-2&1&0&0     &0 &1  &-2&1 &0     &1 &-1&1 &0   & 1&-1& 1& 0&0 &0 \\        
    4      & 1&-1&1&0    &0 &0  &1 &-1&1     &0 &1 &-2&1  & 0& 1&-2& 1&0 &0  \\        
    5      & 0&1&-2&1    &0 &0  &0 &1 &-1    &0 &0 & 1&-2   & 0& 0& 1&-2&1 &0\\     
    6      & 0&0& 1&-1   &1& 0  &0 &0 & 1    &0 &0 & 0& 1   & 0& 0& 0& 1&-1&1\\     
           &  &  & &     & &    & &  &       &  & &  &     & 0& 0& 0& 0&1 &-1\\     
    ex     &  &  &7&     & &    & &  12&     &  & & 4&      &&&&&16&\\   
 
 \end{array}    
\end{equation*} 
 
%% 
%[0, 1, -2, 1, 0, 0, 0] 
%[0, 0, 1, -1, 1, 0, -1] 
%[0, 0, 0, 1, -2, 1, 0],  
%[1, 0, 0, 0, 1, -1, -1],  
%%  
%[-1, 1, 0, 0, 0, 1, -1], 
%[1, -1, 1, 0, 0, 0, -1] 
%[0, 1, -2, 1, 0, 0, 0] 
%[0, 0, 1, -1, 1, 0, -1] 
%[0, 0, 0, 1, -1, 1, -1] 
%% 
%[1, -2, 1, 0, 0, 0, 0],  
%[0, 1, -1, 1, 0, 0, -1] 
%[0, 0, 1, -2, 1, 0, 0],  
%[[0, 0, 0, 1, -2, 1, 0],  
%%[ 
%[1, -2, 1, 0, 0, 0, 0, 0],  
%[0, 1, -1, 1, 0, 0, 0, -1], 
%[0, 0, 1, -2, 1, 0, 0, 0],  
%[0, 0, 0, 1, -2, 1, 0, 0]] 
%[0, 0, 0, 0, 1, -1, 1, -1],  
%[1, 0, 0, 0, 0, 1, -1, -1],   

\begin{equation*} 
 \label{dataf1} 
 \footnotesize 
\begin{array}{c|rrrrrrr|rrrrrrr|rrrrrrrr| } 
   \Delta_B &  \multicolumn{7}{c}{12(5)} & \multicolumn{7}{c}{13(6)}  &  \multicolumn{8}{c}{14(6)} \\ 
   \nu^B_i \!&\!l^{(1)}\!&\!l^{(2)}\!&\!l^{(3)}\!&\!l^{(4)}\!&\! l^{(5)}\!&\!l^{(6)}\!&\! l^{(7)}\!&\! l^{(1)}\!&\!l^{(2)}\!&\!l^{(3)}\!&\!l^{(4)}\!&\! l^{(5)}\!&\!l^{(6)}\!&\! l^{(7)}\!&\!l^{(1)}\!&\!l^{(2)}\!&\!l^{(3)}\!&\!l^{(4)}\!&\! l^{(5)}\!&\!l^{(6)}\!&\! l^{(7)}\!&\!l^{(8)}\\ 
    z      \!&\!-1\!&\!-1 \!&\!-1\!&\!0 \!&\! -1 \!&\!0  \!&\!-1     \!&\!0 \!&\!-1 \!&\!0 \!&\! 0 \!&\!0\!&\!-1 \!&\!0   \!&\!-1\!&\! 0\!&\!-1\!&\!0 \!&\!0\!&\!-1 \!&\!0 \!&\!-1\\    
    1      \!&\!1 \!&\! 1 \!&\!0 \!&\!0 \!&\! 0  \!&\!0  \!&\!0      \!&\!-2\!&\! 1 \!&\!0 \!&\! 0 \!&\!0 \!&\!0 \!&\!0   \!&\!-1\!&\! 1\!&\! 0\!&\!0 \!&\!0\!&\! 0 \!&\!0 \!&\!1 \\        
    2      \!&\!0 \!&\!-1 \!&\!1 \!&\!0 \!&\! 0 \! &\!0 \!&\!0       \!&\! 1\!&\!-1 \!&\!1 \!&\! 0 \!&\!0 \!&\!0 \!&\!0   \!&\! 1\!&\!-2\!&\! 1\!&\!0 \!&\!0\!&\! 0 \!&\!0 \!&\! 0\\        
    3      \!&\!0 \!&\! 1 \!&\!-1\!&\!1 \!&\! 0  \!&\!0  \!&\!0      \!&\! 0\!&\! 1 \!&\!-2\!&\! 1 \!&\!0 \!&\!0 \!&\!0   \!&\! 0\!&\! 1\!&\!-1\!&\!1 \!&\!0\!&\! 0 \!&\!0 \!&\! 0\\        
    4      \!&\!0 \!&\! 0 \!&\!1 \!&\!-2 \!&\!1  \!&\!0 \!&\!0       \!&\! 0\!&\! 0 \!&\!1 \!&\!-2 \!&\!1 \!&\!0 \!&\! 0   \!&\! 0\!&\! 0\!&\! 1\!&\!-2\!&\!1\!&\! 0 \!&\!0 \!&\! 0\\        
    5      \!&\!0 \!&\! 0 \!&\! 0\!&\!1 \!&\! -1 \!&\!1  \!&\!0      \!&\! 0\!&\! 0 \!&\!0 \!&\! 1 \!&\!-2\!&\!1\!&\! 0   \!&\! 0\!&\! 0\!&\! 0\!&\!1 \!&\!-1\!&\! 1 \!&\!0\!&\! 0\\     
    6      \!&\!1 \!&\!0  \!&\! 0\!&\! 0\!&\! 1  \!&\!-2 \!&\!-2     \!&\! 0\!&\! 0 \!&\!0 \!&\! 0 \!&\!1 \!&\!-1\!&\! 1   \!&\! 0\!&\! 0\!&\! 0\!&\!0 \!&\!1\!&\!-1\!&\!1  \!&\! 0\\     
    7      \!&\!-1\!&\!0  \!&\! 0\!&\! 0\!&\! 0 \!&\! 1 \!&\!1       \!&\! 0\!&\! 0 \!&\!0 \!&\! 0 \!&\!0 \!&\!1 \!&\! -2   \!&\! 0\!&\! 0\!&\! 0\!&\!0 \!&\!0\!&\! 1\!&\!-2 \!&\! 1\\     
    8      \!&\!  \!&\!  \!&\!  \!&\!  \!&\!   \!&\!   \!&\!         \!&\! 1\!&\! 0 \!&\!0 \!&\! 0 \!&\!0 \!&\!0 \!&\! 1    \!&\! 1\!&\! 0\!&\! 0\!&\!0 \!&\!0\!&\! 0\!&\! 1 \!&\!-1 \\   
    ex     \!&\!  \!&\!  \!&\!  \!&\!  \!&\! \!&\!   29 \!&\!        \!&\!  \!&\!   \!&\!  \!&\!   \!&\!  \!&\! 20\!&\!    \!&\!  \!&\!  \!&\!  \!&\! \!&\!\!&\! \!&\! 43 \!&\!  \\   
 \end{array}    
\end{equation*} 
 
%[-1, 1, 0, 0, 0, 0, 1, -1],  
%[1, -1, 1, 0, 0, 0, 0, -1], 
%[[0, 1, -1, 1, 0, 0, 0, -1],  
%[0, 0, 1, -2, 1, 0, 0, 0], 
%[0, 0, 0, 1, -1, 1, 0, -1]] 
%[0, 0, 0, 0, 1, -2, 1, 0],  
%[1, 0, 0, 0, 0, 1, -1, -1],  
%29 
%[-2, 1, 0,0, 0, 0, 0, 1, 0],  
%[1, -1, 1, 0, 0, 0, 0, 0, -1],  
%[0, 1, -2, 1, 0, 0, 0, 0, 0],  
%[0, 0, 1, -2, 1, 0, 0, 0, 0],  
%[0, 0, 0, 1, -2, 1, 0, 0, 0]] 
%[0, 0, 0, 0, 1, -1, 1, 0, -1],  
%[0, 0, 0, 0, 0, 1, -2, 1, 0], 
%20 
%[-1, 1, 0, 0, 0, 0, 0, 1, -1],  
%[1, -2, 1, 0, 0, 0, 0, 0, 0],  
%[0, 1, -1, 1, 0, 0, 0, 0, -1],  
%[0, 0, 1, -2, 1, 0, 0, 0, 0],  
%[0, 0, 0, 1, -2, 1, 0, 0, 0]] 
%[0, 0, 0, 0, 1, -1, 1, 0, -1],  
%[[0, 0, 0, 0, 0, 1, -2, 1, 0],  
%[1, 0, 0, 0, 0, 0, 1, -1, -1],  
% 43 

\begin{equation*} 
 \label{dataf1} 
 \footnotesize 
\begin{array}{c|rrrrrrrr|rrrrrrrrr| } 
   \Delta_B &  \multicolumn{8}{c}{15(5)} & \multicolumn{9}{c}{16(7)}    \\ 
   \nu^B_i \!&\!l^{(1)}\!&\!l^{(2)}\!&\!l^{(3)}\!&\!l^{(4)}\!&\! l^{(5)}\!&\!l^{(6)}\!&\! l^{(7)}&\!l^{(8)}\!&\! l^{(1)}\!&\!l^{(2)}\!&\!l^{(3)}\!&\!l^{(4)}\!&\! l^{(5)}\!&\!l^{(6)}\!&\! l^{(7)}\!&\!l^{(8)}\!&\!l^{(9)}\\ 
    z      \!&\!0 \!&\!-1 \!&\!0\!&\!-1 \!&\! 0 \!&\!-1  \!&\!0  \!&\!-1     \!&\! 0 \!&\!0 \!&\!-1 \!&\! 0 \!&\!0\!&\!-1 \!&\!0  \!&\!0 \!&\!-1     \\    
    1      \!&\!-2\!&\! 1 \!&\!0 \!&\!0 \!&\! 0  \!&\!0  \!&\!0 \!&\!1       \!&\!-2\!&\! 1 \!&\!0 \!&\! 0 \!&\!0 \!&\!0 \!&\!0  \!&\!0 \!&\! 1  \\        
    2      \!&\!1 \!&\!-1 \!&\!1 \!&\!0 \!&\! 0 \! &\!0 \!&\!0  \!&\!0        \!&\!1\!&\!-2 \!&\!1 \!&\! 0 \!&\!0 \!&\!0 \!&\!0  \!&\!0 \!&\! 0  \\        
    3      \!&\!0 \!&\! 1 \!&\!-2\!&\!1 \!&\! 0  \!&\!0  \!&\!0 \!&\!0        \!&\ 0\!&\! 1 \!&\!-1\!&\! 1 \!&\!0 \!&\!0 \!&\!0  \!&\!0  \!&\!0 \\        
    4      \!&\!0 \!&\! 0 \!&\!1 \!&\!-1 \!&\!1  \!&\!0 \!&\!0  \!&\!0        \!&\!0\!&\! 0 \!&\!1 \!&\!-2 \!&\!1 \!&\!0 \!&\! 0  \!&\!0  \!&\!0 \\        
    5      \!&\!0 \!&\! 0 \!&\! 0\!&\!1 \!&\! -2 \!&\!1  \!&\!0 \!&\!0       \!&\! 0\!&\! 0 \!&\!0 \!&\! 1 \!&\!-2\!&\!1\!&\! 0   \!&\!0 \!&\! 0\\     
    6      \!&\!0 \!&\! 0  \!&\! 0\!&\! 0\!&\!1  \!&\!-1 \!&\!1 \!&\!0       \!&\! 0\!&\! 0 \!&\!0 \!&\! 0 \!&\!1 \!&\!-1\!&\! 1   \!&\!0 \!&\!0 \\     
    7      \!&\!0\!&\!  0  \!&\! 0\!&\! 0\!&\!0 \!&\! 1 \!&\! -2  \!&\!1     \!&\!0\!&\! 0 \!&\!0 \!&\! 0 \!&\!0 \!&\!1 \!&\! -2 \!&\!1 \!&\!0 \\     
    8      \!&\!1\!&\!  0 \!&\! 0 \!&\! 0 \!&\!0 \!&\! 0 \!&\! 1    \!&\!-1   \!&\!0\!&\! 0 \!&\!0 \!&\! 0 \!&\!0 \!&\!0 \!&\! 1   \!&\!-2 \!&\!1   \\  
    9     \!&\! \!&\!    \!&\!   \!&\!  \!&\!   \!&\!   \!&\!     \!&\!     \!&\! 1\!&\!  0 \!&\!0  \!&\!0 \!&\!0\!&\!0\!&\! 0  \!&\!1   \!&\!-1  \\   
    ex     \!&\! \!&\!    \!&\!   \!&\!  \!&\!   \!&\!   \!&\!     \!&\! 53   \!&\!  \!&\!   \!&\!  \!&\!   \!&\!  \!&\! \!&\!   \!&\!  59 \!&\!\\   
 \end{array}    
\end{equation*}

% 
%[-2, 1, 0, 0, 0, 0, 0, 1, 0],  
%[1, -1, 1, 0, 0, 0, 0, 0, -1], 
%[0, 1, -2, 1, 0, 0, 0, 0, 0],  
%[0, 0, 1, -1, 1, 0, 0, 0, -1],  
%[0, 0, 0, 1, -2, 1, 0, 0, 0]] 
%[0, 0, 0, 0, 1, -1, 1, 0, -1],  
%[[0, 0, 0, 0, 0, 1, -2, 1, 0],  
%[1, 0, 0, 0, 0, 0, 1, -1, -1],  
%53 
%[-2, 1, 0, 0, 0, 0, 0, 0, 1, 0],  
%[1, -2, 1, 0, 0, 0, 0, 0, 0, 0],  
%0, 1, -1, 1, 0, 0, 0, 0, 0, -1],  
%[0, 0, 1, -2, 1, 0, 0, 0, 0, 0]] 
%[0, 0, 0, 1, -2, 1, 0, 0, 0, 0] 
%[0, 0, 0, 0, 1, -1, 1, 0, 0, -1],  
%[0, 0, 0, 0, 0, 1, -2, 1, 0, 0],  
%[0, 0, 0, 0, 0, 0, 1, -2, 1, 0],  
%[1, 0, 0, 0, 0, 0, 0, 1, -1, -1],  
%59 
 
The simplicial mori cone for the model 15 and 16 occur e.g. for the triangulation depicted here 
\begin{figure}[htb] 
\begin{center} 
\includegraphics[width=.4\textwidth]{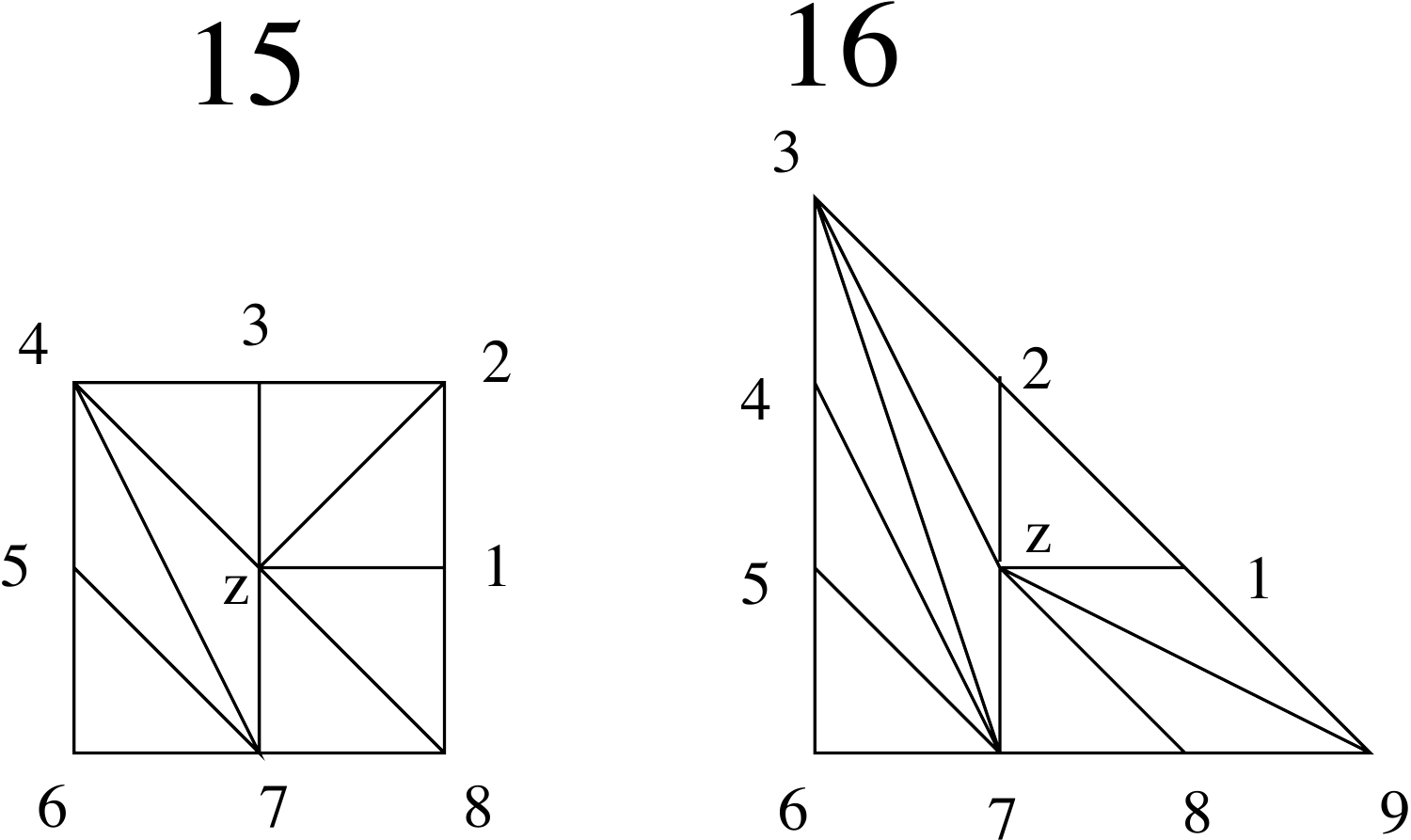} 
\begin{quote} 
\caption{Nonstar  triangulations of the basis of model 15 and 16, which lead to  
simplicial K\"ahler cone for the Calabi-Yau space \vspace{-1.2cm}} \label{flopandblowdown} 
\end{quote} 
\end{center} 
\end{figure} 
 
For the model 15 the moricone reads  
\begin{equation}  
\begin{split} 
l^{(e)}&=(-6, 0, 0, 0, 0, 1, -1, 1, 0, 0, 2, 3), \quad  l^{(1)}=( 0, -2, 1, 0, 0, 0, 0, 0, 1, 0, 0, 0)\\   
l^{(2)}&=(0, 1, -1, 1, 0, 0, 0, 0, 0, -1, 0, 0),\quad  l^{(3)}= (0, 0, 1, -2, 1, 0, 0, 0, 0, 0, 0, 0), \\    
l^{(4)}&=(0, 0, 0, 1, -1,1, 0, 0, 0, -1, 0, 0),  \quad  l^{(5)}=(0, 0, 0, 0, 0, -1, 0, 1, -1, 1, 0, 0), \\   
l^{(6)}&= , 0, 0, 0, 0, 0, 0, 1, -2, 1, 0, 0, 0)\\   
\end{split} 
\end{equation} 
 
This  yields the intersection numbers  
\begin{equation} 
\begin{split} 
{\cal R}=& 
4 J_e^3 + 2 J_e^2 J_2 + 4 J_e^2 J_3 + J_e J_2 J_3 + 2 J_e J_3^2 + 3 J_e^2 J_4 + J_e J_2 J_4 + 2 J_e J_3 J_4 + J_e J_4^2  
+\\ &  2 J_e^2 J_5 + J_e J_2 J_5 + 2 J_e J_3 J_5 + J_e J_4 J_5 + 6 J_e^2 J_6 + 2 J_e J_2 J_6 + 4 J_e J_3 J_6 + J_2 J_3 J_6  
+ \\ & 2 J_3^2 J_6 + 3 J_e J_4 J_6 + J_2 J_4 J_6 + 2 J_3 J_4 J_6 + J_4^2 J_6 + 2 J_e J_5 J_6 + J_2 J_5 J_6 + 2 J_3 J_5 J_6  
+ \\ & J_4 J_5 J_6 + 6 J_e J_6^2 + 2 J_2 J_6^2 + 4 J_3 J_6^2 + 3 J_4 J_6^2 + 2 J_5 J_6^2 + 6 J_6^3 + 5 J_e^2 J_7 + 2 J_e J_2 J_7  
+\\ &  4 J_e J_3 J_7 + J_2 J_3 J_7 +  2 J_3^2 J_7 + 3 J_e J_4 J_7 + J_2 J_4 J_7 + 2 J_3 J_4 J_7 + J_4^2 J_7 + 2 J_e J_5 J_7  
+\\ &  J_2 J_5 J_7 +  2 J_3 J_5 J_7 + J_4 J_5 J_7 + 6 J_e J_6 J_7 + 2 J_2 J_6 J_7 + 4 J_3 J_6 J_7 + 3 J_4 J_6 J_7 + 2 J_5 J_6 J_7  
+ \\ & 6 J_6^2 J_7 + 5 J_e J_7^2 + 2 J_2 J_7^2 +  4 J_3 J_7^2 + 3 J_4 J_7^2 + 2 J_5 J_7^2 + 6 J_6 J_7^2 +  5 J_7^3 
\end{split} 
\end{equation} 
and  the evaluation of $c_2$ on the basis $J_i$  
\begin{equation} 
\begin{split}  
c_2J_e&=52,\quad c_2J_1=24,\quad c_2J_2=48, \quad c_2J_3=36,\\ 
c_2J_4&=24,\quad c_2J_5=72,\quad c_2J_6=62.\\ 
\end{split} 
\end{equation}

The same data for the model 16  
\begin{equation}  
\begin{split} 
l^{(e)}&=(-6, 0, 0, 0, 0, 1, -1, 1, 0, 0, 0, 2, 3), \quad  l^{(1)}=(0, -2, 1, 0, 0, 0, 0, 0, 0, 1, 0, 0, 0), \\   
l^{(2)}&=(0, 1, -2, 1, 0, 0, 0, 0, 0, 0, 0, 0, 0),  \quad  l^{(3)}=(0, 0, 0, -1, 1, 0, 0, -1, 0, 0, 1, 0, 0),\\    
l^{(4)}&=(0, 0, 0, 1, -2, 1, 0, 0, 0, 0, 0, 0, 0), \quad  l^{(5)}=(0, 0, 0, 0, 1, -2, 1, 0, 0, 0, 0, 0, 0), \\   
l^{(6)}&=(0, 0, 0, 0, 0, 0, 0, 1, -2, 1, 0, 0, 0),  \quad  l^{(7)}=(0, 1, 0, 0, 0, 0, 0, 0, 1, -1, -1, 0, 0),\\   
\end{split} 
\end{equation} 
and the intersection  by  
\begin{equation}  
\begin{split} 
{\cal R}=& 
3 J_e^3 + 4 J_e^2 J_2 + 2 J_e J_2^2 + 2 J_e^2 J_3 + J_e J_2 J_3 + 6 J_e^2 J_4 + 4 J_e J_2 J_4 + 2 J_2^2 J_4 + 2 J_e J_3 J_4  
+\\ & J_2 J_3 J_4 + 6 J_e J_4^2 + 4 J_2 J_4^2 + 2 J_3 J_4^2 + 6 J_4^3 + 5 J_e^2 J_5 + 4 J_e J_2 J_5 + 2 J_2^2 J_5 + 2 J_e J_3 J_5  
+ \\ &J_2 J_3 J_5 +  6 J_e J_4 J_5 + 4 J_2 J_4 J_5 + 2 J_3 J_4 J_5 + 6 J_4^2 J_5 + 5 J_e J_5^2 + 4 J_2 J_5^2 + 2 J_3 J_5^2 + 6 J_4 J_5^2  
+ \\ &5 J_5^3 + 4 J_e^2 J_6 + 4 J_e J_2 J_6 + 2 J_2^2 J_6 + 2 J_e J_3 J_6 + J_2 J_3 J_6 + 6 J_e J_4 J_6 + 4 J_2 J_4 J_6 + 2 J_3 J_4 J_6  
+ \\ & 6 J_4^2 J_6 + 5 J_e J_5 J_6 + 4 J_2 J_5 J_6 + 2 J_3 J_5 J_6 + 6 J_4 J_5 J_6 + 5 J_5^2 J_6 + 4 J_e J_6^2 + 4 J_2 J_6^2 + 2 J_3 J_6^2 +  
\end{split} 
\end{equation}  
\begin{equation}  
\begin{split} 
\phantom{{\cal R}=}&\\ &6 J_4 J_6^2 +  5 J_5 J_6^2 + 4 J_6^3 + 3 J_e^2 J_7 + 2 J_e J_2 J_7 + J_e J_3 J_7 + 3 J_e J_4 J_7 + 2 J_2 J_4 J_7 + J_3 J_4 J_7  
+ \\ &3 J_4^2 J_7 + 3 J_e J_5 J_7 +  2 J_2 J_5 J_7 + J_3 J_5 J_7 + 3 J_4 J_5 J_7 + 3 J_5^2 J_7 + 3 J_e J_6 J_7 + 2 J_2 J_6 J_7 + J_3 J_6 J_7  
+ \\ &3 J_4 J_6 J_7 + 3 J_5 J_6 J_7 +  3 J_6^2 J_7 + J_e J_7^2 + J_4 J_7^2 + J_5 J_7^2 + J_6 J_7^2 + 6 J_e^2 J_8 + 4 J_e J_2 J_8  
+ \\ &2 J_e J_3 J_8 + 6 J_e J_4 J_8 + 4 J_2 J_4 J_8 + 2 J_3 J_4 J_8 + 6 J_4^2 J_8 + 6 J_e J_5 J_8 + 4 J_2 J_5 J_8 + 2 J_3 J_5 J_8  
+ \\ &6 J_4 J_5 J_8 + 6 J_5^2 J_8 +  6 J_e J_6 J_8 + 4 J_2 J_6 J_8 + 2 J_3 J_6 J_8 + 6 J_4 J_6 J_8 + 6 J_5 J_6 J_8 + 6 J_6^2 J_8 +  
3 \\ &J_e J_7 J_8 + 3 J_4 J_7 J_8 + 3 J_5 J_7 J_8 +  3 J_6 J_7 J_8 + 6 J_e J_8^2 + 6 J_4 J_8^2 + 6 J_5 J_8^2 +  6 J_6 J_8^2 
\end{split} 
\end{equation}  
and the evaluation on $c_2$ is 
\begin{equation} 
\begin{split}  
c_2J_e&=42,\quad c_2J_1=48,\quad c_2J_2=24, \quad c_2J_3=72,\\ 
c_2J_4&=62,\quad c_2J_5=52,\quad c_2J_6=36, \quad c_2J_7=72.\\ 
\end{split} 
\end{equation}

\sectiono{Results for the other fibre types with $\mathbb{F}_1$ base} 
\label{app:fibref1} 
We give some results of the periods for the different fibre types with base $\mathbb{F}_1$. The corresponding Picard-Fuchs operators read \cite{Lian-Yau} 
\begin{equation} 
\begin{split} 
{\cal L}_{E7} &= \theta^2-4z(4\theta +3)(4 \theta+1), \\ 
{\cal L}_{E6} &= \theta^2-3z(3\theta +2)(3 \theta+1)\\ 
{\cal L}_{D5} &= \theta^2-4z(2\theta +1)^2 
\end{split} 
\end{equation} 
The solutions read as follows 
\begin{equation} 
\label{eq:periodexps}
\begin{split} 
\phi_{E7} &= \sum_{n \geq 0}^\infty \frac{(4n)! }{(n!)^2 (2n)!} z^n =\,  _2F_1(\frac{3}{4}, \frac{1}{4}, 1, 64 z),\\ 
\phi_{E6} &= \sum_{n \geq 0}^\infty \frac{(3n)! }{(n!)^3} z^n=\,  _2F_1(\frac{2}{3}, \frac{1}{3}, 1, 27 z),\\ 
\phi_{D5} &= \sum_{n \geq 0}^\infty \frac{(2n)!^2 }{(n!)^4} z^n= \,  _2F_1(\frac{1}{2}, \frac{1}{2}, 1, 16 z), 
\end{split} 
\end{equation} 
with:
\begin{equation} 
{}_2 F_1(a,b,c;x)=\sum_{n=0}^\infty\frac{(a)_n(b)_n}{(c)_n}\frac{x^n}{n!}, 
\end{equation} 
where $(a)_n=a(a+1)\dots(a+n-1)$ denotes the Pochhammer symbol.

The $j$-functions read for these read 
\begin{equation} 
\begin{split} 
1728 j_{E7} &= \frac{(1+192 z)^3}{z(1-64 z)^2} \\ 
1728 j_{E6} &= \frac{(1+216 z)^3}{z(1-27 z)^3}\\ 
1728 j_{D5} &= \frac{(1+244 z+256 z^2)}{z(-1+16z)^4} 
\end{split} 
\end{equation} 
We collect the expressions for the solutions in terms of modular forms 
\begin{equation} 
\begin{split} 
\phi_{E7} (z(q))^2&=1+24 q+24 q^2+96 q^3 +\dots  = -E_2(\tau) + 2 E_2(2 \tau)\\ 
\phi_{E6} (z(q))&= 1+6 q+6 q^3+ \dots = \sum_{m,n \in \mathbb{Z}} q^{m^2 + n^2 + m n} = \theta_2(\tau) \theta_2(3\tau) +  \theta_3(\tau) \theta_3(3\tau)\\ 
\phi_{D5} (z(q))&= 1+4q+4q^2+\dots=\theta_3(2\tau)^2 
\end{split} 
\end{equation} 
Following analogous steps presented in section \ref{sec:bmodel}, one can again proof the holomorphic anomaly equation for genus 0.

\section{Modular functions} 
\label{app:modfunctions} 
 
This appendix lists various modular functions, which appear in the 
generating functions in the main text. Define $q:=e^{2\pi i \tau}$, 
$w:= e^{2\pi i z}$, with $\tau \in \mathbb{H}$ and $z\in \mathbb{C}$.  
The Dedekind eta and Jacobi theta functions are defined by:  
\begin{eqnarray} 
\label{eq:etatheta} 
&&\eta(\tau)\quad \,\,:=q^{\frac{1}{24}}\prod_{n=1}^\infty (1-q^n),\non\\ 
&&\theta_1(z,\tau):=i\sum_{r\in \mathbb{Z}+\frac{1}{2}} (-1)^{r-\frac{1}{2}} q^{\frac{r^2}{2}}w^{r},\\ 
&&\theta_2(z,\tau):=\sum_{r\in \mathbb{Z}+\frac{1}{2}} q^{r^2/2}w^{r},\non\\ 
&&\theta_3(z,\tau):=\sum_{n\in\mathbb{Z}}q^{n^2/2}w^n\non. 
\end{eqnarray} 

We define the indefinite theta function $F(\tau,u,v)$ for $0<-\im\, 
u/\im\, \tau <1$ and $0<\im\, 
v/\im\, \tau <1$\cite{Gottsche:1996} 
\begin{eqnarray} 
\label{eq:F} 
F(\tau,u,v)&=&\sum_{n\geq 0, m>0}q^{mn}e^{2\pi i u n+2\pi i v 
  m}-\sum_{n> 0, m\geq 0} q^{mn}e^{-2\pi i u n-2\pi i v m} \\ 
&=&\sum_{n\geq 0,m>0}-\sum_{n<0,m\leq 0}q^{nm}e^{2\pi i u n+2\pi i v 
  m}. \non 
\end{eqnarray} 
Analytic extension of this function gives: 
\be 
F(\tau,u,v)=-i \frac{\eta(\tau)^3\,\theta_1(\tau,u+v)}{\theta_1(\tau,u)\,\theta_1(\tau,v)}. 
\ee

\providecommand{\href}[2]{#2}\begingroup\raggedright\endgroup

\end{document}